\documentclass[10pt,aps,preprintnumbers,prd,noshowpacs,nofootinbib,noshowkeys,floatfix,superscriptaddress]{revtex4}
\usepackage[dvips]{graphics,graphicx}
\usepackage[colorlinks=true,linktocpage=true,linkcolor=blue,citecolor=blue]{hyperref}
\usepackage[usenames,dvipsnames]{color}
\usepackage{amsmath, amssymb,oldgerm}
\usepackage{multirow}
\usepackage{xcolor}
\usepackage[normalem]{ulem}  
\usepackage{braket}
\usepackage{slashed}
\usepackage[mathscr]{eucal}
\usepackage[dvips]{graphics,graphicx}
 \usepackage{caption}
\usepackage{subcaption}
\usepackage[version=4]{mhchem}

\newcommand{\n}{\nonumber}

\newcommand{\Pc}{\mathcal{P}}

\newcommand{\mC}{\mathfrak{C}}

\newcommand{\ms}{\mathfrak{s}}

\newcommand{\beq}{\begin{equation}}
\newcommand{\eeq}{\end{equation}}

\newcommand{\eqs}{Eqs.~}
\newcommand{\eq}{Eq.~}

\newcommand{\proj}{\Delta}

\newcommand{\bbb}{\textcolor{black}}

\newcommand{\p}{\mathbf{p}}

\newcommand{\mJ}{\mathcal{J}}

\newcommand{\mG}{\mathcal{G}}
\newcommand{\mI}{\mathcal{I}}

\newcommand{\bxi}{\bar{\xi}}

\begin{document}

\title{Spin polarization of an expanding and rotating system}

\author{Nora Weickgenannt}

\affiliation{Institut de Physique Th\'eorique, Universit\'e Paris Saclay, CEA, CNRS, F-91191 Gif-sur-Yvette, France}

\author{Jean-Paul Blaizot}

\affiliation{Institut de Physique Th\'eorique, Universit\'e Paris Saclay, CEA, CNRS, F-91191 Gif-sur-Yvette, France}

\begin{abstract}
We study the longitudinal spin polarization of a relativistic fluid of massive spin-1/2 particles undergoing a boost-invariant expansion in the longitudinal direction and rotating in the transverse plane. We express the polarization vector in terms of spin moments and  derive closed equations of motion for the latter using  spin kinetic theory with a nonlocal relaxation time approximation. These equations of motion are valid at any time of the evolution, from the free-streaming regime to the hydrodynamic regime. At late time, the polarization features contributions from gradients of the fluid velocity and of the temperature, that emerge from the nonlocal part of the collision term. Our results can be used to explore polarization phenomena in the context of heavy-ion collisions.
\end{abstract}

\maketitle

\section{Introduction}

Spin polarization of relativistic fluids attracted a lot of attention during the past years. Mainly motivated by the observation of 
 polarization phenomena in heavy-ion collisions~\cite{Becattini:2013vja,Becattini:2013fla,Becattini:2015ska,Becattini:2016gvu,Karpenko:2016jyx,Pang:2016igs,Xie:2017upb,STAR:2017ckg,Adam:2018ivw,ALICE:2019aid,STAR:2019erd}, recent efforts include studies of relativistic fluids with spin, e.g., thermodynamic calculations~\cite{Liu:2021uhn,Fu:2021pok,Becattini:2021suc,Becattini:2021iol}, spin kinetic theory~\cite{Weickgenannt:2019dks,Gao:2019znl,Hattori:2019ahi,Wang:2019moi,Weickgenannt:2020aaf,Liu:2020flb,Weickgenannt:2021cuo,Sheng:2021kfc,Wagner:2022amr,Wagner:2023cct,Weickgenannt:2024ibf,Wang:2024lis}, and spin hydrodynamics~\cite{Florkowski:2017ruc,Florkowski:2017dyn,Florkowski:2018fap,Montenegro:2018bcf,Hattori:2019lfp,Bhadury:2020puc,Singh:2020rht,Montenegro:2020paq,Gallegos:2021bzp,Fukushima:2020ucl,Li:2020eon,Wang:2021ngp,Hu:2021pwh,Hongo:2021ona,Daher:2022xon,Weickgenannt:2022zxs,Weickgenannt:2022jes,Gallegos:2022jow,Cao:2022aku,Weickgenannt:2022qvh,Biswas:2023qsw,Weickgenannt:2023btk,Weickgenannt:2023bss,Daher:2024bah,Wagner:2024fhf,Drogosz:2024gzv,Wagner:2024fry,Drogosz:2024lkx}. While different approaches commonly agree on the fact  that gradients of the fluid velocity and of the temperature can generate spin polarization~\cite{Liang:2004ph,Voloshin:2004ha,Betz:2007kg,Becattini:2007sr,Fu:2021pok,Becattini:2021iol,Weickgenannt:2020aaf,Wang:2024lis}, the dynamics of this process is not yet well understood. In particular, open questions remain in the description of the longitudinal Lambda polarization as a function of momentum. Although recent progress has been reported in Refs.~\cite{Liu:2021uhn,Fu:2021pok,Becattini:2021suc,Becattini:2021iol}, those works include only ideal contributions to the polarization, while neglecting dissipative terms. A priory, it is not clear how to order these dissipative contibutions in terms of a gradient expansion, while a consistent treatment should take into account all the terms that are of the same order in the gradients. A useful tool to address this issue is spin kinetic theory, which has the advantage over thermodynamic or hydrodynamic approaches to be valid even away from equilibrium,  and to be independent of the so-called pseudo-gauge ambiguity. However, solving the Boltzmann equation exactly is highly challenging. It is therefore desirable to obtain a set of simpler equations of motion that capture the crucial dynamics of the exact kinetic theory, while being more tractable than the Boltzmann equation. It is the goal of this paper to provide such equations for an expanding and rotating system.

In Ref.~\cite{Weickgenannt:2023bss} we have studied the transverse spin polarization of a boost invariant expanding system using a standard relaxation time approximation. We have shown that, even in the absence of vorticity, the polarization at freeze-out may be nonzero, since parts of an initial polarization can survive during the full evolution of the system. In the present work, we extend the efforts of Ref.~\cite{Weickgenannt:2023bss} in several aspects. In particular, we relax the restriction of translational invariance in the transverse plane, and impose only rotational symmetry. Thus, in the present paper, the fluid vorticity around the $z$-axis (the longitudinal direction) is in general nonzero. Since this vorticity is expected to generate a spin polarization in the longitudinal direction, we concentrate on the longitudinal polarization in our discussion, and make no additional assumptions for the polarization in other directions. The starting point of our analysis is kinetic theory with spin. It is known that to account for the conversion between orbital angular momentum and spin, a nonlocal collision term has to be considered~\cite{Weickgenannt:2020aaf}. For this reason, we use the recently developed nonlocal relaxation time approximation (NLRTA)~\cite{Weickgenannt:2024ibf} to model the collision term. 

Analogously to Ref.~\cite{Weickgenannt:2023bss}, we express the polarization in terms of so-called spin moments and derive equations of motion for the latter from the Boltzmann equation. In the next step, we close the equations of motion employing established methods to take into account the dynamics both in the early-time free-streaming regime and the late-time hydrodynamic regime~\cite{Jaiswal:2022udf,Weickgenannt:2023nge}. Here, the spin moments do not converge to their local-equilibrium values in the late-time limit due to the nonlocality of the collision term. Instead, they approach the so-called asymptotic spin moments, which consist of local-equilibrium parts and nonlocal parts,  the latter depending on the gradients of the fluid velocity. Through these terms, the polarization acquires contributions from the thermal vorticity and the thermal shear, respectively. Furthermore, the asymptotic spin moments depend on the spin potential, which is the Lagrange multiplier conjugate to the dipole-moment tensor in the local-equilibrium distribution function. Through the matching condition, the spin potential is expressed as a function of the total angular momentum, which we treat as an additional dynamical variable following its own equations of motion.

A priori, the polarization vector is a sum over an infinite number of spin moments. To truncate this sum, we note that a measurement of the polarization in heavy-ion collisions will take place at freeze-out. Therefore, it is sufficient to take into account the spin moments which are most relevant at late time, and truncate the sum by neglecting all faster decaying spin moments. On the other hand, the early-time dynamics of the relevant spin moments may leave an imprint on their values at late time. For this reason, we take into account the behavior of the dynamical variables both at early and late time when closing the equations of motion. Our main result is an expression for the polarization vector as a function of the azimuthal momentum angle and a finite number of dynamical spin moments, together with the corresponding closed set of equations of motion.

This paper is organized as follows. In Sec.~\ref{boltzsec} we introduce the Boltzmann equation for an expanding and rotating system including the NLRTA. In Sec.~\ref{polsec} we express the longitudinal polarization vector as a series of spin moments. The equations of motion for the spin moments are derived from the Boltzmann equation in Sec.~\ref{eomsec}. In Sec.~\ref{asympsec} we study the asymptotic spin moments. Section \ref{tamsec} deals with the equations of motion for the total angular momentum, as well as with the relations between the latter and the asymptotic spin moments. In Sec.~\ref{clossec} we outline how to truncate the expansion of the polarization vector in spin moments and close the equations of motion. The final form of the longitudinal polarization is presented in Sec.~\ref{finsec}. We provide conclusions in Sec.~\ref{concsec}.
As usual in spin kinetic theory, we work up to first order in $\hbar$ throughout the paper. We use natural units, but keep $\hbar$ explicit in most places, since it serves as a power-counting parameter. Furthermore, we sum over repeated indices and use the following notations and conventions: $\tilde{A}^{\mu\nu}\equiv \epsilon^{\mu\nu\lambda\rho}A_{\lambda\rho}$, $a\cdot b\equiv a_\mu b^\mu$, $a_{[\mu}b_{\nu]}\equiv a_\mu b_\nu-a_\nu b_\mu$, $a_{(\mu}b_{\nu)}\equiv a_\mu b_\nu+a_\nu b_\mu$, $\mathbf{a}\cdot\mathbf{b}\equiv a^i b^i$, $\mathbf{a}_\bot\equiv (a^x,a^y,0)$,  $g_{\mu\nu}=\text{diag}(+,-,-,-)$, $\epsilon^{0123}=-\epsilon_{0123}=1$.

\section{Boltzmann equation for boost-invariant expansion and rotation}
\label{boltzsec}

We consider a fluid of massive particles with spin which expands in $z$-direction and rotates in the $x$-$y$-plane. The rotation of the fluid around the $z$-axis induces a polarization in that direction through nonlocal particle collisions. However, this polarization is also influenced by the expansion. In the following, we will derive an expression for the longitudinal polarization in terms of spin moments and obtain equations of motion for the latter. Hence, by solving these equations, the polarization can be determined as a function of time. To simplify the discussion, we assume boost invariance in longitudinal direction and  purely vortical flow in the transverse plane, i.e., the distribution function is constant along the lines of particle motion in the plane,
\begin{equation}
  \p_\bot \cdot \boldsymbol{\partial} f(x,p,\ms)=0 \; ,
  \label{vortonly}
\end{equation}
where $\mathbf{p}_\bot\equiv (p^x,p^y)$, $f$ is the distribution function and $\ms$ is the spin four-vector in phase space.
This allows us to use the free-streaming part, i.e., the left-hand side, of the Boltzmann equation from Ref.~\cite{Weickgenannt:2023bss} without modifications,
\begin{equation}
    \left(\partial_\tau-\frac{p_z}{\tau}\partial_{p_z}-\frac{p_z}{E_p^2}\frac{\mathbf{p}\cdot \boldsymbol{\ms}}{\tau}\partial_{\ms_z}\right)f(\tau,x,y,\p_\bot,p_z,\theta_\ms,\ms_z)=\mC[f]\;,
    \label{boltzbjorkcomp0}
\end{equation}
with $\tau$ being the proper time, $E_p\equiv\sqrt{\p^2+m^2}$, and $\theta_\ms$ being the polar angle of the spin three vector in cylindrical coordinates.
For the collision term $\mC[f]$, we use the nonlocal relaxation time approximation (NLRTA)~\cite{Weickgenannt:2024ibf}
\begin{equation}
    \mC[f]=-p\cdot u \frac{f-f_\text{LE}}{\tau_R}+ \xi\frac{p\cdot u}{\tau_R} \Delta^\mu p^\nu (\partial_\mu \beta_\nu+\Omega_{\mu\nu})f^{(0)}\; ,
    \label{nlrta}
\end{equation}
where the superscript $(0)$ denotes the zeroth order in $\hbar$.
Here $\tau_R$ is the relaxation time, $\xi$ is a parameter which determines the time scale of angular-momentum conversion, and
\begin{equation}
    f_\text{LE}=\frac{1}{(2\pi\hbar)^3}\exp\left(-\beta\cdot p + \frac{\hbar}{4} \Omega_{\mu\nu}\Sigma_\ms^{\mu\nu}\right)\;  \label{eqgen}
\end{equation}
is the local-equilibrium distribution function. Furthermore, we defined $\beta^\mu$ as the fluid velocity divided by the temperature, the spin potential $\Omega_{\mu\nu}$, and the dipole-moment tensor $\Sigma_\ms^{\mu\nu}\equiv -(1/m)\epsilon^{\mu\nu\alpha\beta}p_\alpha \ms_\beta$.  The nonlocality of the collision term is given by
\begin{equation}
\Delta^\mu\equiv-\frac{\hbar}{2m (E_p+m)}\epsilon^{\mu\nu\alpha\beta}p_\nu t_\alpha \ms_\beta \; , 
\label{deltamu}
\end{equation}
where $t^\mu\equiv (1,\mathbf{0})$ is the time unit vector. The first term in \eq\eqref{nlrta} drives the system to local equilibrium, defined as the state where the distribution function takes the form \eqref{eqgen} and the polarization is determined by the spin potential independently of the vorticity. On the other hand, the second term in \eq\eqref{nlrta} provides the nonlocal part of the collision term and is responsible for the contributions to the polarization from fluid gradients. In particular, it prevents the system from reaching local equilibrium unless the conditions of global equilibrium are fulfilled, i.e., $\beta^\mu$ is a Killing vector and the spin potential is equal to the thermal vorticity.
We have shown in Ref.~\cite{Weickgenannt:2024ibf} that, at late time, $\tau/\tau_R\rightarrow \infty$, the NLRTA drives the distribution function to its  asymptotic form given by
\begin{equation}
f_\infty=    f_\text{LE}+\xi\Delta^\mu p^\nu (\Omega_{\mu\nu}+\partial_\mu\beta_\nu)f_\text{LE}^{(0)}\;  \label{finfty1}
\end{equation}
instead of the local-equilibrium form \eqref{eqgen}.
We now make an additional simplification and replace $f^{(0)}$ in the nonlocal part of the collision term \eqref{nlrta} by $f^{(0)}_\text{LE}$. Since the collision term  contributes  significantly only in the hydrodynamic regime, i.e., at late time when the zeroth-order distribution function is close to its local-equilibrium form, this replacement does not change the crucial properties of the NLRTA. Equation \eqref{boltzbjorkcomp} then may be written as
\begin{equation}
    \left(\partial_\tau-\frac{p_z}{\tau}\partial_{p_z}-\frac{p_z}{E_p^2}\frac{\mathbf{p}\cdot \boldsymbol{\ms}}{\tau}\partial_{\ms_z}\right)f(\tau,x,y,\p_\bot,p_z,\theta_\ms,\ms_z)=-\frac{f-f_\infty}{\tau_R}\;. \label{boltzbjorkcomp}
\end{equation}
The right-hand side of \eq\eqref{boltzbjorkcomp} is similar to the collision term in the usual relaxation time approximation, with the only difference that here the distribution relaxes to the asymptotic distribution function $f_\infty$ instead of the local-equilibrium distribution function $f_\text{LE}$.
According to the symmetry, we assume that $\partial^z\beta^x=\partial^z\beta^y=\partial^x\beta^z=\partial^y\beta^z=\partial^x\beta^0=\partial^y\beta^0=0$ and $\partial^{(x}\beta^{y)}=0$. While the left-hand side of \eq\eqref{boltzbjorkcomp} describes the boost-invariant expansion in the early stages of the time evolution, the right-hand side determines the approach to the asymptotic distribution function at late time. Note that the time scale on which the asymptotic distribution function is reached is set by $\tau_R$ both for the spin-dependent and spin-independent parts. We refer to the time regime in which the distribution function is close to its asymptotic form as to the hydrodynamic regime.
As already mentioned, we are interested in the longitudinal polarization. Following the strategy of Ref.~\cite{Weickgenannt:2023bss}, we will express the $z$-component of the Pauli-Lubanski vector in terms of spin moments, derive equations of motion for these moments, and then order the spin moments according to their behavior in the late-time regime.

\section{Longitudinal polarization}
\label{polsec}

In this section, we outline how to expand the momentum dependence of the polarization in terms of spherical harmonics, and express the expansion coefficients as so-called spin moments.
As outlined in Ref.~\cite{Weickgenannt:2023bss}, the polarization vector may be written as
\begin{align}
    \boldsymbol{\Pi}_\star(\phi)
        &= \frac{1}{2\mathcal{N}}\sum_{n=0}^\infty \sum_{\ell=-n}^n \int dp \int d\cos\theta\, p^2  E_p \left[ \mathbf{A}_{n\ell}+\frac{(\mathbf{A}_{n\ell}\cdot \p)\p}{m(m+E_p)} \right] N_{n\ell} Y_n^\ell(\theta,\phi)\; , \label{pistar}
\end{align}
with the sum running only over the values of $\ell$ with $n+\ell$ even, $Y_n^\ell$ being spherical harmonics, and $\mathbf{p}\equiv p(\sin\theta\, \cos\phi,\sin\theta\, \sin\phi,\cos\theta)$ in spherical coordinates. Furthermore, we defined the normalization constants
\begin{equation}
N_{n\ell}\equiv \frac{2n+1}{4\pi} \frac{(n-\ell)!}{(n+\ell!)}\; ,
\end{equation}
the particle density
\begin{equation}
 \mathcal{N}(\tau)\equiv \frac{1}{\int d\Sigma_\tau} \int dp \int d\cos\theta\, p^2 \int d\Sigma_\lambda p^\lambda \int dS(\p)\, f  \; ,
\end{equation}
and the expansion coefficients
\begin{equation}
 A^k_{n\ell}(\tau,p)\equiv\int dS\, \int d\cos\theta \int d\phi\, \ms^k Y_n^\ell(\theta,\phi) f\; .  
\end{equation}
\bbb{Equation \eqref{pistar} is obtained by expanding the distribution function in spherical harmonics, where the integral over spherical harmonics with $n+\ell$ odd vanishes.}
Analogously to what was done for the transverse polarization in Ref.~\cite{Weickgenannt:2023bss}, we express the longitudinal polarization in terms of the following spin moments
\begin{align}
\mG_{n\ell r}^k&\equiv \int_{p\ms} \left(\frac{p}{E_p}\right)^r Y_n^\ell(\theta,\phi) \ms^k f\; ,\n\\
\mI_{n\ell r}^k&\equiv m  \int_{p\ms} \frac{1}{E_p} \left(\frac{p}{E_p}\right)^r Y_n^\ell(\theta,\phi) \ms^k f \; ,  \label{gidef} 
\end{align}
with $\int_{p\ms}\equiv (m/\sqrt{3}\pi)\int d^3p\, d\ms_z\, d\theta_\ms$.
Performing the $dp$-integration over the terms in the square brackets in \eq\eqref{pistar} for the $z$-component, we obtain
\begin{equation}
\int dp\, p^2 E_p A^z_{n\ell}=\int_{p\ms} \ms^z Y_n^\ell(\theta,\phi) f\equiv \mG^z_{n\ell0}\; ,    \label{pi1stterm}
\end{equation}
and
\begin{align}
        \int dp\, p^2 E_p \frac{(\mathbf{A}_{n\ell}\cdot \p)p^z}{m(m+E_p)}
         &=\int_{p\ms} \frac{E_p-m}{m}  \left[\ms^x \sin\theta\, \frac12\left(e^{i(\ell+1)\phi}+e^{i(\ell-1)\phi}\right)+\ms^y \sin\theta\, \frac12\left(e^{i(\ell+1)\phi}-e^{i(\ell-1)\phi}\right)+\ms^z \cos\theta e^{i\ell\phi}\right]\n\\
        &\times\cos\theta\,  \Pc_n^\ell(\cos\theta) f\n\\
        &= \mathfrak{d}^+_{n\ell} \left(\mI^x_{n(\ell+1)0}-\mG^x_{n(\ell+1)0}+\mI^y_{n(\ell+1)0}-\mG^y_{n(\ell+1)0}\right)\n\\
        &+\mathfrak{e}^+_{n\ell} \left(\mI^x_{(n-2)(\ell+1)0}-\mG^x_{(n-2)(\ell+1)0}+\mI^y_{(n-2)(\ell+1)0}-\mG^y_{(n-2)(\ell+1)0} \right)\n\\
         &+\mathfrak{f}^+_{n\ell} \left(\mI^x_{(n+2)(\ell+1)0}-\mG^x_{(n+2)(\ell+1)0}+\mI^y_{(n+2)(\ell+1)0}-\mG^y_{(n+2)(\ell+1)0} \right)\n\\
         &+\mathfrak{d}^-_{n\ell} \left(\mI^x_{n(\ell-1)0}-\mG^x_{n(\ell-1)0}+\mI^y_{n(\ell-1)0}-\mG^y_{n(\ell-1)0}\right)\n\\
         &+\mathfrak{e}^-_{n\ell} \left(\mI^x_{(n-2)(\ell-1)0}-\mG^x_{(n-2)(\ell-1)0}+\mI^y_{(n-2)(\ell-1)0}-\mG^y_{(n-2)(\ell-1)0} \right)\n\\
         &+\mathfrak{f}^-_{n\ell} \left(\mI^x_{(n+2)(\ell-1)0}-\mG^x_{(n+2)(\ell-1)0}+\mI^y_{(n+2)(\ell-1)0}-\mG^y_{(n+2)(\ell-1)0} \right)\n\\
         &+\mathfrak{g}_{n\ell} \left(\mI^z_{n\ell0}-\mG^z_{n\ell0}\right)+\mathfrak{h}_{n\ell} \left(\mI^z_{(n-2)\ell0}-\mG^z_{(n-2)\ell0}\right)+\mathfrak{i}_{n\ell} \left(\mI^z_{(n+2)\ell0}-\mG^z_{(n+2)\ell0}\right)\; , 
         \label{zpollt}
\end{align}
with $\Pc_n^\ell$ being associated Legendre Polynomials and the coefficients defined in App.~\ref{coeffapp}. Inserting \eqs\eqref{pi1stterm} and \eqref{zpollt} into \eq\eqref{pistar}, one obtains an expression for the spin vector which is a sum over an infinite number of spin moments. Analogously to the strategy in Ref.~\cite{Weickgenannt:2023bss}, we will truncate this sum and keep only the spin moments which are most relevant at the time when the polarization is measured. These spin moments are treated dynamically and are to be determined by their equations of motion, which will be presented in the next section.

\section{Equations of motion for the spin moments}
\label{eomsec}

The spin moments appearing in the polarization vector can be determined dynamically by solving their equations of motion. The first step to obtain the latter is to derive the exact infinite set of coupled equations of motion from the Boltzmann equation, which will be done in this section.
Defining the asymptotic spin moments,
\begin{align}
\mG_{n\ell r,\infty}^k&\equiv \int_{p\ms} \left(\frac{p}{E_p}\right)^r Y_n^\ell(\theta,\phi) \ms^k f_\infty\; ,\n\\
\mI_{n\ell r,\infty}^k&\equiv  m \int_{p\ms} \frac{1}{E_p} \left(\frac{p}{E_p}\right)^r Y_n^\ell(\theta,\phi) \ms^k f_\infty \; ,
\label{asspinmom}
\end{align}
with $f_\infty$ given by \eq\eqref{finfty1},
we obtain from \eq\eqref{boltzbjorkcomp} the following equations of motion for the longitudinal spin moments 
\begin{align}
   \partial_\tau \mG^z_{n\ell r}=&  -\frac{1}{\tau}\int_{p\ms}  \ms^z \left(\frac{p}{E_p}\right)^r  \mathcal{P}^\ell_n(\cos\theta) e^{i\ell\phi} f-\frac{1}{\tau}\int_{p\ms}  \ms^z \left(\frac{p}{E_p}\right)^r  \mathcal{P}^\ell_n(\cos\theta) e^{i\ell\phi} \cos^2\theta \left[r+(1-r)\frac{p^2}{E_p^2}\right] f\n\\
    &+\frac{1}{\tau}\int_{p\ms}  \ms^z \left(\frac{p}{E_p}\right)^r  e^{i\ell\phi}\cos\theta\left[ n\cos\theta\, \mathcal{P}_n^\ell(\cos\theta)-(n+\ell)\mathcal{P}^\ell_{(n-1)}(\cos\theta)\right] f\n\\
   & -\frac{1}{\tau}  \int_{p\ms} \left(\frac{p}{E_p}\right)^r  \mathcal{P}^\ell_n(\cos\theta) e^{i\ell\phi}  \frac{p^2}{E_p^2} \cos\theta \left( \ms^x \sin\theta\, \cos\phi +\ms^y \sin\theta\, \sin\phi +\ms^z \cos\theta\right)f-\frac{1}{\tau_R}(\mG^z_{n\ell r}-\mG^z_{n\ell r,\infty})\; . \label{eomgz}
\end{align}
The derivation is outlined in App.~\ref{eomapp}. 
There, we also show that the equations of motion for the transverse spin moments are independent of the longitudinal spin moments, i.e., the equations of motion derived in Ref.~\cite{Weickgenannt:2023bss} with the polarization restricted to the transverse plane for the transverse spin moments are still valid even if we allow for longitudinal polarization. Including the NLRTA, they read
\begin{align}
    \partial_\tau \mG^x_{n\ell r}
    =& -\frac{1}{\tau}\int_{p\ms}  \ms^x  \left(\frac{p}{E_p}\right)^r  \mathcal{P}^\ell_n(\cos\theta) e^{i\ell\phi} f-\frac{1}{\tau}\int_{p\ms}  \ms^x \left(\frac{p}{E_p}\right)^r  \mathcal{P}^\ell_n(\cos\theta) e^{i\ell\phi} \cos^2\theta \left[r+(1-r)\frac{p^2}{E_p^2}\right] f\n\\
    &+\frac{1}{\tau}\int_{p\ms}  \ms^x \left(\frac{p}{E_p}\right)^r  e^{i\ell\phi}\cos\theta\left[ n\cos\theta\, \mathcal{P}_n^\ell(\cos\theta)-(n+\ell)\mathcal{P}^\ell_{(n-1)}(\cos\theta)\right]  f-\frac{1}{\tau_R}(\mG^x_{n\ell r}-\mG^x_{n\ell r,\infty})\; .
  \label{eomgxgenpol}
\end{align}
Before we express the right-hand sides of \eqs\eqref{eomgz} and \eqref{eomgxgenpol} in terms of spin moments, let us first discuss their structure, which is actually very simple.  There are two main contributions: the terms $\sim 1/\tau$, describing the free expansion, and the term $\sim 1/\tau_R$, which drives the spin moments to their asymptotic values. At early time $\tau\ll\tau_R$, the evolution is determined by the terms $\sim 1/\tau$, corresponding to free streaming. On the other hand, at late time $\tau\gg\tau_R$, the collision term $\sim 1/\tau_R$ drives the system to the asymptotic solution. Note that according to \eqs\eqref{pi1stterm} and \eqref{zpollt} only spin moments $\mG^z_{n\ell 0}$ with $n+\ell$ even and spin moments $\mG^x_{n\ell 0}$ and $\mG^y_{n\ell 0}$ with $n+\ell$ odd enter the polarization \eqref{pistar}. Consider the equations of motion for the former, \eq\eqref{eomgz}. We notice that it couples spin moments with different $n$, $\ell$ and $r$. In addition, a coupling to the transverse spin moments appears in the last line. The free-streaming dynamics with Bjorken symmetry is determined by the longitudinal expansion, which drives the system to a state with $\cos\theta=0$. One can understand this limit intuitively, since all particles with $p^z\neq0$ will have left the $z=0$-slice after some time. We assume in this paper that this state is reached well before  the collision-dominated regime sets in, which allows for a nontrivial dynamics in between the two regimes. For $\cos\theta=0$, all the free-streaming terms except the first one in \eq\eqref{eomgz} vanish, and therefore the longitudinal spin moments with $n+\ell$ even decay as $\tau^{-1}$ in the late free-streaming regime. At even later time, the last term in \eq\eqref{eomgz} starts to control the dynamics. The decay of the spin moments in this regime is determined by a power law. Note that if all the asymptotic spin moments were nonzero, all spin moments would decay as $ \tau^{-1}$. On the other hand, if all asymptotic spin moments were zero, all spin moments would decay exponentially. However, as we will see in the next section, a small number of asymptotic spin moments $\mG^z_{n\ell r,\infty}$ is nonzero, leading to a power-law behavior with  an exponent depending on $n$ and $\ell$.  Turning to \eq\eqref{eomgxgenpol} for the transverse spin moments, we note that $\Pc_n^\ell(0)=0$ for $n+\ell$ odd. Hence, the transverse spin moments which enter \eq\eqref{pistar} decay as $ \tau^{-2}$ in the late free-streaming regime. The decay in the collision-dominated regime is similar to that of the longitudinal spin moments. The equations of motion for the spin moments $\mI^k_{n\ell r}$ are completely analogous to those for $\mG^k_{n\ell r}$, they are displayed in App.~\ref{eomapp}.

\section{Asymptotic spin moments}
\label{asympsec}

Consider now the asymptotic spin moments in \eq\eqref{asspinmom}, corresponding to the zeroth-order spin moments in a gradient expansion in terms of powers of $w^{-1}\equiv \tau_R/\tau$. They consist of two contributions from the two terms in $f_\infty$ given by \eq\eqref{finfty1}. We will refer to the first term in \eq\eqref{finfty1} as to the local-equilibrium contribution, and to the second term as to the nonlocal contribution. Note that, by using \eqs\eqref{deltamu} and \eqref{intds}, one gets
\begin{align}
    \int_\ms \ms^\lambda f_\infty&= \int_\ms \ms^\lambda \left[ -\frac{\hbar}{4m}\tilde{\Omega}_{\mu\nu} p^\mu \ms^\nu -\xi \frac{\hbar}{2m(E_p+m)}\epsilon^{\mu\rho\alpha\beta}p_\rho t_\alpha \ms_\beta p^\nu (\Omega_{\mu\nu}+\partial_\mu \beta_\nu)\right] f_\text{LE}^{(0)}\n\\
    &= -E_p\left[ \frac{\hbar}{m}\tilde{\Omega}^{\lambda\mu} p_\mu  -\xi\frac{2\hbar}{m(E_p+m)}\epsilon^{\mu\rho\alpha\lambda}p_\rho t_\alpha p^\nu (\Omega_{\mu\nu}+\partial_\mu \beta_\nu)\right] f_\text{LE}^{(0)} \; . \label{intdsfinfty}
\end{align}
For the local-equilibrium parts of the longitudinal spin moments, we obtain
\begin{align}
    \mG^z_{n\ell r,\text{eq}}&=-\frac{\hbar}{m}\int d^3p\, \left(\frac{p}{E_p}\right)^r Y_n^\ell(\theta,\phi) E_p (\tilde{\Omega}^{z0} E_p-\tilde{\Omega}^{zy} p \sin\theta \sin \phi-\tilde{\Omega}^{zx} p\sin\theta\cos\phi) f_0\n\\
    &=-\frac{\hbar}{m}\int d^3p\, \left(\frac{p}{E_p}\right)^rY_n^\ell(\theta,\phi) E_p \n\\
    &\times\left[\tilde{\Omega}^{z0} E_p Y_0^0(\theta,\phi)-i\tilde{\Omega}^{zy} p \left(\frac12Y_1^1(\theta,\phi)-Y_1^{-1}(\theta,\phi)\right)+\tilde{\Omega}^{zx} p \left(\frac12Y_1^1(\theta,\phi)+Y_1^{-1}(\theta,\phi)\right)\right] f_0\n\\
    &= \delta_{n0}\delta_{\ell0} \mG^z_{00 r,\text{eq}}+\delta_{n1}\delta_{\ell1} \mG^z_{11 r,\text{eq}}+\delta_{n1}\delta_{\ell(-1)} \mG^z_{1(-1)r,\text{eq}}\; .
    \label{gzeq}
\end{align}
Here we use the short-hand notation $f_0\equiv f_\text{LE}^{(0)}$.
In addition, we gain the following contributions from the nonlocal collision term, i.e., from the second term in \eq\eqref{finfty1},
\begin{align}
& -2\int_{p} \left(\frac{p}{E_p}\right)^r Y_n^\ell(\theta,\phi) E_p \bxi \epsilon^{ijz} p^i p^\nu (\Omega_{j\nu}+\partial_j \beta_\nu)f^{(0)}_\text{LE}\n\\
 &= -2\int_{p} \left(\frac{p}{E_p}\right)^r Y_n^\ell(\theta,\phi) E_p \bxi  p_\nu p^{[x} (\Omega^{y]\nu}+\partial^{y]} \beta^\nu)f^{(0)}_\text{LE}\n\\
  &= -2\int_{p} \left(\frac{p}{E_p}\right)^r Y_n^\ell(\theta,\phi) E_p^2 \bxi  p^{[x} \kappa_0^{y]}f^{(0)}_\text{LE}-2\int_{p} \left(\frac{p}{E_p}\right)^r Y_n^\ell(\theta,\phi) E_p \bxi  \left[p_\bot^2 \left(\Omega^{yx}-\varpi^{yx}\right)+p^zp^{[x}\left(\Omega^{y]z}-\varpi^{y]z}\right)\right]f^{(0)}_\text{LE} \; .
  \label{gzinftynl}
\end{align}
Here we used the fact that, for $t=(1,\mathbf{0})$, \eq\eqref{deltamu} becomes
\begin{equation}
\Delta^j=-\frac{\hbar}{2m (E_p+m)}\epsilon^{jik}p^i  \ms^k \; ,
\label{delta}
\end{equation}
and we defined $\kappa_0^\mu\equiv -\Omega^{\mu\nu}u_\nu$ and $\varpi^{\mu\nu}\equiv -(1/2)\partial^{[\mu}\beta^{\nu]}$.
We also absorbed coefficients into $\xi(E_p)$ by defining 
\begin{equation}
\bar{\xi}\equiv \frac{E_p}{(E_p+m)m} \xi\; .
\end{equation}
We remember that the longitudinal spin moments which are odd in $p_z$ do not contribute to the longitudinal polarization. For the spin moments even in $p_z$, we find from \eq\eqref{gzinftynl} that the nonlocal collision term contributes to the following asymptotic spin moments (with positive $\ell$),
\begin{equation}
  \mG^z_{00r}\; ,\quad \mG^z_{11r}\; , \quad \mG^z_{20r}\; . 
   \label{gzinf}
\end{equation}
For the $x$-component we have the local-equilibrium contributions
\begin{align}
    \mG^x_{n\ell r,\text{eq}}&=-\hbar\int d^3p\, \left(\frac{p}{E_p}\right)^r Y_n^\ell(\theta,\phi) E_p (\tilde{\Omega}^{x0} E_p-\tilde{\Omega}^{xy} p \sin\theta \sin \phi-\tilde{\Omega}^{xz} p\cos\theta) f_0\n\\
    &=-\hbar\int d^3p\, \left(\frac{p}{E_p}\right)^r Y_n^\ell(\theta,\phi) E_p \left[\tilde{\Omega}^{x0} E_p Y_0^0(\theta,\phi)-i\tilde{\Omega}^{xy} p \left(\frac12Y_1^1(\theta,\phi)-Y_1^{-1}(\theta,\phi)\right)+\tilde{\Omega}^{xz} p Y_1^0(\theta,\phi)\right] f_0\n\\
    &= \delta_{n0}\delta_{\ell0} \mG^x_{00 r,\text{eq}}+\delta_{n1}\delta_{\ell1} \mG^x_{11 r,\text{eq}}+\delta_{n1} \delta_{\ell(-1)}\mG^x_{1(-1)r,\text{eq}}+\delta_{n1}\delta_{\ell0}\mG^x_{10r,\text{eq}}\; 
\end{align}
and the nonlocal contributions
\begin{align}
& -2\int_{p} \left(\frac{p}{E_p}\right)^r Y_n^\ell(\theta,\phi) E_p \bxi \epsilon^{ijx} p^i p^\nu (\Omega_{j\nu}+\partial_j \beta_\nu)f^{(0)}_\text{LE}\n\\
 &= -2\int_{p} \left(\frac{p}{E_p}\right)^r Y_n^\ell(\theta,\phi) E_p \bxi  p_\nu p^{[y} (\Omega^{z]\nu}+\partial^{z]} \beta^\nu)f^{(0)}_\text{LE}\n\\
  &= -2\int_{p} \left(\frac{p}{E_p}\right)^r Y_n^\ell(\theta,\phi) E_p^2 \bxi  p^{[y} \kappa_0^{z]}f^{(0)}_\text{LE}-2\int_{p} \left(\frac{p}{E_p}\right)^r Y_n^\ell(\theta,\phi) E_p \bxi  \left[(p_y^2+p_z^2) \left(\Omega^{yx}-\varpi^{yx}\right)+p^xp^{[y}\left(\Omega^{z]x}-\varpi^{z]x}\right)\right]f^{(0)}_\text{LE}\n\\
  \n\\
  &+\int_{p} \left(\frac{p}{E_p}\right)^r Y_n^\ell(\theta,\phi) E_p \bxi p^z p^y \partial^z\beta^z f^{(0)}_\text{LE}\; .
\end{align}
Here only transverse spin moments with $n+\ell$ odd will contribute to the longitudinal polarization, see \eq\eqref{zpollt}. The nonlocal part of the collision term contributes to the following of these spin moments,
\begin{equation}
     \mG^x_{10r}\; , \quad \mG^x_{21r}\; . 
    \label{gxinf}
\end{equation}
As already discussed in Ref.~\cite{Weickgenannt:2023bss}, the polarization is measured at freeze-out, and therefore the relevant spin moments in \eq\eqref{pistar} are those which survive in the late-time regime, i.e., the asymptotic spin moments \eqref{gzinf} and \eqref{gxinf}, and spin moments whose equations of motion directly couple to these quantities. Note that the asymptotic spin moments are functions of the spin potential $\Omega^{\mu\nu}$. This means that we need to obtain equations of motion for the latter, which will be done in the next section.

\section{Total angular momentum}
\label{tamsec}

In this section, we express the spin potential through the total angular momentum, and obtain equations of motion for the latter. To this end, we employ the matching condition~\cite{Weickgenannt:2024ibf}
\begin{equation}
    \int d\Gamma\, \left( \Delta^{[\mu} p^{\nu]}+\frac\hbar2 \Sigma_\ms^{\mu\nu} \right) (f-f_\infty)=0\; , \label{match}
\end{equation}
which ensures that the collision term \eqref{nlrta} conserves the total angular momentum 
\begin{equation}
\mathcal{J}^{\mu\nu}\equiv \int d\Gamma\, \left( \Delta^{[\mu} p^{\nu]}+\frac\hbar2 \Sigma_\ms^{\mu\nu} \right) f\;
\label{jmunu}
\end{equation}
\bbb{in the collision.}
\bbb{Note that $\Delta^\mu$ is the space-like separation between the particle position and the center of the collision~\footnote{\bbb{By ``center of the collision'' we mean here the space-time point where the wordlines of the colliding particles converge at zeroth order in $\hbar$. In general, it does not coincide with the center of mass, nor with the center of inertia at first order in $\hbar$. However, for the orbital angular momentum in the collision, only the relative distance between the particles is relevant, and this does not depend on the precise location of the ``center" of the collision.}} $x$, and therefore $\Delta^{[\mu} p^{\nu]}$ is the orbital angular momentum of the particle in the collision, which can be converted into spin.}
Due to the matching conditions, the total angular momentum is equal to its asymptotic value
\begin{equation}
\mathcal{J}^{\mu\nu}=\mathcal{J}^{\mu\nu}_\infty\equiv \int d\Gamma\, \left( \Delta^{[\mu} p^{\nu]}+\frac\hbar2 \Sigma_\ms^{\mu\nu} \right) f_\infty \; .
\label{jinfty}
\end{equation}
Using \eq\eqref{finfty1} in \eq\eqref{jinfty}, it is clear that the components of $\Omega^{\mu\nu}$ can be expressed as a function of the components of $\mathcal{J}^{\mu\nu}$ with the coefficients being thermodynamic integrals. We may therefore equivalently calculate $\tilde{\mathcal{J}}^{\mu\nu}\equiv \epsilon^{\mu\nu\alpha\beta}\mathcal{J}_{\alpha\beta}$ to obtain $\Omega^{\mu\nu}$. Contracting \eq\eqref{jmunu} with $\epsilon^{\mu\nu\alpha\beta}$, we obtain
\begin{align}
    \tilde{\mathcal{J}}^{\mu\nu}&= \frac\hbar4 \int_{p\ms} \frac{1}{E_p^2} \left[ \frac{m}{2(E_p+m)} \ms^{[\mu} t^{\nu]}+\frac{1}{m} \left(1+\frac{E_p}{2(E_p+m)}\right)p^{[\mu}\ms^{\nu]} \right] f\n\\
&= \frac\hbar4 \int_{p\ms} \frac{1}{E_p^2} \left[  \frac{m(E_p-m)}{2p^2} \ms^{[\mu} t^{\nu]}+\frac{1}{m} \left(1+\frac{E_p(E_p-m)}{2p^2}\right)p^{[\mu}\ms^{\nu]} \right]f\; ,
\label{jtilde}
\end{align}
where we used
\begin{equation}
\epsilon^{\mu\nu\alpha\beta} p_\alpha \epsilon_{\beta\lambda\rho\sigma} p_\lambda \ms_\rho t_\sigma=p_\alpha\left(p^\mu \ms^{[\nu} t^{\alpha]}+p^\alpha \ms^{[\mu} t^{\nu]}+p^\nu \ms^{[\alpha}t^{\mu]}\right)=E_pp^{[\mu} \ms^{\nu]}+m^2 \ms^{[\mu} t^{\nu]}\; .
\end{equation}
We see that \eq\eqref{jtilde} cannot be expressed through the already defined spin moments $\mG^k_{n\ell r}$ or $\mI^k_{n\ell r}$. Therefore, we determine its equations of motion separately in App.~\ref{eomapp}. 
The result reads
\begin{align}
    \partial_\tau \tilde{\mathcal{J}}^{\mu\nu}&= -\frac{1}{\tau}\frac{\hbar}{4} \int_{p\ms} \frac{1}{E_p}\left\{\left(\frac{1}{E_p}-\frac{p_z^2}{E_p^3}\right) \left[  -\frac{m}{2(E_p+m)}  t^{[\mu}+\frac{1}{m} \left(1+\frac{E_p}{2(E_p+m)}\right)p^{[\mu} \right]+\frac{p_z}{E_p}\right.  \n\\
  &\left. \times\left[\frac{m}{2(E_p+m)^2}  \frac{p_z}{E_p}t^{[\mu}+\frac{1}{m} \left(1+\frac{E_p}{2(E_p+m)}\right)\left(\delta^{z[\mu}+\frac{p_z}{E_p}\delta^{0[\mu}\right)+\frac{p_z}{2m(E_p+m)}\left(\frac{1}{E_p}-\frac{1}{E_p+m}\right)p^{[\mu}\right]\right\} \ms^{\nu]} f\n\\ 
   & +\frac{\hbar}{4} \frac{1}{\tau} t^{[\mu} \delta^{\nu]z} \int_{p\ms} \frac{p_z}{E_p^4}  \frac{m}{2(E_p+m)} \left( p^x\ms^x+p^y \ms^y+p^z \ms^z\right)f\n\\ 
    & -\frac{\hbar}{4} \frac{1}{\tau}\int_{p\ms} \frac{p_z}{E_p^2} \frac{1}{m} \left(1+\frac{E_p}{2(E_p+m)}\right)p^{[\mu} \bigg\{\left[\frac{1}{m^2}\delta^{\nu]z}+\left(\frac{1}{m^2}-\frac{1}{E_p^2}\right)\delta^{\nu]0}\frac{p_z}{E_p}\right]\left( p^x\ms^x+p^y \ms^y+p^z \ms^z\right)
    +\frac{\delta^{\nu]0}}{E_p}\ms^z\bigg\} f\; .
    \label{angmomeom}
    \end{align}
According to relation \eqref{jinfty}, we may replace $\tilde{\mJ}^{\mu\nu}$ on the left-hand side of \eq\eqref{angmomeom} by $\tilde{\mJ}^{\mu\nu}_\infty$, and the asymptotic spin moments can be expressed as a function of the latter.
From the second identity in \eq\eqref{jinfty}, we obtain
\begin{align}
\tilde{\mathcal{J}}^{i0}_\infty&= \frac\hbar4 \int_{p\ms} \frac{1}{E_p^2} \left[  \frac{m}{2(E_p+m)} \ms^i +\frac{1}{m} \left(1+\frac{E_p}{2(E_p+m)}\right)p^{[i}\ms^{0]} \right] f_\infty\n\\
&= \frac\hbar4 \int_{p\ms} \frac{1}{E_p^2} \left[  \left(\frac{m}{2(E_p+m)}-\frac{E_p}{m}-\frac{E_p^2}{2m(E_p+m)}\right) \ms^i +\frac{1}{m} \left(1+\frac{E_p}{2(E_p+m)}\right)p^{i}\ms^{0} \right] f_\infty
\end{align}
and
\begin{align}
\tilde{\mathcal{J}}^{ij}_\infty&= \frac\hbar4 \int_{p\ms} \frac{1}{E_p^2} \frac{1}{m} \left(1+\frac{E_p}{2(E_p+m)}\right)p^{[i}\ms^{j]}  f_\infty\: .   
\end{align}
Since all components contain contributions independent of $\theta$ and/or contributions linear in $\cos\theta$ or $\sin\theta$, all of them are nonzero in the long-time limit. Expressing each component of $\tilde{\mJ}^{\mu\nu}=\tilde{\mathcal{J}}_\infty^{\mu\nu}$ as a function of the spin potential by inserting the asymptotic distribution function \eqref{finfty1}, we find
\begin{align}
\tilde{\mathcal{J}}^{xy}
    &=-\hbar^2 \kappa_0^{z} \int_{p} \frac{1}{E_p} \frac{1}{m^2} \left(1+\frac{E_p}{2(E_p+m)}\right) (1+m\bxi)  p_z^2 f^{(0)}_\text{LE}\; ,\n\\
    \tilde{\mathcal{J}}^{xz} 
    &=\hbar^2 \kappa_0^{y} \int_{p} \frac{1}{E_p} \frac{1}{m^2} \left(1+\frac{E_p}{2(E_p+m)}\right) (1+m\bxi)   p_z^2 f^{(0)}_\text{LE}\; ,\n
    \\
    \tilde{\mathcal{J}}^{z0}
    &= \frac{\hbar^2}{2}  \Omega^{xy} \int_p  \left[\frac{m^2-2E_p(E_p+m)-E_p^2}{2m^2(E_p+m)}\left(1-\frac{p_z^2}{E_p^2}m\bxi\right)-\frac{p_z^2}{E_pm^2}\left(1+\frac{E_p}{2(E_p+m)}\right) \right]  f^{(0)}_\text{LE}\n\\
    &+ \frac{\hbar^2}{2} \varpi^{xy} \int_{p}  \frac{m^2-2E_p(E_p+m)-E_p^2}{2m(E_p+m)} \bxi  \frac{p_z^2}{E_p^2} f^{(0)}_\text{LE}\; ,
    \label{alljinf}
\end{align}
see App.~\ref{relapp} for the calculation.
Analogous relations are found for $\tilde{\mathcal{J}}^{yz}_\infty$, $\tilde{\mathcal{J}}^{x0}_\infty$ and $\tilde{\mathcal{J}}^{y0}_\infty$.  

Finally, using \eqs\eqref{asspinmom} and \eqref{alljinf}, we may express the asymptotic longitudinal spin moments through $\tilde{\mathcal{J}}^{\mu\nu}$ as follows,
\begin{subequations}
\begin{align}
    \mG^z_{000,\infty}
  & = \lambda_{00}^{00} \tilde{\mathcal{J}}^{z0}+\kappa_{00}^{00} \varpi^{xy}\; ,\label{g000infty}\\
 \mG^z_{200,\infty}&=\lambda_{20}^{00} \tilde{\mathcal{J}}^{z0}+\kappa_{20}^{00} \varpi^{xy}\; ,\label{g200infty}\\
 \mG^z_{110,\infty}&=  \lambda_{11}^{00} \tilde{\mathcal{J}}^{xz}+\bar{\lambda}_{11}^{00} \tilde{\mathcal{J}}^{yz}\; ,
 \label{gz11inf}
\end{align}
\label{gzinfty}
\end{subequations}
where the calculation and the coefficients are shown in App.~\ref{relapp}.

Furthermore we obtain for the transverse spin moments
\begin{subequations}
\begin{align}
    \mG^x_{100,\infty}&= \lambda_{10}^{00} \tilde{\mathcal{J}}^{xz} \; ,\label{gx10inf}\\
   \mG^x_{210,\infty}
   &= \lambda_{21}^{00} \tilde{\mathcal{J}}^{z0}+\kappa_{21}^{00}\varpi^{xy}-\tilde{\kappa}_{21}^{00}\sigma\; , \label{gx21inf}\\
   \mG^y_{100,\infty}
    &= \lambda_{10}^{00} \tilde{\mathcal{J}}^{yz} \; , \label{gy10inf}\\
    \mG^y_{210,\infty}
    &= \bar{\lambda}_{21}^{00} \tilde{\mathcal{J}}^{z0}+\bar{\kappa}_{21}^{00}\varpi^{xy}+\hat{\kappa}_{21}^{00}\sigma \label{gy21inf}
\end{align}
\label{gxyinfty}
    \end{subequations}
with $\sigma\equiv \partial^z\beta^z$ and the calculation and coefficients again displayed in App.~\ref{relapp}. Relations \eqref{gzinfty} and \eqref{gxyinfty} are inserted into the equations of motion for the spin moments, \eqref{eomgz} and \eqref{eomgxgenpol}, and the nonzero components of $\mJ^{\mu\nu}$ are treated as additional dynamical variables with the equations of motion \eqref{angmomeom}. What remains to be done now is to truncate the free-streaming parts of each equation of motion to obtain a closed set of equations.

\section{Closing the equations of motion}
\label{clossec}

In this section, we show how to truncate the infinite sum in the polarization vector \eqref{pistar}, and then close the equations of motion \eqref{eomgz}, \eqref{eomgxgenpol}, and \eqref{angmomeom} for the spin moments and the total angular momentum, respectively. We will use two different truncation methods for \eq\eqref{pistar} and for the equations of motion for the spin moments which remain in \eq\eqref{pistar} after the first truncation. The polarization in heavy-ion collisions is measured at freeze-out. It is therefore sufficient to keep in \eq\eqref{pistar} only the spin moments which are relevant in the late-time regime.
Consider, e.g., \eq\eqref{eomgz}. To figure out the behavior of the spin moments in the hydrodynamic regime, we need to find out how the spin moments couple to the asymptotic spin moments ($n=0$ and $\ell=0$, $n=1$ and $\ell=\pm1$, $n=2$ and $\ell=0$ for the longitudinal components or $n=1$ and $\ell=0$ or $n=2$ and $\ell=\pm 1$ for the transverse components). Using the properties of the associated Legendre polynomials, it is easy to show that the terms in the first two lines of \eq\eqref{eomgz} will couple moments with indices $n$ and $\ell$ to those with equal $\ell$ and $n\pm 2$. The last line will contribute with terms $\sim \cos\theta\, \sin\theta\, \sin\phi$, $\cos\theta\, \sin\theta\, \cos\phi$, and $\cos^2\theta$. This means that in addition it will couple moments with $k=z$, $n$, $\ell$ to those with $k=x$/ $k=y$, $n$ or $n\pm2$, and $\ell\pm1$. Thus, since equations for different $\ell$ are coupled, all equations of motion implicitly depend on asymptotic quantities, and all moments decay with power laws. To keep the discussion as simple as possible, we will restrict the sum in \eq\eqref{pistar} to spin moments which are nonvanishing at first order in $w^{-1}\equiv \tau_R/\tau$ in the following, and neglect all faster decaying spin moments in that equation. Furthermore, we note that only longitudinal spin moments with $n+\ell$ even and transverse spin moments with $n+\ell$ odd appear in the transverse polarization, and their equations of motion decouple from the longitudinal spin moments with $n+\ell$ odd and the transverse spin moments with $n+\ell$ even. For this reason, we do not consider the latter two in the following.

The spin moments which are nonzero at first order in $w^{-1}$ are those which explicitly contain asymptotic spin moments in their equations of motion. These are
\begin{align}
  &\mG^z_{00r}\; ,\quad \mG^z_{1(\pm1)r}\; , \quad \mG^z_{20r}\; , \quad \mG^z_{40r}\; , \quad \mG^z_{3(\pm1)r}\; , \quad \mG^z_{2(\pm2)r}\; , \quad \mG^z_{4(\pm2)r}\;, \n\\
  &\mG^x_{10r}\; , \quad \mG^x_{30r}\; , \quad \mG^x_{2(\pm1)r}\; , \quad \mG^x_{4(\pm1)r}\; ,
  \label{1stordernonzero}
\end{align}
and the same spin moments with $x\rightarrow y$.
We will now obtain closed equations of motion for the spin moments \eqref{1stordernonzero} with $r=0$ from \eqs\eqref{eomgz} and \eqref{eomgxgenpol} following a strategy similar to Refs.~\cite{Jaiswal:2022udf,Weickgenannt:2023nge}. In contrast to the truncation procedure for \eq\eqref{pistar}, we cannot simply neglect the early-time behavior of the equations of motion, since the dynamics at early time determines the spin moments at the time of the onset of the hydrodynamic regime and thus may leave an imprint on their values at freeze-out. Therefore, our truncation strategy takes into account both the early- and late-time regimes.
We replace in \eqs\eqref{eomgz} and \eqref{eomgxgenpol} all spin moments $\mG^k_{n\ell0}$, which do not appear in \eqref{1stordernonzero}, by their value in the late free-streaming regime with $\cos\theta=0$, expressed in terms of the moment on the left-hand side of each equation of motion,
\begin{equation}
\mG^z_{(n_\text{max}+2)\ell0}\rightarrow \frac{\Pc_{n_\text{max}+2}^\ell(0)}{\Pc_{n_\text{max}}^\ell(0)} \mG_{n_\text{max}\ell0}^z\; ,
\label{replgz}
\end{equation}
and
\begin{equation}
\mG^x_{(n_\text{max}+2)\ell0}\rightarrow \frac{(n_\text{max}+2+\ell)\Pc_{n_\text{max}+1}^\ell(0)}{(n_\text{max}+\ell)\Pc_{n_\text{max}-1}^\ell(0)} \mG_{n_\text{max}\ell0}^x\; ,
\label{replgx}
\end{equation}
where $n_\text{max}$ is the maximal value of $n$ appearing in \eqref{1stordernonzero} for the respective component $z$ or $x$ and the corresponding value of $\ell$. Note that in the long-time limit $w\rightarrow\infty$ both sides of \eqs\eqref{replgz} and \eqref{replgx} vanish, respectively, such that the replacement is justified in both regimes and an interpolation is not needed.
On the other hand, spin moments with $r\neq0$, which are nonzero in the late-time limit, but do not enter the polarization, are approximated by an interpolation of the form
\begin{equation}
    \mG^k_{n\ell r}\rightarrow e^{-w/2} \mG^k_{n\ell r,\circ}+\left(1-e^{-w/2}\right)\mG^k_{n\ell r,\infty}\; , \label{r2intpol}
\end{equation}
i.e., we use an interpolation between the late free-streaming regime, denoted by $\mG^k_{n\ell r,\circ}$, where all terms proportional to spin moments with $r\neq0$ in \eq\eqref{eomgz} vanish, and the late-time regime, where the spin moments are given by their asymptotic values. The interpolation is needed since the terms proportional to spin moments with $r\neq0$ vanish in the late free-streaming regime, but reappear in the hydrodynamic regime. \bbb{The form of the interpolation \eqref{r2intpol} has been introduced and checked in Ref.~\cite{Jaiswal:2022udf}.} This truncation significantly simplifies the equations of motion compared to the case where the full dynamical moments are used, which would induce many additional coupling terms. Since these terms are suppressed by at least $\cos\theta$ in free streaming, they do not contribute significantly before the onset of the late-time regime. The simplest closed equations of motion which can be obtained from \eq\eqref{eomgz} then read
\begin{align}
  & \partial_w  \mG^z_{n\ell0}= -\frac1w\left( \zeta_{n\ell} \mG^z_{n\ell0}+\chi_{n\ell} \mG^z_{(n-2)\ell0}+\xi_{n\ell} \mG^z_{(n+2)\ell0} \right)-\frac1w\left(1-e^{-w/2}\right)\left( \mathcal{K}^z_{n\ell,\infty}+ \mathcal{D}^x_{n\ell,\infty}+ \bar{\mathcal{D}}^y_{n\ell,\infty} \right) -\left(\mG^z_{n\ell0}-\mG^z_{n\ell0,\infty}\right)\; ,\n\\
  &\text{ for } n=0,2\; , \ell=0,2\; ,\n\\
 & \partial_w  \mG^z_{4\ell0}= -\frac1w\left( \hat{\zeta}_{4\ell} \mG^z_{4\ell0}+\chi_{4\ell} \mG^z_{2\ell0} \right)
   -\frac1w\left(1-e^{-w/2}\right)\left( \mathcal{K}^z_{4\ell,\infty}+  \mathcal{D}^x_{4\ell,\infty}+ \bar{\mathcal{D}}^y_{4\ell,\infty} \right)-\mG^z_{4\ell0}\; ,\n\\ &\text{ for } \ell=0,2\; ,\n\\
  & \partial_w  \mG^z_{110}= -\frac1w\left( \zeta_{11} \mG^z_{110}+\xi_{11} \mG^z_{310} \right)-\frac1w\left(1-e^{-w/2}\right)\left( \mathcal{K}^z_{11,\infty}+ \mathcal{D}^x_{11,\infty}+ \bar{\mathcal{D}}^y_{11,\infty} \right)-\left(\mG^z_{110}-\mG^z_{110,\infty}\right)\; ,\n\\
  &\partial_w  \mG^z_{310}= -\frac1w\left( \hat{\zeta}_{31} \mG^z_{310}+\chi_{31} \mG^z_{110} \right)
   -\frac1w\left(1-e^{-w/2}\right)\left(  \mathcal{K}^z_{31,\infty}+\mathcal{D}^x_{31,\text{eq}}+ \bar{\mathcal{D}}^y_{31,\text{eq}} \right)-\mG^z_{310}\; .
   \label{eomgzfin}
\end{align}
Here we defined the following quantities, which depend only on $\Omega_{\mu\nu}$ and thermodynamic integrals,
\begin{align}
    \mathcal{K}^z_{n\ell,\infty}&\equiv 2\int_{p\ms} \ms^z \left(\frac{p}{E_p}\right)^2 \Pc_n^\ell(\cos\theta)\cos^2\theta\, e^{i\ell\phi} f_\infty\; ,\n\\
    \mathcal{D}^x_{n\ell,\infty}&\equiv \int_{p\ms} \ms^x \Pc_n^\ell(\cos\theta)\left(\frac{p}{E_p}\right)^2 \cos\theta\, \sin\theta\, \cos\phi\, e^{i\ell\phi} f_\infty\; ,\n\\
    \bar{\mathcal{D}}^y_{n\ell,\infty}&\equiv \int_{p\ms} \ms^y \Pc_n^\ell(\cos\theta)\left(\frac{p}{E_p}\right)^2 \cos\theta\, \sin\theta\, \sin\phi\, e^{i\ell\phi} f_\infty\; ,
    \label{asint1}
\end{align}
The coefficients in \eqs\eqref{eomgzfin} are defined in App.~\ref{coeffapp}. With the help of App.~\ref{asympapp}, \eqs\eqref{asint1} can be expressed through spin moments analogous to $\mG^k_{n\ell}$, whose long-time limits are similar to \eqs\eqref{gzinfty} and \eqref{gxyinfty}, up to the factor $(p/E_p)^2$. The combinations appearing in \eqs\eqref{eomgzfin} read
\begin{align}
\mathcal{K}^z_{n0,\infty}+\mathcal{D}^x_{n0,\infty}+\bar{\mathcal{D}}^y_{n0,\infty} &= a_{n0} \tilde{\mathcal{J}}^{z0}+b_{n0}\varpi^{xy}+c_{n0}\sigma
\; , \qquad \text{for }n=0,2,4\; ,\n\\
\mathcal{K}^z_{n2,\infty}+\mathcal{D}^x_{n2,\infty}+\bar{\mathcal{D}}^y_{n2,\infty} &= a_{n2} \tilde{\mathcal{J}}^{z0}+b_{n2}\varpi^{xy}+c_{n2}\sigma  \; , \qquad \text{for } n=2,4\; ,\n\\
\mathcal{K}^z_{n1,\infty}+\mathcal{D}^x_{n1,\infty}+\bar{\mathcal{D}}^y_{n1,\infty}&= \mathfrak{a}_{n1} \tilde{\mathcal{J}}^{xz}+\bar{\mathfrak{a}}_{n1}\tilde{\mathcal{J}}^{yz}\; , \quad  \qquad \qquad \text{for } n=1,3 \label{kdd}
\end{align}
where the coefficients are again defined in App.~\ref{coeffapp}.

For the longitudinal spin moments, we follow the same strategy and find from \eq\eqref{eomgxgenpol}
\begin{align}
 \partial_w \mG_{100}^x&=-\frac{1}{w} \left( \bar{a}_{100} \mG^x_{100}+\bar{c}_{100} \mG^x_{300}\right)-\frac1w\left(1-e^{-w/2}\right)\frac12\mathcal{K}^x_{10,\infty}-\left(\mG_{100}^x-\mG_{100,\infty}^x \right)\; ,\n\\
 \partial_w \mG_{300}^x&=-\frac{1}{w} \left( \hat{a}_{300} \mG^x_{300}+\bar{b}_{300}\mG^x_{100}\right)-\frac1w\left(1-e^{-w/2}\right) \frac12\mathcal{K}^x_{30,\infty}-\mG_{300}^x \; ,\n\\
 \partial_w \mG_{210}^x&=-\frac{1}{w} \left( \bar{a}_{210} \mG^x_{210}+\bar{c}_{210} \mG^x_{410}\right)-\frac1w\left(1-e^{-w/2}\right) \frac12\mathcal{K}^x_{21,\infty}-(\mG_{210}^x-\mG_{210,\infty}^x) \; ,\n\\
 \partial_w \mG_{410}^x&=-\frac{1}{w} \left( \hat{a}_{410} \mG^x_{410}+\bar{b}_{410}\mG^x_{210}\right)-\frac1w\left(1-e^{-w/2}\right) \frac12\mathcal{K}^x_{41,\infty}-\mG_{410}^x \; ,
 \label{eomgxfin}
\end{align}
where the coefficients are given in App.~\ref{coeffapp} and we defined the asymptotic quantity
\begin{align}    \mathcal{K}^x_{n\ell,\infty}&\equiv  2\int_{p\ms} \ms^x  \frac{p^2}{E_p^2} \cos^2\theta\, \Pc_n^\ell(\cos\theta)  f_\infty\; . \label{asint2}
\end{align}
For \eq\eqref{asint2} and the respective $y$-component we obtain, again using App.~\ref{asympapp},
\begin{align}
    \mathcal{K}^x_{n0}&= \alpha_{n0} \tilde{\mathcal{J}}^{xz}\; ,  \qquad \qquad  \qquad \qquad \qquad  \text{for } n=1,3\; ,\n\\
   \mathcal{K}^x_{n1}&= \alpha_{n1} \tilde{\mathcal{J}}^{z0}+\eta_{n1}\varpi^{xy}-\tilde{\alpha}_{n1}\sigma \; , \qquad \text{for } n=2,4 \; ,\n\\
       \mathcal{K}^y_{n0}&= \alpha_{n0} \tilde{\mathcal{J}}^{yz}\; ,  \qquad \qquad  \qquad \qquad \qquad  \text{for } n=1,3\; ,\n\\
   \mathcal{K}^y_{n1}&= \bar{\alpha}_{n1} \tilde{\mathcal{J}}^{z0}+\bar{\eta}_{n1}\varpi^{xy}+\hat{\alpha}_{n1}\sigma \; , \qquad \text{for } n=2,4 \; , \label{kxy}
\end{align}
where the coefficients are also defined in App.~\ref{coeffapp}.

Finally, we consider the equations of motion for the total angular momentum. Note that $\tilde{\mathcal{J}}^{x0}$, $\tilde{\mathcal{J}}^{y0}$, and $\tilde{\mathcal{J}}^{xy}$ do not appear in \eqs\eqref{gzinfty}, \eqref{gxyinfty}, \eqref{kdd}, and \eqref{kxy}. Therefore, we need only equations of motion for $\tilde{\mathcal{J}}^{xz}$, $\tilde{\mathcal{J}}^{yz}$, and $\tilde{\mathcal{J}}^{z0}$, which will be derived from \eq\eqref{angmomeom} in the following.
First, we note that the two terms from the antisymmetrization of the Lorentz indices in \eq\eqref{angmomeom} decay differently in the late free-streaming regime with $\cos\theta=0$. To be able to consider these two contributions separately, we split, e.g., $\tilde{\mathcal{J}}^{zx}$ into
\begin{equation}
\tilde{\mathcal{J}}^{zx}\equiv \tilde{j}^{zx}-\tilde{j}^{xz}
\label{splitjzx}
\end{equation}
with
\begin{equation}
    \tilde{j}^{\mu\nu}\equiv \frac{\hbar}{4} \int_{p\ms} \frac{1}{E_p^2} \frac{1}{m} \left(1+\frac{E_p}{2(E_p+m)}\right)p^{\mu}  \ms^{\nu} f \; .
    \label{littlej}
\end{equation}
In App.~\ref{littlejapp} we analyze the equations of motion for the relevant components of $\tilde{j}^{\mu\nu}$. For $\mu=z$, $\nu=x$ we find
\begin{align}
\partial_\tau \tilde{j}^{zx} 
    &\simeq-2 \frac{1}{\tau} \tilde{j}^{zx}\; ,
\end{align}
for free streaming with $\cos\theta\simeq 0$. On the other hand, for $\mu=x$, $\nu=z$ we obtain
\begin{align}
  \tilde{j}^{xz} 
     &\simeq -\frac{1}{\tau} \tilde{j}^{xz}\; .
\end{align}
The $z$-$y$-components show the same behavior. Furthermore, we split $\tilde{\mathcal{J}^{z0}}$ into
\begin{equation}
  \tilde{\mathcal{J}^{z0}}\equiv \tilde{j}^{z0}-\tilde{j}^{0z}_+  
\end{equation}
with
\begin{equation}
    \tilde{j}_+^{0z}\equiv \tilde{j}^{0z}-\frac\hbar4 \int_{p\ms} \frac{1}{E_p^2}   \frac{m(E_p-m)}{2p^2} \ms^{z} f \; .
    \label{littlej+}
\end{equation}
Using these definitions, we show in App.~\ref{littlejapp} that the decay in the late free-streaming regime is determined by
\begin{align}
\partial_\tau \tilde{j}^{z0} 
    &\simeq-\frac{1}{\tau} 2 \tilde{j}^{z0}\; ,\n\\
\partial_\tau \tilde{j}^{0z}_+    
     &\simeq-\frac{1}{\tau} \tilde{j}^{0z}_+\; .
\end{align}
The complete equations of motion for $\tilde{j}^{\mu\nu}$ and $\tilde{j}^{0z}_+$ are then again obtained by an interpolation between the late free-streaming and the hydrodynamic regime. The results can be found in App.~\ref{littlejapp}. Note that $\tilde{j}^{zx}-\tilde{j}^{zx}_\infty$ in general is nonvanishing, while $\tilde{j}^{[zx]}-\tilde{j}^{[zx]}_\infty=0$ according to \eq\eqref{jinfty}. However, due to the symmetry of $f_\infty$, we have 
\begin{equation}
\tilde{j}^{zx}_\infty=-\tilde{j}^{xz}_\infty=\frac12 \tilde{\mathcal{J}}^{zx}_\infty \; .
\end{equation}
We now define
\begin{equation}
    \tilde{\mathcal{J}}_+^{zx}\equiv \tilde{j}^{zx}+\tilde{j}^{xz}\; ,
    \label{j+}
\end{equation}
and treat this quantity as a dynamical variable in addition to $\tilde{\mathcal{J}}^{zx}$. Using \eqs\eqref{eomsmjxzapp} and \eqref{eomsmjzxapp} with $\tilde{j}^{zx}=(1/2)(\tilde{\mathcal{J}}^{zx}+\tilde{\mathcal{J}}^{zx}_+)$, $\tilde{j}^{xz}=(1/2)(-\tilde{\mathcal{J}}^{zx}+\tilde{\mathcal{J}}^{zx}_+)$, we obtain
\begin{align}
    \partial_w \tilde{\mathcal{J}}^{zx}&=-\frac{1}{w} \frac12\left(3 \tilde{\mathcal{J}}^{zx}+ \tilde{\mathcal{J}}^{zx}_+ \right)+\frac{1}{w}\left(1-e^{-w/2}\right) \Lambda_{x} \tilde{\mathcal{J}}^{zx}\; ,\n\\
    \partial_w \tilde{\mathcal{J}}^{zx}_+
&=-\frac{1}{w} \frac12\left( \tilde{\mathcal{J}}^{zx}+ 3\tilde{\mathcal{J}}^{zx}_+ \right)- \tilde{\mathcal{J}}^{zx}_+  \; ,
\label{eomjplus}
\end{align}
where the calculation and the coefficient $\Lambda_x$ are shown in App.~\ref{littlejapp}.
 The equations of motion for the $z$-$y$-components are analogous. For the 0-$z$-components we define
\begin{equation}
    \tilde{\mathcal{J}}_+^{z0}\equiv \tilde{j}^{z0}+\tilde{j}^{0z}_+ \; .
    \label{jz0+}
\end{equation}
Note that, as any asymptotic quantity, $\tilde{\mJ}_{+,\infty}^{z0}$ can be expressed as a function of $\tilde{\mJ}^{z0}$ and gradients of $\beta^\mu$. We obtain
\begin{align}
    \tilde{\mathcal{J}}^{z0}_{+,\infty}
     &=\Gamma_\Omega \tilde{\mathcal{J}}^{z0}+\Gamma_\varpi \varpi^{xy} \; ,
\end{align}
see App.~\ref{littlejapp} for details and the definitions of the coefficients. There, we also show that, following the same procedure as before, the closed equations of motion are 
\begin{align}
  \partial_w  \tilde{\mathcal{J}}^{z0} 
&= -\frac{1}{w} \frac12\left(3 \tilde{\mathcal{J}}^{z0}+ \tilde{\mathcal{J}}^{z0}_+ \right)+\frac{1}{w}\left(1-e^{-w/2}\right) \left(\Lambda_0  \tilde{\mathcal{J}}^{z0}+\Lambda_\varpi \varpi^{xy} \right)\; ,\n\\
  \partial_w  \tilde{\mathcal{J}}_+^{z0}
&= -\frac{1}{w} \frac12\left( \tilde{\mathcal{J}}^{z0}+3 \tilde{\mathcal{J}}^{z0}_+ \right)+\frac{1}{w}\left(1-e^{-w/2}\right) \left(\bar{\Lambda}_0  \tilde{\mathcal{J}}^{z0}+\bar{\Lambda}_\varpi \varpi^{xy} \right)-\left(\tilde{\mathcal{J}}^{z0}_+- \Gamma_\Omega \tilde{\mathcal{J}}^{z0}-\Gamma_\varpi \varpi^{xy}\right)\; 
\label{eomtjz0}
\end{align}
with the coefficients also defined in App.~\ref{littlejapp}.

\section{Longitudinal polarization from closed moment equation}
\label{finsec}

Finally, we obtain the longitudinal polarization by inserting \eqs\eqref{pi1stterm} and \eqref{zpollt} into \eq\eqref{pistar} and keeping only the spin moments \eqref{1stordernonzero},
\begin{align}
    \Pi_\star^z(\phi)&= \frac{1}{\mathcal{N}} \Bigg\{ \frac12\sum_{n=0,2} \left(\aleph_n\mG_{n00}^z+\bar{\aleph}_n\mI^z_{n00}\right)+\sum_{n=2,4}\left[\tilde{\aleph}_n \text{Re}  \left(\mI^x_{n10}-\mG^x_{n10}+\mI^y_{n10}-\mG^y_{n10}\right)\right]\n\\
    &+\sum_{n=1,3} \text{Re}\left[\left(\beth_n  \mG^z_{n10}+\bar{\beth}_n \mI^z_{n10}\right) e^{i\phi} \right]+\sum_{n=1,3}\tilde{\beth}_n \text{Re} \left[ \left(\mI^x_{n00}-\mG^x_{n00}+\mI^y_{n00}-\mG^y_{n00}\right)e^{i\phi}\right] \n\\
    &+\sum_{n=2,4}  \text{Re}\left[\left(\gimel_n \mG^z_{n20}+\bar{\gimel}_n \mI^z_{n20}\right) e^{2i\phi}\right]+\sum_{n=2,4} \tilde{\gimel}_n \text{Re} \left[ \left(\mI^x_{n10}-\mG^x_{n10}+\mI^y_{n10}-\mG^y_{n10}\right)e^{2i\phi}\right]\Bigg\}
   \label{polfin}
\end{align}
with the coefficients defined in App.~\ref{coeffapp}. 
For convenience, we collect the closed equations of motion for the spin moments appearing in \eq\eqref{polfin} in App.~\ref{finapp}. In particular, we show also the equations of motion for the $\mI$-moments in App.~\ref{finapp}, which can be obtained from \eqs\eqref{eomiz} and \eqref{eomix} following the same steps as for the $\mG$-moments. We then obtain a closed set of 36 equations for the spin moments. Although they are many, these are simple linear equations whose solution should present no particular difficulty.  Once the zeroth-order equations of motion for the background fluid, which are unaffected by the spin-dependent terms and which determine the fluid velocity and the temperature, are solved, one can compute the solution of the equations of motion in App.~\ref{finapp} on top to explicitly obtain the spin moments in \eqref{polfin}. We also remark that the polarization \eqref{polfin} is a simple function of the momentum angle $\phi$. 

Note that for $w\rightarrow\infty$ the polarization \eqref{polfin} reduces to the Lorentz transform to the particle rest frame of
\begin{align}
\Pi^z_{\infty}  &= -\frac{1}{\mathcal{N}} E_p\left[ \frac{\hbar}{m}\tilde{\Omega}^{z\mu} p_\mu  +\xi\frac{2\hbar}{m(E_p+m)}\epsilon^{\mu\rho\alpha z}p_\rho t_\alpha p^\nu (\Omega_{\mu\nu}+\partial_\mu \beta_\nu)\right] f_\text{LE}^{(0)}
\label{piinf}
\end{align}
where we used \eq\eqref{intdsfinfty}. As already discussed in Ref.~\cite{Weickgenannt:2024ibf}, this expression agrees with the local-equilibrium polarization derived from the Zubarev approach in Refs.~\cite{Becattini:2021suc,Buzzegoli:2021wlg} for an appropriate choice of $\xi$. Equation \eqref{polfin} provides the first-order correction to \eq\eqref{piinf} for an expanding and rotating system. We remark that in our approach, we obtain the results of Refs.~\cite{Becattini:2021suc,Buzzegoli:2021wlg} at the leading order, while the correction terms are of the next order in a gradient expansion, and thus are expected to be smaller, while their effect is still to be quantified.
Also note that the choice $\bxi=1/m$ corresponds to the canonical pseudo-gauge in the Zubarev formalism~\cite{Weickgenannt:2024ibf}. 
Although the system considered in this paper is highly idealized, we believe that the mechanism of generating spin polarization from fluid gradients through nonlocal collisions is universal, and we expect that \eq\eqref{polfin} can provide a useful  orientation regarding  the role of first-order effects in this context.

\section{Conclusions}
\label{concsec}

In this paper, we derived an expression for the longitudinal spin polarization of an expanding and rotating system in terms of dynamical spin moments, i.e., moments of the distribution function weighted with a linear power of the spin three vector ($\ms^z$ for longitudinal spin moments and $\ms^x$ or $\ms^y$ for transverse spin moments) and arbitrary powers of momentum. Furthermore, we obtained closed equations of motion for the spin moments. We assumed that the expansion in the longitudinal direction is boost invariant, and that the fluid motion in the transverse plane is purely vortical. Both the longitudinal expansion and the rotation in the transverse plane influence the polarization in $z$-direction. This paper can be regarded as a follow-up of the work done in Ref.~\cite{Weickgenannt:2023bss}, where we studied the transverse polarization of an expanding system without rotation, neglecting nonlocal effects in particle collisions  and assuming that the longitudinal polarization vanishes. In contrast, in this work we allow for nonzero vorticity, polarization in any direction, and nonlocal collisions.  While the explicit form of the polarization vector derived in this paper is valid at first order in the ratio of the relaxation time $\tau_R$ to the proper time $\tau$, $w^{-1}$, the equations of motion capture the dynamics of the full evolution of the system from free streaming to hydrodynamics. Since we work up to first order in $\hbar$, the fluid motion is not affected by the presence of spin polarization and can be regarded as a standard  background.

Analogously to what was done in Ref.~\cite{Weickgenannt:2023bss}, we expanded the polarization in terms of spherical harmonics and spin moments. In heavy-ion collisions, one is interested in the dependence of the longitudinal polarization on $\phi$, the azimuthal angle  of the momentum vector. Within the imposed symmetry, we found that up to first order in $w^{-1}$ the polarization vector is a superposition of terms independent of $\phi$, proportional to $\exp(i\phi)$, and proportional to $\exp(2i\phi)$, each weighted with a linear combination of spin moments. Spherical harmonics of higher orders do not appear in the polarization up to first order in $w^{-1}$. In turn, this implies that a potential measurement of higher spherical harmonics would indicate that higher powers of $w^{-1}$ are relevant.

Since the polarization is measured in the particle rest frame, one has to transform the polarization vector from the lab frame to the particle rest frame. Since the corresponding Lorentz transformation mixes longitudinal and transverse components, the longitudinal polarization in the particle rest frame also depends on transverse spin moments, which are defined in the lab frame.  However, in the late free-streaming regime, the transverse spin moments which are part of the longitudinal polarization decay as $\tau^{-2}$, whereas the longitudinal spin moments decay as $ \tau^{-1}$. Therefore, in the initial hydrodynamic regime the longitudinal spin moments provide the major contribution to the longitudinal polarization. At later time, the collision term determines the dynamics of the system. We considered nonlocal effects in particle collisions by using the nonlocal relaxation time approximation in the equations of motion for the spin moments. Through the nonlocal part of the collision term, the long-time limits of the spin moments, and hence also of the polarization vector, gain contributions from the thermal vorticity and the thermal shear. These contributions are in agreement with Refs.~\cite{Becattini:2021suc,Buzzegoli:2021wlg} when taking into account that, in the present work, certain terms vanish due to symmetry restrictions. In addition, the asymptotic spin moments depend on the spin potential, which is expressed as a function of the total angular momentum through the matching condition. The components of the total angular momentum are treated as additional dynamical variables which follow equations of motion similar to those for the spin moments. In the late-time limit, the total angular momentum is proportional to the thermal vorticity, as expected. To model the transition from the late free-streaming to the hydrodynamic regime in all equations of motion, we followed an already established strategy and used an interpolation which has been shown to successfully reproduce the exact solution for the same type of equations of motion in Refs.~\cite{Jaiswal:2022udf,Weickgenannt:2023nge}. Since we consistently include all terms of first order in gradients in the polarization vector, our work can help to understand whether the dissipative contributions play a significant role compared to the ideal ones and hence need to be added to the results of Refs.~\cite{Liu:2021uhn,Fu:2021pok,Becattini:2021suc,Becattini:2021iol}.

The closed equations of motion for a finite number of dynamical variables provided in this work can be solved for a certain choice of initial conditions  with much less numerical effort than solving the Boltzmann equation would require. Thus, one can obtain the longitudinal polarization as a function of $\tau$ and $\phi$. The results may shed new light on the longitudinal polarization beyond local equilibrium in heavy-ion collisions, and in particular on the effects of the early-time dynamics and on the role of nonlocal particle collisions.

\section*{Acknowledgements}

N.W.\ acknowledges support by the German National Academy of Sciences Leopoldina through the Leopoldina fellowship program with funding code LPDS 2022-11.

\begin{appendix}

\section{Coefficients}
\label{coeffapp}

In this appendix, we collect the definitions of various coefficients.
The coefficients in \eq\eqref{zpollt} read
\begin{align}
    \mathfrak{d}^+_{n\ell}&\equiv-\frac12\frac{1}{2n+1}\left(\frac{n+\ell+2}{2n+3}-\frac{n-\ell-1}{2n-1}\right)\; ,\n\\
    \mathfrak{e}^+_{n\ell}&\equiv\frac12\frac{1}{2n+1}\frac{n+\ell}{2n-1}\; ,\n\\
    \mathfrak{f}^+_{n\ell}&\equiv -\frac12\frac{1}{2n+1} \frac{n-\ell+1}{2n+3}\; ,\n\\
    \mathfrak{d}^-_{n\ell}&\equiv \frac12\frac{1}{2n+1}\left[(n-\ell+1)(n-\ell+2)\frac{n+\ell}{2n+3}-(n+\ell-1)(n+\ell)\frac{n-\ell+1}{2n-1} \right]\; ,\n\\
    \mathfrak{e}^-_{n\ell}&\equiv -\frac12\frac{1}{2n+1} (n+\ell-1)(n+\ell)\frac{n+\ell-2}{2n-1}\; ,\n\\
    \mathfrak{f}^-_{n\ell}&\equiv \frac12\frac{1}{2n+1} (n-\ell+1)(n-\ell+2)\frac{n-\ell+3}{2n+3}\; ,\n\\
    \mathfrak{g}_{n\ell}&\equiv \frac{(n+1-\ell)(n+1+\ell)}{(2n+1)(2n+3)}+\frac{(n+\ell)(n-\ell)}{(2n+1)(2n-1)}\; ,\n\\
    \mathfrak{h}_{n\ell}&\equiv \frac{(n+\ell)(n-1+\ell)}{(2n+1)(2n-1)}\; ,\n\\
    \mathfrak{i}_{n\ell}&\equiv \frac{(n+1-\ell)(n+2-\ell)}{(2n+1)(2n+3)}\; .
    \label{somecoeffis}
\end{align}
The coefficients in \eq\eqref{eomgzfin} are given by
\begin{align}
     \zeta_{n\ell} &\equiv 1-n\left(\frac{(n+1-\ell)(n+1+\ell)}{(2n+1)(2n+3)}+\frac{(n+\ell)(n-\ell)}{(2n+1)(2n-1)}\right)+\frac{(n+\ell)(n-\ell)}{2n-1}\; ,\n\\
 \chi_{n\ell} & \equiv -n \frac{(n+\ell)(n-1+\ell)}{(2n+1)(2n-1)}+\frac{(n+\ell)(n-1+\ell)}{2n-1}\; ,\n\\
 \xi_{n\ell} &\equiv -n \frac{(n+1-\ell)(n+2-\ell)}{(2n+1)(2n+3)}\; ,\n\\
 \hat{\zeta}_{n\ell}&\equiv \zeta_{n\ell}+\frac{\Pc_{n+2}^\ell(0)}{\Pc_n^\ell(0)} \xi_{n\ell}\; .
 \label{zetachixi}
\end{align}

The coefficients in \eq\eqref{kdd} are defined as
\begin{align}
    a_{n\ell}&\equiv 2\left(\mathfrak{i}_{n\ell} \lambda_{(n+2)\ell}^{20}+\mathfrak{g}_{n\ell}\lambda_{n\ell}^{20}+\mathfrak{h}_{n\ell}\lambda_{(n-2)\ell}^{20}\right)+\mathfrak{f}^+_{n\ell}\left(\lambda_{(n+2)(\ell+1)}^{20}+i\bar{\lambda}_{(n+2)(\ell+1)}^{20}\right)+\mathfrak{d}^+_{n\ell} \left(\lambda_{n(\ell+1)}^{20}+i\bar{\lambda}_{n(\ell+1)}^{20}\right)\n\\
    &+\mathfrak{e}^+_{n\ell}\left(\lambda_{(n-2)(\ell+1)}^{20}+i\bar{\lambda}_{(n-2)(\ell+1)}^{20}\right)+\mathfrak{f}^-_{n\ell}\left(\lambda_{(n+2)(\ell-1)}^{20}+i\bar{\lambda}_{(n+2)(\ell-1)}^{20}\right)+\mathfrak{d}^-_{n\ell} \left(\lambda_{n(\ell-1)}^{20}+i\bar{\lambda}_{n(\ell-1)}^{20}\right)\n\\
    &+\mathfrak{e}^-_{n\ell}\left(\lambda_{(n-2)(\ell-1)}^{20}+i\bar{\lambda}_{(n-2)(\ell-1)}^{20}\right)\; ,\n\\
b_{n\ell}&\equiv 2\left(\mathfrak{i}_{n\ell} \kappa_{(n+2)\ell}^{20}+\mathfrak{g}_{n\ell}\kappa_{n\ell}^{20}+\mathfrak{h}_{n\ell}\kappa_{(n-2)\ell}^{20}\right)+\mathfrak{f}^+_{n\ell}\left(\kappa_{(n+2)(\ell+1)}^{20}+i\bar{\kappa}_{(n+2)(\ell+1)}^{20}\right)+\mathfrak{d}^+_{n\ell} \left(\kappa_{n(\ell+1)}^{20}+i\bar{\kappa}_{n(\ell+1)}^{20}\right)\n\\
    &+\mathfrak{e}^+_{n\ell}\left(\kappa_{(n-2)(\ell+1)}^{20}+i\bar{\kappa}_{(n-2)(\ell+1)}^{20}\right)+\mathfrak{f}^-_{n\ell}\left(\kappa_{(n+2)(\ell-1)}^{20}+i\bar{\kappa}_{(n+2)(\ell-1)}^{20}\right)+\mathfrak{d}^-_{n\ell} \left(\kappa_{n(\ell-1)}^{20}+i\bar{\kappa}_{n(\ell-1)}^{20}\right)\n\\
    &+\mathfrak{e}^-_{n\ell}\left(\kappa_{(n-2)(\ell-1)}^{20}+i\bar{\kappa}_{(n-2)(\ell-1)}^{20}\right)\; ,\n\\
c_{n\ell}&\equiv -\left(\mathfrak{f}^+_{n\ell}\tilde{\kappa}_{(n+2)(\ell+1)}^{20}+\mathfrak{d}^+_{n\ell} \tilde{\kappa}_{n(\ell+1)}^{20}+\mathfrak{e}^+_{n\ell}\tilde{\kappa}_{(n-2)(\ell+1)}^{20}+\mathfrak{f}^-_{n\ell}\tilde{\kappa}_{(n+2)(\ell-1)}^{20}+\mathfrak{d}^-_{n\ell} \tilde{\kappa}_{n(\ell-1)}^{20}+\mathfrak{e}^-_{n\ell}\tilde{\kappa}_{(n-2)(\ell-1)}^{20}\right)\; ,\n\\
& +i\left(\mathfrak{f}^+_{n\ell}\hat{\kappa}_{(n+2)(\ell+1)}^{20}+\mathfrak{d}^+_{n\ell} \hat{\kappa}_{n(\ell+1)}^{20}+\mathfrak{e}^+_{n\ell}\hat{\kappa}_{(n-2)(\ell+1)}^{20}+\mathfrak{f}^-_{n\ell}\hat{\kappa}_{(n+2)(\ell-1)}^{20}+\mathfrak{d}^-_{n\ell} \hat{\kappa}_{n(\ell-1)}^{20}+\mathfrak{e}^-_{n\ell}\hat{\kappa}_{(n-2)(\ell-1)}^{20}\right)\; ,\n\\
\mathfrak{a}_{n\ell}&\equiv 2\left(\mathfrak{i}_{n\ell} \lambda_{(n+2)\ell}^{20}+\mathfrak{g}_{n\ell}\lambda_{n\ell}^{20}+\mathfrak{h}_{n\ell}\lambda_{(n-2)\ell}^{20}\right)+\mathfrak{f}^+_{n\ell}\lambda_{(n+2)(\ell+1)}^{20}+\mathfrak{d}^+_{n\ell} \lambda_{n(\ell+1)}^{20}+\mathfrak{e}^+_{n\ell}\lambda_{(n-2)(\ell+1)}^{20}+\mathfrak{f}^-_{n\ell}\lambda_{(n+2)(\ell-1)}^{20}\n\\
&+\mathfrak{d}^-_{n\ell} \lambda_{n(\ell-1)}^{20}+\mathfrak{e}^-_{n\ell}\lambda_{(n-2)(\ell-1)}^{20}\; ,\n\\
\bar{\mathfrak{a}}_{n\ell}&\equiv 2\left(\mathfrak{i}_{n\ell} \bar{\lambda}_{(n+2)\ell}^{20}+\mathfrak{g}_{n\ell}\bar{\lambda}_{n\ell}^{20}+\mathfrak{h}_{n\ell}\bar{\lambda}_{(n-2)\ell}^{20}\right)-i\left(\mathfrak{f}^+_{n\ell}\lambda_{(n+2)(\ell+1)}^{20}+\mathfrak{d}^+_{n\ell} \lambda_{n(\ell+1)}^{20}+\mathfrak{e}^+_{n\ell}\lambda_{(n-2)(\ell+1)}^{20}+\mathfrak{f}^-_{n\ell}\lambda_{(n+2)(\ell-1)}^{20}\right.\n\\
&\left.+\mathfrak{d}^-_{n\ell} \lambda_{n(\ell-1)}^{20}+\mathfrak{e}^-_{n\ell}\lambda_{(n-2)(\ell-1)}^{20} \right)\; . 
\label{abc}
\end{align}
The coefficients in \eq\eqref{eomizfinfin} are defined as
\begin{align}
    a_{n\ell}^\prime&\equiv \left(\mathfrak{i}_{n\ell} \lambda_{(n+2)\ell}^{21}+\mathfrak{g}_{n\ell}\lambda_{n\ell}^{21}+\mathfrak{h}_{n\ell}\lambda_{(n-2)\ell}^{21}\right)+\mathfrak{f}^+_{n\ell}\left(\lambda_{(n+2)(\ell+1)}^{21}+i\bar{\lambda}_{(n+2)(\ell+1)}^{21}\right)+\mathfrak{d}^+_{n\ell} \left(\lambda_{n(\ell+1)}^{21}+i\bar{\lambda}_{n(\ell+1)}^{21}\right)\n\\
    &+\mathfrak{e}^+_{n\ell}\left(\lambda_{(n-2)(\ell+1)}^{21}+i\bar{\lambda}_{(n-2)(\ell+1)}^{21}\right)+\mathfrak{f}^-_{n\ell}\left(\lambda_{(n+2)(\ell-1)}^{21}+i\bar{\lambda}_{(n+2)(\ell-1)}^{21}\right)+\mathfrak{d}^-_{n\ell} \left(\lambda_{n(\ell-1)}^{21}+i\bar{\lambda}_{n(\ell-1)}^{21}\right)\n\\
    &+\mathfrak{e}^-_{n\ell}\left(\lambda_{(n-2)(\ell-1)}^{21}+i\bar{\lambda}_{(n-2)(\ell-1)}^{21}\right)\; ,\n\\
b_{n\ell}^\prime&\equiv \left(\mathfrak{i}_{n\ell} \kappa_{(n+2)\ell}^{21}+\mathfrak{g}_{n\ell}\kappa_{n\ell}^{21}+\mathfrak{h}_{n\ell}\kappa_{(n-2)\ell}^{21}\right)+\mathfrak{f}^+_{n\ell}\left(\kappa_{(n+2)(\ell+1)}^{21}+i\bar{\kappa}_{(n+2)(\ell+1)}^{21}\right)+\mathfrak{d}^+_{n\ell} \left(\kappa_{n(\ell+1)}^{21}+i\bar{\kappa}_{n(\ell+1)}^{21}\right)\n\\
    &+\mathfrak{e}^+_{n\ell}\left(\kappa_{(n-2)(\ell+1)}^{21}+i\bar{\kappa}_{(n-2)(\ell+1)}^{21}\right)+\mathfrak{f}^-_{n\ell}\left(\kappa_{(n+2)(\ell-1)}^{21}+i\bar{\kappa}_{(n+2)(\ell-1)}^{21}\right)+\mathfrak{d}^-_{n\ell} \left(\kappa_{n(\ell-1)}^{21}+i\bar{\kappa}_{n(\ell-1)}^{21}\right)\n\\
    &+\mathfrak{e}^-_{n\ell}\left(\kappa_{(n-2)(\ell-1)}^{21}+i\bar{\kappa}_{(n-2)(\ell-1)}^{21}\right)\; ,\n\\
c_{n\ell}^\prime&\equiv -\left(\mathfrak{f}^+_{n\ell}\tilde{\kappa}_{(n+2)(\ell+1)}^{21}+\mathfrak{d}^+_{n\ell} \tilde{\kappa}_{n(\ell+1)}^{21}+\mathfrak{e}^+_{n\ell}\tilde{\kappa}_{(n-2)(\ell+1)}^{21}+\mathfrak{f}^-_{n\ell}\tilde{\kappa}_{(n+2)(\ell-1)}^{21}+\mathfrak{d}^-_{n\ell} \tilde{\kappa}_{n(\ell-1)}^{21}+\mathfrak{e}^-_{n\ell}\tilde{\kappa}_{(n-2)(\ell-1)}^{21}\right)\n\\
&+i\left(\mathfrak{f}^+_{n\ell}\hat{\kappa}_{(n+2)(\ell+1)}^{21}+\mathfrak{d}^+_{n\ell} \hat{\kappa}_{n(\ell+1)}^{21}+\mathfrak{e}^+_{n\ell}\hat{\kappa}_{(n-2)(\ell+1)}^{21}+\mathfrak{f}^-_{n\ell}\hat{\kappa}_{(n+2)(\ell-1)}^{21}+\mathfrak{d}^-_{n\ell} \hat{\kappa}_{n(\ell-1)}^{21}+\mathfrak{e}^-_{n\ell}\hat{\kappa}_{(n-2)(\ell-1)}^{21}\right)\; ,\n\\
\mathfrak{a}_{n\ell}^\prime&\equiv \left(\mathfrak{i}_{n\ell} \lambda_{(n+2)\ell}^{21}+\mathfrak{g}_{n\ell}\lambda_{n\ell}^{21}+\mathfrak{h}_{n\ell}\lambda_{(n-2)\ell}^{21}\right)+\mathfrak{f}^+_{n\ell}\lambda_{(n+2)(\ell+1)}^{21}+\mathfrak{d}^+_{n\ell} \lambda_{n(\ell+1)}^{21}+\mathfrak{e}^+_{n\ell}\lambda_{(n-2)(\ell+1)}^{21}+\mathfrak{f}^-_{n\ell}\lambda_{(n+2)(\ell-1)}^{21}\n\\
&+\mathfrak{d}^-_{n\ell} \lambda_{n(\ell-1)}^{21}+\mathfrak{e}^-_{n\ell}\lambda_{(n-2)(\ell-1)}^{21}\; ,\n\\
\bar{\mathfrak{a}}_{n\ell}^\prime&\equiv \left(\mathfrak{i}_{n\ell} \bar{\lambda}_{(n+2)\ell}^{21}+\mathfrak{g}_{n\ell}\bar{\lambda}_{n\ell}^{21}+\mathfrak{h}_{n\ell}\bar{\lambda}_{(n-2)\ell}^{21}\right)-i\left(\mathfrak{f}^+_{n\ell}\lambda_{(n+2)(\ell+1)}^{21}+\mathfrak{d}^+_{n\ell} \lambda_{n(\ell+1)}^{21}+\mathfrak{e}^+_{n\ell}\lambda_{(n-2)(\ell+1)}^{21}+\mathfrak{f}^-_{n\ell}\lambda_{(n+2)(\ell-1)}^{21}\right.\n\\
&\left.+\mathfrak{d}^-_{n\ell} \lambda_{n(\ell-1)}^{21}+\mathfrak{e}^-_{n\ell}\lambda_{(n-2)(\ell-1)}^{21} \right)\; .
\label{abcprime}
\end{align}

The coefficients in \eq\eqref{eomgxfin} are
\begin{align}
 \bar{a}_{n\ell r} &\equiv 1-(n-r)\left(\frac{(n+1-\ell)(n+1+\ell)}{(2n+1)(2n+3)}+\frac{(n+\ell)(n-\ell)}{(2n+1)(2n-1)}\right)+\frac{(n+\ell)(n-\ell)}{2n-1}\; ,\n\\
 \bar{b}_{n\ell r} & \equiv -(n-r) \frac{(n+\ell)(n-1+\ell)}{(2n+1)(2n-1)}+\frac{(n+\ell)(n-1+\ell)}{2n-1}\; ,\n\\
 \bar{c}_{n\ell r} &\equiv -(n-r) \frac{(n+1-\ell)(n+2-\ell)}{(2n+1)(2n+3)}\; ,\n\\
  \bar{d}_{n\ell r} &\equiv -(r-1)\left(\frac{(n+1-\ell)(n+1+\ell)}{(2n+1)(2n+3)}+\frac{(n+\ell)(n-\ell)}{(2n+1)(2n-1)}\right)\; ,\n\\
 \bar{e}_{n\ell r} & \equiv -(r-1) \frac{(n+\ell)(n-1+\ell)}{(2n+1)(2n-1)}\; ,\n\\
 \bar{f}_{n\ell r} &\equiv -(r-1) \frac{(n+1-\ell)(n+2-\ell)}{(2n+1)(2n+3)}\; .\label{abcdefbar}
\end{align}
We also introduced
\begin{equation}
    \hat{a}_{n\ell r} \equiv \bar{a}_{n\ell r}+ \frac{(n+2)\Pc_{n+1}^{\ell}(0)}{n\Pc_{n-1}^\ell(0)}\bar{c}_{n\ell r} \; . 
\end{equation}

The coefficients in \eq\eqref{kxy} read
\begin{align}
    \alpha_{n\ell}&\equiv 2\left(\mathfrak{i}_{n\ell} \lambda_{(n+2)\ell}^{20}+\mathfrak{g}_{n\ell}\lambda_{n\ell}^{20}+\mathfrak{h}_{n\ell}\lambda_{(n-2)\ell}^{20}\right)\; ,\n\\
    \eta_{n\ell}&= 2\left(\mathfrak{i}_{n\ell} \kappa_{(n+2)\ell}^{20}+\mathfrak{g}_{n\ell}\kappa_{n\ell}^{20}+\mathfrak{h}_{n\ell}\kappa_{(n-2)\ell}^{20}\right)\; ,\n\\
    \hat{\alpha}_{n\ell}&= 2\left(\mathfrak{i}_{n\ell} \hat{\kappa}_{(n+2)\ell}^{20}+\mathfrak{g}_{n\ell}\hat{\kappa}_{n\ell}^{20}+\mathfrak{h}_{n\ell}\hat{\kappa}_{(n-2)\ell}^{20}\right)\; ,\n\\
    \tilde{\alpha}_{n\ell}&= 2\left(\mathfrak{i}_{n\ell} \tilde{\kappa}_{(n+2)\ell}^{20}+\mathfrak{g}_{n\ell}\tilde{\kappa}_{n\ell}^{20}+\mathfrak{h}_{n\ell}\tilde{\kappa}_{(n-2)\ell}^{20}\right)\; ,\n\\
     \bar{\alpha}_{n\ell}&\equiv 2\left(\mathfrak{i}_{n\ell} \bar{\lambda}_{(n+2)\ell}^{20}+\mathfrak{g}_{n\ell}\bar{\lambda}_{n\ell}^{20}+\mathfrak{h}_{n\ell}\bar{\lambda}_{(n-2)\ell}^{20}\right)\; ,\n\\
    \bar{\eta}_{n\ell}&= 2\left(\mathfrak{i}_{n\ell} \bar{\kappa}_{(n+2)\ell}^{20}+\mathfrak{g}_{n\ell}\bar{\kappa}_{n\ell}^{20}+\mathfrak{h}_{n\ell}\bar{\kappa}_{(n-2)\ell}^{20}\right)\; .
    \label{alphaeta}
\end{align}

The coefficients in \eq\eqref{polfin} are given by
\begin{align}
    \aleph_n &\equiv N_{n0} \mathfrak{I}_{n0} (1-\mathfrak{g}_{n0})-N_{(n+2)0} \mathfrak{I}_{(n+2)0}\mathfrak{h}_{(n+2)0}-N_{(n-2)0} \mathfrak{I}_{(n-2)0}\mathfrak{i}_{(n-2)0}=\delta_{n0}\frac23-\delta_{n2}\frac{12}{35}\; ,\n\\
    \bar{\aleph}_n& \equiv N_{n0} \mathfrak{I}_{n0}\mathfrak{g}_{n0}+N_{(n+2)0} \mathfrak{I}_{(n+2)0}\mathfrak{h}_{(n+2)0}+N_{(n-2)0} \mathfrak{I}_{(n-2)0}\mathfrak{i}_{(n-2)0}=\delta_{n0}\frac13+\delta_{n2}\frac{12}{35}\; ,\n\\
    \tilde{\aleph}_n&\equiv N_{n0} \mathfrak{I}_{n0}\mathfrak{d}^+_{n0}+N_{(n+2)0} \mathfrak{I}_{(n+2)0}\mathfrak{e}^+_{(n+2)0}+N_{(n-2)0} \mathfrak{I}_{(n-2)0}\mathfrak{f}^+_{(n-2)0}\n\\
    \beth_n &\equiv  N_{n1} \mathfrak{I}_{n1} (1-\mathfrak{g}_{n1})-N_{(n+2)1} \mathfrak{I}_{(n+2)1}\mathfrak{h}_{(n+2)1}-N_{(n-2)1} \mathfrak{I}_{(n-2)1}\mathfrak{i}_{(n-2)1}\; ,\n\\
    \bar{\beth}_n&\equiv N_{n1} \mathfrak{I}_{n1}\mathfrak{g}_{n1}+N_{(n+2)1} \mathfrak{I}_{(n+2)1}\mathfrak{h}_{(n+2)1}+N_{(n-2)1} \mathfrak{I}_{(n-2)1}\mathfrak{i}_{(n-2)1}\; ,\n\\
    \tilde{\beth}_n&\equiv N_{n1} \mathfrak{I}_{n1}\mathfrak{d}_{n1}^-+N_{(n+2)1} \mathfrak{I}_{(n+2)1}\mathfrak{e}^-_{(n+2)1}+N_{(n-2)1} \mathfrak{I}_{(n-2)1}\mathfrak{f}^-_{(n-2)1}\n\\
    \gimel_n &\equiv  N_{n2} \mathfrak{I}_{n2} (1-\mathfrak{g}_{n2})-N_{(n+2)2} \mathfrak{I}_{(n+2)2}\mathfrak{h}_{(n+2)2}-N_{(n-2)2} \mathfrak{I}_{(n-2)2}\mathfrak{i}_{(n-2)2}\; ,\n\\
    \bar{\gimel}_n&\equiv N_{n2} \mathfrak{I}_{n2}\mathfrak{g}_{n2}+N_{(n+2)1} \mathfrak{I}_{(n+2)2}\mathfrak{h}_{(n+2)2}+N_{(n-2)2} \mathfrak{I}_{(n-2)2}\mathfrak{i}_{(n-2)2}\; ,\n\\
    \tilde{\gimel}_n&\equiv N_{n2} \mathfrak{I}_{n2}\mathfrak{d}^-_{n2}+N_{(n+2)1} \mathfrak{I}_{(n+2)2}\mathfrak{e}^-_{(n+2)2}+N_{(n-2)2} \mathfrak{I}_{(n-2)2}\mathfrak{f}^-_{(n-2)2}\; .\n\\
\end{align}
Here we defined
\begin{equation}
    \mathfrak{I}_{n\ell}\equiv \int d\cos\theta\, \Pc_n^\ell(\cos\theta)
\end{equation}
Note that $\mathfrak{I}_{n\ell}=0$ for $n+\ell$ odd.

    \section{Calculation of the equations of motion}
    \label{eomapp}

    In this appendix, we present the derivation of the exact equations of motion for the spin moments and the total angular momentum.
    The free-streaming part of \eq\eqref{eomgz} is obtained from the Boltzmann equation \eqref{boltzbjorkcomp} as follows,
\begin{align}
    \partial_\tau \mG^z_{n\ell r}=& \int_{p\ms} \left(\frac{p}{E_p}\right)^r \ms^z \mathcal{P}^\ell_n(\cos\theta) e^{i\ell\phi} \left(\frac{p_z}{\tau}\partial_{p_z}+\frac{p_z}{E_p^2}\frac{\mathbf{p}\cdot \boldsymbol{\ms}}{\tau}\partial_{\ms_z}\right)f\n\\
    =& \int_{p\ms} \left(\frac{p}{E_p}\right)^r \ms^z \mathcal{P}^\ell_n(\cos\theta) e^{i\ell\phi} \left(\frac{p_z}{\tau}\partial_{p_z}+\frac{p_z}{E_p^2}\frac{\mathbf{p}\cdot \boldsymbol{\ms}}{\tau}\partial_{\ms_z}\right)(F+\ms^0 A^0-\ms^x A^x-\ms^y A^y-\ms^z A^z)\n\\
    =& \int_{p\ms} \left(\frac{p}{E_p}\right)^r \ms^z \mathcal{P}^\ell_n(\cos\theta) e^{i\ell\phi} \frac{p_z}{\tau}\left[\ms^x\partial_{p_z} \left(\frac{p_x}{E_p} A^0-A^x\right)+\ms^y\partial_{p_z} \left(\frac{p_y}{E_p}A^0-A^y\right)
    \right.\n\\
   & \left.+\ms^z \partial_{p_z} \left(\frac{p_z}{E_p}A^0-A^z\right) +\frac{\ms^0}{E_p}\left(\frac{p_z}{E_p} A^0-A^z\right)\right]\n\\
    =& 4 \int_{p} {E_p} \left(\frac{p}{E_p}\right)^r  \mathcal{P}^\ell_n(\cos\theta) e^{i\ell\phi} \frac{p_z}{\tau} \left[\frac{p_xp_z}{m^2}\partial_{p_z} \left(\frac{p_x}{E_p} A^0-A^x\right)+\frac{p_z p_y}{m^2}\partial_{p_z} \left(\frac{p_y}{E_p}A^0-A^y\right)\right.\n\\
    &\left. + \left(1+\frac{p_z^2}{m^2}\right) \partial_{p_z} \left(\frac{p_z}{E_p}A^0-A^z\right)+\frac{p_z}{m^2}\left(\frac{p_z}{E_p} A^0-A^z\right)\right]\n\\
=& -4 \int_{p} {E_p} \left(\frac{p}{E_p}\right)^r  \mathcal{P}^\ell_n(\cos\theta) e^{i\ell\phi} \frac{1}{\tau} \left[2\frac{p_xp_z}{m^2} \left(\frac{p_x}{E_p} A^0-A^x\right)+2\frac{p_z p_y}{m^2} \left(\frac{p_y}{E_p}A^0-A^y\right)\right.\n\\
    &\left. + \left(1+3\frac{p_z^2}{m^2}\right) \left(\frac{p_z}{E_p}A^0-A^z\right)-\frac{p^2_z}{m^2}\left(\frac{p_z}{E_p} A^0-A^z\right)\right]\n\\
   & -4\frac{1}{\tau} \int_{p} {E_p} \left(\frac{p}{E_p}\right)^r  \mathcal{P}^\ell_n(\cos\theta) e^{i\ell\phi} \left[-(r-1)\frac{p_z^2}{E_p^2}+r\frac{p_z^2}{p^2}\right]  \left[\frac{p_xp_z}{m^2} \left(\frac{p_x}{E_p} A^0-A^x\right)+\frac{p_x p_z}{m^2} \left(\frac{p_y}{E_p}A^0-A^y\right)\right.\n\\
    &\left.+\left(1+\frac{p_z^2}{m^2}\right)   \left(\frac{p_z}{E_p}A^0-A^z\right)\right]\n\\
    &-4\frac{1}{\tau}\int_{p} {E_p} \left(\frac{p}{E_p}\right)^r  (\mathcal{P}^\ell_n)^\prime(\cos\theta) e^{i\ell\phi}  \cos\theta \left(1-\cos^2\theta\right)  \left[\frac{p_xp_z}{m^2} \left(\frac{p_x}{E_p} A^0-A^x\right)+\frac{p_x p_z}{m^2} \left(\frac{p_y}{E_p}A^0-A^y\right)\right.\n\\
    &\left. + \left(1+\frac{p_z^2}{m^2}\right)  \left(\frac{p_z}{E_p}A^0-A^z\right)\right]\n\\
=& -4 \int_{p} {E_p} \left(\frac{p}{E_p}\right)^r  \mathcal{P}^\ell_n(\cos\theta) e^{i\ell\phi} \frac{1}{\tau} \left[2\frac{p_xp_z}{m^2} \left(\frac{p_x}{E_p} A^0-A^x\right)+2\frac{p_z p_y}{m^2} \left(\frac{p_y}{E_p}A^0-A^y\right)\right.\n\\
    &\left. + 2\left(1+\frac{p_z^2}{m^2}\right) \left(\frac{p_z}{E_p}A^0-A^z\right)-\left(\frac{p_z}{E_p} A^0-A^z\right)\right]\n\\ 
     & -4\frac{1}{\tau} \int_{p} {E_p} \left(\frac{p}{E_p}\right)^r  \mathcal{P}^\ell_n(\cos\theta) e^{i\ell\phi} \left[-(r-1)\frac{p_z^2}{E_p^2}+r\frac{p_z^2}{p^2}\right]  \left[\frac{p_xp_z}{m^2} \left(\frac{p_x}{E_p} A^0-A^x\right)+\frac{p_x p_z}{m^2} \left(\frac{p_y}{E_p}A^0-A^y\right)\right.\n\\
    &\left.+\left(1+\frac{p_z^2}{m^2}\right)   \left(\frac{p_z}{E_p}A^0-A^z\right)\right]\n\\
    &-4\frac{1}{\tau}\int_{p} {E_p} \left(\frac{p}{E_p}\right)^r  (\mathcal{P}^\ell_n)^\prime(\cos\theta) e^{i\ell\phi}  \cos\theta \left(1-\cos^2\theta\right)  \left[\frac{p_xp_z}{m^2} \left(\frac{p_x}{E_p} A^0-A^x\right)+\frac{p_x p_z}{m^2} \left(\frac{p_y}{E_p}A^0-A^y\right)\right.\n\\
    &\left. + \left(1+\frac{p_z^2}{m^2}\right)  \left(\frac{p_z}{E_p}A^0-A^z\right)\right]\n\\
    =& -\frac{1}{\tau}  \int_{p\ms} \left(\frac{p}{E_p}\right)^r  \mathcal{P}^\ell_n(\cos\theta) e^{i\ell\phi}  \left(\ms^z+\frac{p_z}{E_p}\ms^0\right)f
    -\frac{1}{\tau}\int_{p\ms}  \ms^z \left(\frac{p}{E_p}\right)^r  \mathcal{P}^\ell_n(\cos\theta) e^{i\ell\phi} \cos^2\theta \left[r+(1-r)\frac{p^2}{E_p^2}\right] f\n\\
    &+\frac{1}{\tau}\int_{p\ms}  \ms^z \left(\frac{p}{E_p}\right)^r  e^{i\ell\phi}\cos\theta\left[ n\cos\theta\, \mathcal{P}_n^\ell(\cos\theta)-(n+\ell)\mathcal{P}^\ell_{(n-1)}(\cos\theta)\right] f\; .
\end{align}
Here we used
\begin{equation}
    \left(-\frac{p_z}{\tau}\partial_{p_z}-\frac{p_z}{E_p^2}\frac{\mathbf{p}\cdot \boldsymbol{\ms}}{\tau}\partial_{\ms_z}\right)\ms_\bot=0\; 
\end{equation}
and $p\cdot A=0$. We also made use of
\begin{equation}
  \int_{\ms} \ms^\mu \ms^\nu= 2 E_p \int dS(p)\, \ms^\mu \ms^\nu=-4 E_p\left(g^{\mu\nu}-\frac{p^\mu p^\nu}{m^2}\right)\;  \label{intds}
\end{equation}
to find
\begin{align}
    & \int_{p\ms} \left(\frac{p}{E_p}\right)^r \ms^z \mathcal{P}^\ell_n(\cos\theta) e^{i\ell\phi} \frac{p_z}{\tau}\left[\ms^x\partial_{p_z} \left(\frac{p_x}{E_p} A^0-A^x\right)+\ms^y\partial_{p_z} \left(\frac{p_y}{E_p}A^0-A^y\right)
    \right.\n\\
   & \left.+\ms^z \partial_{p_z} \left(\frac{p_z}{E_p}A^0-A^z\right) +\frac{\ms^0}{E_p}\left(\frac{p_z}{E_p} A^0-A^z\right)\right]\n\\
   =& -2\int_{p} 2E_p \left(\frac{p}{E_p}\right)^r  \mathcal{P}^\ell_n(\cos\theta) e^{i\ell\phi} \frac{p_z}{\tau}\left[-\frac{p^zp^x}{m^2}\partial_{p_z} \left(\frac{p_x}{E_p} A^0-A^x\right)-\frac{p^zp^y}{m^2}\partial_{p_z} \left(\frac{p_y}{E_p}A^0-A^y\right)
    \right.\n\\
   & \left.+\left(g^{zz}-\frac{p_z^2}{m^2}\right) \partial_{p_z} \left(\frac{p_z}{E_p}A^0-A^z\right) -\frac{1}{E_p}\frac{p^z p^0}{m^2}\left(\frac{p_z}{E_p} A^0-A^z\right)\right] \; .
\end{align}
Analogously, we obtain for the free-streaming part of \eq\eqref{eomgxgenpol}
\begin{align}
    \partial_\tau \mG^x_{n\ell r}=& \int_{p\ms} \left(\frac{p}{E_p}\right)^r \ms^x \mathcal{P}^\ell_n(\cos\theta) e^{i\ell\phi} \left(\frac{p_z}{\tau}\partial_{p_z}+\frac{p_z}{E_p^2}\frac{\mathbf{p}\cdot \boldsymbol{\ms}}{\tau}\partial_{\ms_z}\right)f\n\\
    =& \int_{p\ms} \left(\frac{p}{E_p}\right)^r \ms^x \mathcal{P}^\ell_n(\cos\theta) e^{i\ell\phi} \left(\frac{p_z}{\tau}\partial_{p_z}+\frac{p_z}{E_p^2}\frac{\mathbf{p}\cdot \boldsymbol{\ms}}{\tau}\partial_{\ms_z}\right)(F+\ms^0 A^0-\ms^x A^x-\ms^y A^y-\ms^z A^z)\n\\
    =& \int_{p\ms} \left(\frac{p}{E_p}\right)^r \ms^x \mathcal{P}^\ell_n(\cos\theta) e^{i\ell\phi} \frac{p_z}{\tau}\left[\ms^x\partial_{p_z} \left(\frac{p_x}{E_p} A^0-A^x\right)+\ms^y\partial_{p_z} \left(\frac{p_y}{E_p}A^0-A^y\right)
    \right.\n\\
   & \left.+\ms^z \partial_{p_z} \left(\frac{p_z}{E_p}A^0-A^z\right) +\frac{\ms^0}{E_p}\left(\frac{p_z}{E_p} A^0-A^z\right)\right]\n\\
    =& 4 \int_{p} {E_p} \left(\frac{p}{E_p}\right)^r  \mathcal{P}^\ell_n(\cos\theta) e^{i\ell\phi} \frac{p_z}{\tau} \left[\left(1+\frac{p_x^2}{m^2}\right)\partial_{p_z} \left(\frac{p_x}{E_p} A^0-A^x\right)+\frac{p_x p_y}{m^2}\partial_{p_z} \left(\frac{p_y}{E_p}A^0-A^y\right)\right.\n\\
    &\left. + \frac{p_z p_x}{m^2} \partial_{p_z} \left(\frac{p_z}{E_p}A^0-A^z\right)+\frac{p_x}{m^2}\left(\frac{p_z}{E_p} A^0-A^z\right)\right]\n\\
    =& -4\frac{1}{\tau} \int_{p} {E_p} \left(\frac{p}{E_p}\right)^r  \mathcal{P}^\ell_n(\cos\theta) e^{i\ell\phi} \left[1-(r-1)\frac{p_z^2}{E_p^2}+r\frac{p_z^2}{p^2}\right]  \left[\left(1+\frac{p_x^2}{m^2}\right) \left(\frac{p_x}{E_p} A^0-A^x\right)+\frac{p_x p_y}{m^2} \left(\frac{p_y}{E_p}A^0-A^y\right)\right.\n\\
    &\left. + \frac{p_z p_x}{m^2}  \left(\frac{p_z}{E_p}A^0-A^z\right)\right]\n\\
    &-4\frac{1}{\tau}\int_{p} {E_p} \left(\frac{p}{E_p}\right)^r  (\mathcal{P}^\ell_n)^\prime(\cos\theta) e^{i\ell\phi}  \cos\theta \left(1-\cos^2\theta\right)  \left[\left(1+\frac{p_x^2}{m^2}\right) \left(\frac{p_x}{E_p} A^0-A^x\right)+\frac{p_x p_y}{m^2} \left(\frac{p_y}{E_p}A^0-A^y\right)\right.\n\\
    &\left. + \frac{p_z p_x}{m^2}  \left(\frac{p_z}{E_p}A^0-A^z\right)\right]\n\\
    =& -\frac{1}{\tau}\int_{p\ms}  \ms^x  \left(\frac{p}{E_p}\right)^r  \mathcal{P}^\ell_n(\cos\theta) e^{i\ell\phi} f-\frac{1}{\tau}\int_{p\ms}  \ms^x \left(\frac{p}{E_p}\right)^r  \mathcal{P}^\ell_n(\cos\theta) e^{i\ell\phi} \cos^2\theta \left[r+(1-r)\frac{p^2}{E_p^2}\right] f\n\\
    &+\frac{1}{\tau}\int_{p\ms}  \ms^x \left(\frac{p}{E_p}\right)^r  e^{i\ell\phi}\cos\theta\left[ n\cos\theta\, \mathcal{P}_n^\ell(\cos\theta)-(n+\ell)\mathcal{P}^\ell_{(n-1)}(\cos\theta)\right]  f\; .\n\\
  \label{eomgxgenpolapp}
\end{align}

The $\mI$-moments are also treated analogously. We obtain the following equations of motion for free streaming
\begin{align}
    \partial_\tau \mI^z_{n\ell r}=& m\int_{p\ms} \frac{1}{E_p} \left(\frac{p}{E_p}\right)^r \ms^z \mathcal{P}^\ell_n(\cos\theta) e^{i\ell\phi} \left(\frac{p_z}{\tau}\partial_{p_z}+\frac{p_z}{E_p^2}\frac{\mathbf{p}\cdot \boldsymbol{\ms}}{\tau}\partial_{\ms_z}\right)f\n\\
    =&m \int_{p\ms} \frac{1}{E_p} \left(\frac{p}{E_p}\right)^r \ms^z \mathcal{P}^\ell_n(\cos\theta) e^{i\ell\phi} \left(\frac{p_z}{\tau}\partial_{p_z}+\frac{p_z}{E_p^2}\frac{\mathbf{p}\cdot \boldsymbol{\ms}}{\tau}\partial_{\ms_z}\right)(F+\ms^0 A^0-\ms^x A^x-\ms^y A^y-\ms^z A^z)\n\\
    =& m \int_{p\ms} \frac{1}{E_p} \left(\frac{p}{E_p}\right)^r \ms^z \mathcal{P}^\ell_n(\cos\theta) e^{i\ell\phi} \frac{p_z}{\tau}\left[\ms^x\partial_{p_z} \left(\frac{p_x}{E_p} A^0-A^x\right)+\ms^y\partial_{p_z} \left(\frac{p_y}{E_p}A^0-A^y\right)
    \right.\n\\
   & \left.+\ms^z \partial_{p_z} \left(\frac{p_z}{E_p}A^0-A^z\right) +\frac{\ms^0}{E_p}\left(\frac{p_z}{E_p} A^0-A^z\right)\right]\n\\
    =& 4 m \int_{p} \left(\frac{p}{E_p}\right)^r  \mathcal{P}^\ell_n(\cos\theta) e^{i\ell\phi} \frac{p_z}{\tau} \left[\frac{p_xp_z}{m^2}\partial_{p_z} \left(\frac{p_x}{E_p} A^0-A^x\right)+\frac{p_z p_y}{m^2}\partial_{p_z} \left(\frac{p_y}{E_p}A^0-A^y\right)\right.\n\\
    &\left. + \left(1+\frac{p_z^2}{m^2}\right) \partial_{p_z} \left(\frac{p_z}{E_p}A^0-A^z\right)+\frac{p_z}{m^2}\left(\frac{p_z}{E_p} A^0-A^z\right)\right]\n\\
=& -4 m\int_{p} \left(\frac{p}{E_p}\right)^r  \mathcal{P}^\ell_n(\cos\theta) e^{i\ell\phi} \frac{1}{\tau} \left[2\frac{p_xp_z}{m^2} \left(\frac{p_x}{E_p} A^0-A^x\right)+2\frac{p_z p_y}{m^2} \left(\frac{p_y}{E_p}A^0-A^y\right)\right.\n\\
    &\left. + \left(1+3\frac{p_z^2}{m^2}\right) \left(\frac{p_z}{E_p}A^0-A^z\right)-\frac{p^2_z}{m^2}\left(\frac{p_z}{E_p} A^0-A^z\right)\right]\n\\
   & -4m\frac{1}{\tau} \int_{p} \left(\frac{p}{E_p}\right)^r  \mathcal{P}^\ell_n(\cos\theta) e^{i\ell\phi} \left[-r\frac{p_z^2}{E_p^2}+r\frac{p_z^2}{p^2}\right]  \left[\frac{p_xp_z}{m^2} \left(\frac{p_x}{E_p} A^0-A^x\right)+\frac{p_x p_z}{m^2} \left(\frac{p_y}{E_p}A^0-A^y\right)\right.\n\\
    &\left.+\left(1+\frac{p_z^2}{m^2}\right)   \left(\frac{p_z}{E_p}A^0-A^z\right)\right]\n\\
    &-4m\frac{1}{\tau}\int_{p} \left(\frac{p}{E_p}\right)^r  (\mathcal{P}^\ell_n)^\prime(\cos\theta) e^{i\ell\phi}  \cos\theta \left(1-\cos^2\theta\right)  \left[\frac{p_xp_z}{m^2} \left(\frac{p_x}{E_p} A^0-A^x\right)+\frac{p_x p_z}{m^2} \left(\frac{p_y}{E_p}A^0-A^y\right)\right.\n\\
    &\left. + \left(1+\frac{p_z^2}{m^2}\right)  \left(\frac{p_z}{E_p}A^0-A^z\right)\right]\n\\
=& -4m \int_{p} \left(\frac{p}{E_p}\right)^r  \mathcal{P}^\ell_n(\cos\theta) e^{i\ell\phi} \frac{1}{\tau} \left[2\frac{p_xp_z}{m^2} \left(\frac{p_x}{E_p} A^0-A^x\right)+2\frac{p_z p_y}{m^2} \left(\frac{p_y}{E_p}A^0-A^y\right)\right.\n\\
    &\left. + 2\left(1+\frac{p_z^2}{m^2}\right) \left(\frac{p_z}{E_p}A^0-A^z\right)-\left(\frac{p_z}{E_p} A^0-A^z\right)\right]\n\\ 
     & -4m\frac{1}{\tau} \int_{p}  \left(\frac{p}{E_p}\right)^r  \mathcal{P}^\ell_n(\cos\theta) e^{i\ell\phi} \left[-r\frac{p_z^2}{E_p^2}+r\frac{p_z^2}{p^2}\right]  \left[\frac{p_xp_z}{m^2} \left(\frac{p_x}{E_p} A^0-A^x\right)+\frac{p_x p_z}{m^2} \left(\frac{p_y}{E_p}A^0-A^y\right)\right.\n\\
    &\left.+\left(1+\frac{p_z^2}{m^2}\right)   \left(\frac{p_z}{E_p}A^0-A^z\right)\right]\n\\
    &-4m\frac{1}{\tau}\int_{p} \left(\frac{p}{E_p}\right)^r  (\mathcal{P}^\ell_n)^\prime(\cos\theta) e^{i\ell\phi}  \cos\theta \left(1-\cos^2\theta\right)  \left[\frac{p_xp_z}{m^2} \left(\frac{p_x}{E_p} A^0-A^x\right)+\frac{p_x p_z}{m^2} \left(\frac{p_y}{E_p}A^0-A^y\right)\right.\n\\
    &\left. + \left(1+\frac{p_z^2}{m^2}\right)  \left(\frac{p_z}{E_p}A^0-A^z\right)\right]\n\\
    =& -\frac{1}{\tau}  m\int_{p\ms} \frac{1}{E_p} \left(\frac{p}{E_p}\right)^r  \mathcal{P}^\ell_n(\cos\theta) e^{i\ell\phi}  \left(\ms^z+\frac{p_z}{E_p}\ms^0\right)f
    -r\, m\frac{1}{\tau}\int_{p\ms} \frac{1}{E_p} \ms^z \left(\frac{p}{E_p}\right)^r  \mathcal{P}^\ell_n(\cos\theta) e^{i\ell\phi} \cos^2\theta \left[1-\frac{p^2}{E_p^2}\right] f\n\\
    &+\frac{1}{\tau}m\int_{p\ms} \frac{1}{E_p} \ms^z \left(\frac{p}{E_p}\right)^r  e^{i\ell\phi}\cos\theta\left[ n\cos\theta\, \mathcal{P}_n^\ell(\cos\theta)-(n+\ell)\mathcal{P}^\ell_{(n-1)}(\cos\theta)\right] f\; .
    \label{eomiz}
\end{align}
Note that for spin moments with $r=0$, the second term in the next-to-last line vanishes. Similarly we find
\begin{align}
    \partial_\tau \mI^x_{n\ell r}&= -\frac{1}{\tau}m\int_{p\ms}  \ms^x \frac{1}{E_p} \left(\frac{p}{E_p}\right)^r  \mathcal{P}^\ell_n(\cos\theta) e^{i\ell\phi} f- r\, m\frac{1}{\tau}\int_{p\ms} \frac{1}{E_p} \ms^x \left(\frac{p}{E_p}\right)^r  \mathcal{P}^\ell_n(\cos\theta) e^{i\ell\phi} \cos^2\theta \left[1-\frac{p^2}{E_p^2}\right] f\n\\
    &+\frac{1}{\tau}m\int_{p\ms} \frac{1}{E_p} \ms^x \left(\frac{p}{E_p}\right)^r  e^{i\ell\phi}\cos\theta\left[ n\cos\theta\, \mathcal{P}_n^\ell(\cos\theta)-(n+\ell)\mathcal{P}^\ell_{(n-1)}(\cos\theta)\right]  f \; .
    \label{eomix}
\end{align}
Note that the equations of motion for $\mI_{n\ell0}^x$ do not couple to spin moments with different $r$.

The equation of motion \eqref{angmomeom} for the total angular momentum is calculated from \eq\eqref{boltzbjorkcomp} as follows
\begin{align}
    \partial_\tau \tilde{\mathcal{J}}^{\mu\nu}&= \frac\hbar4 \int_{p\ms} \frac{1}{E_p^2} \left[  \frac{m}{2(E_p+m)} \ms^{[\mu} t^{\nu]}+\frac{1}{m} \left(1+\frac{E_p}{2(E_p+m)}\right)p^{[\mu}\ms^{\nu]} \right]\n\\
    &\times\left(\frac{p_z}{\tau}\partial_{p_z}+\frac{p_z}{E_p^2}\frac{\mathbf{p}\cdot \boldsymbol{\ms}}{\tau}\partial_{\ms_z}\right)(F+\ms^0 A^0-\ms^x A^x-\ms^y A^y-\ms^z A^z)\n\\
  &= \frac\hbar4 \int_{p\ms} \frac{1}{E_p^2} \left[  -\frac{m}{2(E_p+m)}  t^{[\mu}+\frac{1}{m} \left(1+\frac{E_p}{2(E_p+m)}\right)p^{[\mu} \right] \ms^{\nu]} \frac{p_z}{\tau}\n\\
   & \times \left[\ms^x\partial_{p_z} \left(\frac{p_x}{E_p} A^0-A^x\right)+\ms^y\partial_{p_z} \left(\frac{p_y}{E_p}A^0-A^y\right)
    +\ms^z \partial_{p_z} \left(\frac{p_z}{E_p}A^0-A^z\right) +\frac{\ms^0}{E_p}\left(\frac{p_z}{E_p} A^0-A^z\right)\right]\n\\
    &= -\hbar \frac{1}{\tau}\int_{p} \frac{p_z}{E_p} \left[  -\frac{m}{2(E_p+m)}  t^{[\mu}+\frac{1}{m} \left(1+\frac{E_p}{2(E_p+m)}\right)p^{[\mu} \right]  \n\\
   & \times \left[\proj^{\nu]x}\partial_{p_z} \left(\frac{p_x}{E_p} A^0-A^x\right)+\proj^{\nu]y}\partial_{p_z} \left(\frac{p_y}{E_p}A^0-A^y\right)
    +\proj^{\nu]z} \partial_{p_z} \left(\frac{p_z}{E_p}A^0-A^z\right) +\frac{\proj^{\nu]0}}{E_p}\left(\frac{p_z}{E_p} A^0-A^z\right)\right]\n\\
  &= \hbar \frac{1}{\tau}\int_{p} \left\{\left(\frac{1}{E_p}-\frac{p_z^2}{E_p^3}\right) \left[  -\frac{m}{2(E_p+m)}  t^{[\mu}+\frac{1}{m} \left(1+\frac{E_p}{2(E_p+m)}\right)p^{[\mu} \right]\right.  \n\\
  &\left.+\frac{p_z}{E_p} \left[\frac{m}{2(E_p+m)^2}  \frac{p_z}{E_p}t^{[\mu}+\frac{1}{m} \left(1+\frac{E_p}{2(E_p+m)}\right)\left(\delta^{z[\mu}+\frac{p_z}{E_p}\delta^{0[\mu}\right)+\frac{p_z}{2m(E_p+m)}\left(\frac{1}{E_p}-\frac{1}{E_p+m}\right)p^{[\mu}\right]\right\}\n\\
   & \times \left[\proj^{\nu]x} \left(\frac{p_x}{E_p} A^0-A^x\right)+\proj^{\nu]y} \left(\frac{p_y}{E_p}A^0-A^y\right)
    +\proj^{\nu]z}  \left(\frac{p_z}{E_p}A^0-A^z\right) \right]\n\\ 
   & +\hbar \frac{1}{\tau}\int_{p} \frac{p_z}{E_p} \left[  -\frac{m}{2(E_p+m)}  t^{[\mu}+\frac{1}{m} \left(1+\frac{E_p}{2(E_p+m)}\right)p^{[\mu} \right]  \n\\
   & \times \bigg\{\frac{1}{m^2}\left(\delta^{\nu]z}+\delta^{\nu]0}\frac{p_z}{E_p}\right)\left[-p^x \left(\frac{p_x}{E_p} A^0-A^x\right)-p^y \left(\frac{p_y}{E_p}A^0-A^y\right)
    -p^z \left(\frac{p_z}{E_p}A^0-A^z\right)\right] -\frac{p^{\nu]}}{m^2}\left(\frac{p_z}{E_p}A^0-A^z\right)\n\\
    &-\frac{\proj^{\nu]0}}{E_p}\left(\frac{p_z}{E_p} A^0-A^z\right)\bigg\}\n\\
   &=-\frac{1}{\tau}\frac{\hbar}{4} \int_{p\ms} \frac{1}{E_p}\left\{\left(\frac{1}{E_p}-\frac{p_z^2}{E_p^3}\right) \left[  -\frac{m}{2(E_p+m)}  t^{[\mu}+\frac{1}{m} \left(1+\frac{E_p}{2(E_p+m)}\right)p^{[\mu} \right]+\frac{p_z}{E_p}\right.  \n\\
  &\left. \times\left[\frac{m}{2(E_p+m)^2}  \frac{p_z}{E_p}t^{[\mu}+\frac{1}{m} \left(1+\frac{E_p}{2(E_p+m)}\right)\left(\delta^{z[\mu}+\frac{p_z}{E_p}\delta^{0[\mu}\right)+\frac{p_z}{2m(E_p+m)}\left(\frac{1}{E_p}-\frac{1}{E_p+m}\right)p^{[\mu}\right]\right\} \ms^{\nu]} f\n\\ 
   & +\hbar \frac{1}{\tau}\int_{p} \frac{p_z}{E_p} \left[  -\frac{m}{2(E_p+m)}  t^{[\mu}+\frac{1}{m} \left(1+\frac{E_p}{2(E_p+m)}\right)p^{[\mu} \right]\n\\
   &\times\bigg\{\frac{1}{m^2}\left(\delta^{\nu]z}+\delta^{\nu]0}\frac{p_z}{E_p}\right)\left[-p^x \left(\frac{p_x}{E_p} A^0-A^x\right)-p^y \left(\frac{p_y}{E_p}A^0-A^y\right)-p^z\left(\frac{p_z}{E_p}A^0-A^z\right)\right]\n\\
    &
    -\frac{\delta^{\nu]0}}{E_p}\left(\frac{p_z}{E_p}A^0-A^z\right)\bigg\}\n\\
  &=-\frac{1}{\tau}\frac{\hbar}{4} \int_{p\ms} \frac{1}{E_p}\left\{\left(\frac{1}{E_p}-\frac{p_z^2}{E_p^3}\right) \left[  -\frac{m}{2(E_p+m)}  t^{[\mu}+\frac{1}{m} \left(1+\frac{E_p}{2(E_p+m)}\right)p^{[\mu} \right]+\frac{p_z}{E_p}\right.  \n\\
  &\left. \times\left[\frac{m}{2(E_p+m)^2}  \frac{p_z}{E_p}t^{[\mu}+\frac{1}{m} \left(1+\frac{E_p}{2(E_p+m)}\right)\left(\delta^{z[\mu}+\frac{p_z}{E_p}\delta^{0[\mu}\right)+\frac{p_z}{2m(E_p+m)}\left(\frac{1}{E_p}-\frac{1}{E_p+m}\right)p^{[\mu}\right]\right\} \ms^{\nu]} f\n\\ 
   & +\frac{\hbar}{4} \frac{1}{\tau} t^{[\mu} \delta^{\nu]z} \int_{p\ms} \frac{p_z}{E_p^4}  \frac{m}{2(E_p+m)} \left( p^x\ms^x+p^y \ms^y+p^z \ms^z\right)f\n\\ 
    & -\frac{\hbar}{4} \frac{1}{\tau}\int_{p\ms} \frac{p_z}{E_p^2} \frac{1}{m} \left(1+\frac{E_p}{2(E_p+m)}\right)p^{[\mu} \bigg\{\left[\frac{1}{m^2}\delta^{\nu]z}+\left(\frac{1}{m^2}-\frac{1}{E_p^2}\right)\delta^{\nu]0}\frac{p_z}{E_p}\right]\left( p^x\ms^x+p^y \ms^y+p^z \ms^z\right)
    +\frac{\delta^{\nu]0}}{E_p}\ms^z\bigg\} f
    \label{angmomeomapp}
    \end{align}
where we defined
\begin{equation}
    \proj^{\mu\nu}\equiv g^{\mu\nu}-\frac{p^\mu p^\nu}{m^2} \; ,
\end{equation}
and replaced $A^0=\p\cdot\mathbf{A}/E_p$ in the next-to-last step.

\section{Relations between spin potential, total angular momentum and asymptotic spin moments}
\label{relapp}

In this appendix, we outline how to express the asymptotic spin moments in terms of the total angular momentum and gradients of the fluid velocity. To this end, we first determine the total angular momentum as a function of equilibrium quantities.
Inserting \eq\eqref{finfty1} into \eq\eqref{jinfty} and using \eq\eqref{intdsfinfty}, we obtain \eqs\eqref{alljinf} as follows,
\begin{align}
\tilde{\mathcal{J}}^{xy}_\infty&= \frac\hbar4 \int_{p\ms} \frac{1}{E_p^2} \frac{1}{m} \left(1+\frac{E_p}{2(E_p+m)}\right)p^{[x}\ms^{y]}  f_\infty\n\\
    &= \frac{\hbar^2}{4} \int d^3p\,  \frac{1}{E_p} \frac{1}{m^2} \left(1+\frac{E_p}{2(E_p+m)}\right)(\tilde{\Omega}^{yx} p_x^2-\tilde{\Omega}^{xy} p_y^2)  f_0\n\\
    &-\frac{\hbar^2}{2} \int_{p} \frac{1}{E_p^2} \frac{1}{m} \left(1+\frac{E_p}{2(E_p+m)}\right) \bxi  p^0 p_x^2 (\kappa_0^{z}+\partial^{z} \beta u^0)f^{(0)}_\text{LE}-\frac{\hbar^2}{2} \int_{p} \frac{1}{E_p^2} \frac{1}{m} \left(1+\frac{E_p}{2(E_p+m)}\right) \bxi  p^0 p_y^2 (\kappa_0^{z}+\partial^{z} \beta u^0)f^{(0)}_\text{LE}\n\\
    &=-\hbar^2 \kappa_0^{z} \int_{p} \frac{1}{E_p} \frac{1}{m^2} \left(1+\frac{E_p}{2(E_p+m)}\right) (1+m\bxi)  p_z^2 f^{(0)}_\text{LE}+\hbar^2 \int_{p} \frac{1}{E_p} \frac{1}{m} \left(1+\frac{E_p}{2(E_p+m)}\right) \bxi  p_z^2 \partial^{z} \beta u^0 f^{(0)}_\text{LE}\; ,\\
    \tilde{\mathcal{J}}^{xz}_\infty &= \frac\hbar4 \int_{p\ms} \frac{1}{E_p^2} \frac{1}{m} \left(1+\frac{E_p}{2(E_p+m)}\right)p^{[x}\ms^{z]}  f_\infty\n\\
    &= \frac{\hbar^2}{4} \int d^3p\,  \frac{1}{E_p} \frac{1}{m^2} \left(1+\frac{E_p}{2(E_p+m)}\right)(\tilde{\Omega}^{zx} p_x^2-\tilde{\Omega}^{xz} p_z^2)  f_0+\frac{\hbar^2}{2} \int_{p} \frac{1}{E_p} \frac{1}{m} \left(1+\frac{E_p}{2(E_p+m)}\right) \bxi  p^0 p_x^2 (\kappa_0^{y}+\partial^{y} \beta u^0)f^{(0)}_\text{LE}\n\\
    &+\frac{\hbar^2}{2} \int_{p} \frac{1}{E_p} \frac{1}{m} \left(1+\frac{E_p}{2(E_p+m)}\right) \bxi  p^0 p_z^2 (\kappa_0^{y}+\partial^{y} \beta u^0)f^{(0)}_\text{LE}\n\\
    &=\hbar^2 \kappa_0^{y} \int_{p} \frac{1}{E_p} \frac{1}{m^2} \left(1+\frac{E_p}{2(E_p+m)}\right) (1+m\bxi)   p_z^2 f^{(0)}_\text{LE}+\hbar^2 \int_{p} \frac{1}{E_p} \frac{1}{m} \left(1+\frac{E_p}{2(E_p+m)}\right) \bxi  p_z^2 \partial^{y} \beta u^0 f^{(0)}_\text{LE}\; ,
    \label{jxzinfapp}\\
    \tilde{\mathcal{J}}^{z0}_\infty
    &= \frac\hbar4 \int_{p\ms} \frac{1}{E_p^2} \left[  \frac{m^2-2E_p(E_p+m)-E_p^2}{2m(E_p+m)} \ms^z +\frac{1}{m} \left(1+\frac{E_p}{2(E_p+m)}\right)p^{z}\ms^0 \right] f_\infty\n\\
    &=   -\frac\hbar4 \int_{p} \frac{1}{E_p}   \frac{m^2-2E_p(E_p+m)-E_p^2}{2m(E_p+m)} \left[ \frac{\hbar}{m}\tilde{\Omega}^{z0} E_p  -\xi\frac{2\hbar}{m(E_p+m)}\epsilon^{ij0 z}p_j  p^\nu (\Omega_{i\nu}+\partial_i \beta_\nu)\right] f_\text{LE}^{(0)}  \n\\
     &- \frac{\hbar^2}{4} \int_p \frac{1}{m^2} \frac{1}{E_p} \left(1+\frac{E_p}{2(E_p+m)}\right)p_z^2 \tilde{\Omega}^{0z}  f^{(0)}_\text{LE}\n\\
         &=   \frac{\hbar^2}{2} \int_{p} \frac{1}{E_p}   \left[\frac{m^2-2E_p(E_p+m)-E_p^2}{2m^2(E_p+m)}E_p-\frac{p_z^2}{m^2}\left(1+\frac{E_p}{2(E_p+m)}\right)\right] \Omega^{xy}   f_\text{LE}^{(0)}  \n\\
     &+\frac{\hbar^2}{2} \int_{p} \frac{1}{E_p^2}   \frac{m^2-2E_p(E_p+m)-E_p^2}{2m(E_p+m)} \bxi \left[p_y  p^y (\Omega_{xy}+\partial_x \beta_y)-p_x  p^x (\Omega_{yx}+\partial_y \beta_x)\right] f_\text{LE}^{(0)} \n\\
    &= \frac{\hbar^2}{2}  \Omega^{xy} \int_p  \left[\frac{m^2-2E_p(E_p+m)-E_p^2}{2m^2(E_p+m)}\left(1-\frac{p_z^2}{E_p^2}m\bxi\right)-\frac{p_z^2}{E_pm^2}\left(1+\frac{E_p}{2(E_p+m)}\right) \right]  f^{(0)}_\text{LE}\n\\
    &+ \frac{\hbar^2}{2} \varpi^{xy} \int_{p}  \frac{m^2-2E_p(E_p+m)-E_p^2}{2m(E_p+m)} \bxi  \frac{p_z^2}{E_p^2} f^{(0)}_\text{LE}\; .
\end{align}
Here we used that $\Delta^\mu$ does not depend on $\ms^0$, $\tilde{\Omega}^{yx}=\epsilon^{yx0z}\Omega_{0z}+\epsilon^{yxz0}\Omega_{z0}=-2\kappa_0^z$, $\tilde{\Omega}^{0z}=\epsilon^{0zxy}\Omega^{xy}+\epsilon^{0zxy}\Omega^{yx}=2\Omega^{xy}$, $\kappa_0^i\equiv \Omega^{0i}$, $\tilde{\Omega}^{zy}=2\epsilon^{zyx0}\Omega^{x0}=-2\kappa_0^x$, $\tilde{\Omega}^{zx}=2\epsilon^{zxy0}\Omega^{y0}=2\kappa_0^y$, the fact that $f^{(0)}_\text{LE}$ is symmetric under exchange of $p_x$, $p_y$, and $p_z$, and,  e.g., 
\begin{align}
   & 2\hbar\int d^3p\, \frac{1}{E_p^2} \frac{1}{m} \left(1+\frac{E_p}{2(E_p+m)}\right)  \bxi p^x\epsilon^{ij0y}p_j  p^\nu (\Omega_{i\nu}+\partial_i \beta_\nu) f_\text{LE}^{(0)}\n\\
    &=  2\hbar\int d^3p\,\bxi \frac{1}{E_p^2} \frac{1}{m} \left(1+\frac{E_p}{2(E_p+m)}\right) p^x p_{[x}  p^\nu (\Omega_{{z]}\nu}+\partial_{z]} \beta_\nu) f_\text{LE}^{(0)}\n\\
   &=  2\hbar\int d^3p\,\bxi \frac{1}{E_p^2} \frac{1}{m} \left(1+\frac{E_p}{2(E_p+m)}\right) p_{x}^2 p^0 (\Omega^{{z}0}+\partial^{z} \beta^0)f_\text{LE}^{(0)} \; .  
\end{align}

In the next step, the relations \eqref{gzinfty} between the longitudinal spin moments and the total angular momentum are obtained by using \eq\eqref{intdsfinfty} in \eq\eqref{asspinmom} and then inserting \eqs\eqref{alljinf},
\begin{subequations}
\begin{align}
    \mG^z_{000,\infty}&= -\frac{\hbar}{m}\int d^3p\,  E_p^2\, \tilde{\Omega}^{z0} f_0+2\hbar\int_{p} \bxi  \epsilon^{ij0 z}p_j  p^\nu (\Omega_{i\nu}+\partial_i \beta_\nu)f^{(0)}_\text{LE}
    \n\\
    &= -\frac{\hbar}{m}\int d^3p\,  E_p^2\, \tilde{\Omega}^{z0} f_0-2\hbar\int_{p} \bxi  [p^2_y  (\Omega_{xy}+\partial_x \beta_y)-p_x^2  (\Omega_{yx}+\partial_y \beta_x)]f^{(0)}_\text{LE}\n\\
  &=2\frac{\hbar}{m} \Omega^{xy} \int_p \left(E_p^2-2m\bxi p_z^2\right) f^{(0)}_\text{LE}+4\hbar \varpi^{xy} \int_{p} \bxi p_z^2 f^{(0)}_\text{LE}\n\\
  & \equiv \lambda_{00}^{00} \tilde{\mathcal{J}}^{z0}+\kappa_{00}^{00} \varpi^{xy}\; ,\label{g000inftyapp}\\
 \mG^z_{200,\infty}&=-2\hbar\int_{p} Y_2^0(\theta,\phi) \bxi  [p^2_y  (\Omega_{xy}+\partial_x \beta_y)-p_x^2  (\Omega_{yx}+\partial_y \beta_x)]f^{(0)}_\text{LE}\n\\
 &=2\hbar \left(\Omega^{xy}-\varpi^{xy}\right)\int_p Y_2^0(\theta,\phi) \bxi\, p_z^2 f^{(0)}_\text{LE} \n\\
 & \equiv \lambda_{20}^{00} \tilde{\mathcal{J}}^{z0}+\kappa_{20}^{00} \varpi^{xy}\; ,\label{g200inftyapp}\\
 \mG^z_{110,\infty}&= -\frac{\hbar}{m}\int d^3p\, Y_1^1(\theta,\phi) E_p (\tilde{\Omega}^{zy} p_y+\tilde{\Omega}^{zx} p_x) f_0+2\hbar\int_{p} Y_1^1(\theta,\phi) \bxi  p_{[y}  p^0 (\Omega_{x]0}+\partial_{x]} \beta_0)f^{(0)}_\text{LE}\n\\
 &= -2\frac{\hbar}{m}\kappa_0^{[x}\int_{p} p^{y]} Y_1^1(\theta,\phi) (1+m\bxi)  E_p  f^{(0)}_\text{LE} +2\hbar\int_{p}  Y_1^1(\theta,\phi) \bxi  E_p p^{[y} \partial^{x]} \beta u^0 f^{(0)}_\text{LE}\n\\
 &\equiv \lambda_{11}^{00} \tilde{\mathcal{J}}^{xz}+\bar{\lambda}_{11}^{00} \tilde{\mathcal{J}}^{yz}
\end{align}
\label{gzinftyapp}
\end{subequations}
where we used in that $f^{(0)}_\text{LE}$ is symmetric under exchange of $p^z$, $p^x$ and $p^y$ and in \eq\eqref{g200inftyapp} that $p_\bot^2\equiv p_y^2+p_x^2=p^2-p_z^2$ and $\int_p F(p) Y_n^\ell(\theta,\phi)=0$ for any function $F(p)$ and $n\neq0$. The following identities are useful:
\begin{align}
    \sin^2 \theta\, \cos^2\phi&=\frac12 \sin^2\theta\,  (\cos 2\phi+1)=\frac16 \Pc_2^2(\cos\theta)\ \text{Re}\, e^{2i\phi}-\frac13 \Pc_2^0(\cos\theta)+\frac13\; ,\n\\
    \sin^2\theta\, \cos\phi\, \sin\phi&=\frac12 \sin^2\theta \sin(2\phi)=\frac16 \Pc_2^2(\cos\theta)\, \text{Im}\, e^{2i\phi}\; ,\n\\
    \sin\theta\, \cos\theta\, \cos\phi&= -\frac13 \Pc_2^1(\cos\theta) \text{Re}\, e^{i\phi}\; ,\n\\
    \sin^2\theta\, \sin^2\phi&= -\frac12 \sin^2\theta\,  (\cos 2\phi-1)=-\frac16 \Pc_2^2(\cos\theta)\ \text{Re}\, e^{2i\phi}-\frac13 \Pc_2^0(\cos\theta)+\frac13\; ,\n\\
    \cos^2\theta&= \frac23 \Pc_2^0(\cos\theta)+\frac13 \; .
\end{align}
We also defined the thermodynamic integrals
\begin{align}
    \lambda_{00}^{rs}&\equiv  \left\{ \frac{\hbar}{4}  \int_p  \left[\frac{m^2-4E_p(E_p+m)-2E_p^2}{2m(E_p+m)}- 2 \frac{m^2-2E_p(E_p+m)-E_p^2}{(E_p+m)} \frac{p_z^2}{E_p^2} \bxi \right]  f^{(0)}_\text{LE} \right\}^{-1}\n\\
    &\times 2 \int_p \left(\frac{p}{E_p}\right)^r \left(\frac{m}{E_p}\right)^s \left(E_p^2-2m\bxi p_z^2\right) f^{(0)}_\text{LE} \; ,\n\\
    \kappa_{00}^{rs}&\equiv 4\hbar \int_{p} \left(\frac{p}{E_p}\right)^r \left(\frac{m}{E_p}\right)^s \bxi p_z^2 f^{(0)}_\text{LE}-\lambda_{00} \frac{\hbar^2}{2}  \int_{p} \left(\frac{p}{E_p}\right)^r \left(\frac{m}{E_p}\right)^s \frac{m^2-2E_p(E_p+m)-E_p^2}{m(E_p+m)} \bxi  \frac{p_z^2}{E_p^2} f^{(0)}_\text{LE}\; ,\n\\
    \lambda_{20}^{rs}&\equiv \left\{ \frac{\hbar}{4}  \int_p  \left[\frac{m^2-4E_p(E_p+m)-2E_p^2}{2m(E_p+m)}- 2 \frac{m^2-2E_p(E_p+m)-E_p^2}{(E_p+m)} \frac{p_z^2}{E_p^2} \bxi \right]  f^{(0)}_\text{LE} \right\}^{-1}\n\\
    &\times 2m \int_p \left(\frac{p}{E_p}\right)^r \left(\frac{m}{E_p}\right)^s Y_2^0(\theta,\phi) \bxi\, p_z^2 f^{(0)}_\text{LE}\; ,\n\\
    \kappa_{20}^{rs}&\equiv -2\hbar \int_{p} \left(\frac{p}{E_p}\right)^r \left(\frac{m}{E_p}\right)^s Y_2^0(\theta,\phi) \bxi p_z^2 f^{(0)}_\text{LE}-\lambda_{20} \frac{\hbar^2}{2}  \int_{p} \left(\frac{p}{E_p}\right)^r \left(\frac{m}{E_p}\right)^s \frac{m^2-2E_p(E_p+m)-E_p^2}{m(E_p+m)} \bxi  \frac{p_z^2}{E_p^2} f^{(0)}_\text{LE}\; ,\n\\
    \lambda_{11}^{rs}&\equiv \left[\hbar \int_{p} \frac{1}{E_p} \frac{1}{m} \left(1+\frac{E_p}{2(E_p+m)}\right) (1+m\bxi)   p_z^2 f^{(0)}_\text{LE}\right]^{-1} 2\int_{p} \left(\frac{p}{E_p}\right)^r \left(\frac{m}{E_p}\right)^s p^{x} Y_1^1(\theta,\phi) (1+m\bxi)  E_p  f^{(0)}_\text{LE}\; ,\n\\
    \bar{\lambda}_{11}^{rs}&\equiv \left[\hbar \int_{p} \frac{1}{E_p} \frac{1}{m} \left(1+\frac{E_p}{2(E_p+m)}\right) (1+m\bxi)   p_z^2 f^{(0)}_\text{LE}\right]^{-1} 2\int_{p} \left(\frac{p}{E_p}\right)^r \left(\frac{m}{E_p}\right)^sp^{y} Y_1^1(\theta,\phi) (1+m\bxi)  E_p  f^{(0)}_\text{LE}\; .
    \label{lambkapevapp}
\end{align}
We included the factors $(p/E_p)^r$ and $(m/E_p)^s$ in the integrals since same coefficients with different values for $r$ and $s$ are also needed in this paper. All $\lambda$ or $\kappa$ coefficients with $\ell \geq 0$, which are not defined in \eqs\eqref{lambkapevapp} or \eqref{lambkapodd}, are zero. Those with negative $\ell$ are defined as, e.g.,
\begin{equation}
    \lambda_{n\ell}^{rs}\equiv (-1)^\ell \frac{n+\ell}{n-\ell} \lambda_{n(-\ell)}^{rs}\; , \qquad \qquad \ell<0 \; ,
\end{equation}
and analogously for all other coefficients. Note that we also used the definitions \eqref{somecoeffis}. 

Furthermore, inserting \eq\eqref{intdsfinfty} into \eqref{asspinmom} for the transverse spin moments, we obtain \eqs\eqref{gxyinfty} as
\begin{align}
    \mG^x_{100,\infty}&= -\int_p Y_1^0(\theta,\phi) E_p\left[ \frac{\hbar}{m}\tilde{\Omega}^{xz} p_z  -\xi\frac{2\hbar}{m(E_p+m)}\epsilon^{y z0x}p_z p^0 (\Omega_{y0}+\partial_y \beta_0)\right] f_\text{LE}^{(0)} \n\\
    &=- 2 \frac{\hbar}{m}\kappa_0^{y} \int_{p}  Y_1^0(\theta,\phi) E_p p^{z} (1+m\bxi) f^{(0)}_\text{LE}-2\hbar \int_p Y_1^0(\theta,\phi) E_p p^{z} \bxi \partial^y \beta^0 \n\\
    &= \lambda_{10}^{00} \tilde{\mathcal{J}}^{xz} \; ,\n\\
   \mG^x_{210,\infty}&= \int_p Y_2^1(\theta,\phi) E_p\xi\frac{2\hbar}{m(E_p+m)}[p_z p^\nu (\Omega_{y\nu}+\partial_y \beta_\nu)-p_y p^\nu (\Omega_{z\nu}+\partial_z \beta_\nu)] f_\text{LE}^{(0)}   \n\\
   &=2\hbar \int_p Y_2^1(\theta,\phi) \bxi[p_z p^x (\Omega_{yx}+\partial_y \beta_x)+p_z p^y \partial_y\beta_y-p_y p^z \partial_z \beta_z] f_\text{LE}^{(0)}   \n\\
   &= \lambda_{21}^{00} \tilde{\mathcal{J}}^{z0}+\kappa_{21}^{00}\varpi^{xy}-\tilde{\kappa}_{21}^{00}\sigma\; ,\n\\
   \mG^y_{100,\infty}&= -\int_p Y_1^0(\theta,\phi) E_p\left[ \frac{\hbar}{m}\tilde{\Omega}^{yz} p_z  -\xi\frac{2\hbar}{m(E_p+m)}\epsilon^{xz0y}p_z p^0 (\Omega_{x0}+\partial_x \beta_0)\right] f_\text{LE}^{(0)} \n\\
    &= 2 \frac{\hbar}{m}\kappa_0^{x} \int_{p}  Y_1^0(\theta,\phi) E_p p^{z} (1+m\bxi) f^{(0)}_\text{LE}+2\hbar \int_p Y_1^0(\theta,\phi) E_p p^{z} \bxi \partial^x \beta^0 \n\\
    &= \lambda_{10}^{00} \tilde{\mathcal{J}}^{yz} \; ,\n\\
    \mG^y_{210,\infty}&= \int_p Y_1^2(\theta,\phi) E_p\xi\frac{2\hbar}{m(E_p+m)}[-p_z  p^\nu (\Omega_{x\nu}+\partial_x \beta_\nu)+p_x  p^\nu (\Omega_{z\nu}+\partial_z \beta_\nu)] f_\text{LE}^{(0)} \n\\
    &= 2\hbar \int_p Y_1^2(\theta,\phi) E_p\bxi[-p_z  p^y (\Omega_{xy}+\partial_x \beta_y)-p_z p^x \partial_x\beta_x+p_x  p^z \partial_z \beta_z] f_\text{LE}^{(0)}\n\\
    &= \bar{\lambda}_{21}^{00} \tilde{\mathcal{J}}^{z0}+\bar{\kappa}_{21}^{00}\varpi^{xy}+\hat{\kappa}_{21}^{00}\sigma
    \label{gxyinftyapp}
\end{align}
with
\begin{align}
    \lambda_{10}^{rs}&\equiv -\left[\hbar \int_{p} \frac{1}{E_p} \frac{1}{m} \left(1+\frac{E_p}{2(E_p+m)}\right) (1+m\bxi)   p_z^2 f^{(0)}_\text{LE}\right]^{-1} 2\int_{p} \left(\frac{p}{E_p}\right)^r \left(\frac{m}{E_p}\right)^s p^{z} Y_1^0(\theta,\phi) (1+m\bxi)  E_p  f^{(0)}_\text{LE}\; ,\n\\
    \lambda_{21}^{rs}&\equiv -\left\{ \frac{\hbar}{4}  \int_p  \left[\frac{m^2-4E_p(E_p+m)-2E_p^2}{2m^2(E_p+m)}- 2 \frac{m^2-2E_p(E_p+m)-E_p^2}{m(E_p+m)} \frac{p_z^2}{E_p^2} \bxi \right]  f^{(0)}_\text{LE} \right\}^{-1}\n\\
    &\times 2 \int_p \left(\frac{p}{E_p}\right)^r \left(\frac{m}{E_p}\right)^sY_2^1(\theta,\phi) \bxi p^z p^x f^{(0)}_\text{LE}\; ,\n\\
    \kappa_{21}^{rs}&\equiv  2\hbar \int_p \left(\frac{p}{E_p}\right)^r \left(\frac{m}{E_p}\right)^s Y_2^1(\theta,\phi) \bxi p^z p^x f^{(0)}_\text{LE} -\lambda_{21} \frac{\hbar^2}{2}  \int_{p} \left(\frac{p}{E_p}\right)^r \left(\frac{m}{E_p}\right)^s  \frac{m^2-2E_p(E_p+m)-E_p^2}{m(E_p+m)} \bxi  \frac{p_z^2}{E_p^2} f^{(0)}_\text{LE}\; ,\n\\
    \hat{\kappa}_{21}^{rs}&\equiv -2\hbar \int_p \left(\frac{p}{E_p}\right)^r \left(\frac{m}{E_p}\right)^s Y_2^1(\theta,\phi) \bxi p^z p^x f^{(0)}_\text{LE}\; ,\n\\
    \tilde{\kappa}_{21}^{rs}&\equiv -2 \hbar \int_p \left(\frac{p}{E_p}\right)^r \left(\frac{m}{E_p}\right)^s Y_2^1(\theta,\phi) \bxi p^z p^y f^{(0)}_\text{LE} \; ,\n\\
    \bar{\lambda}_{21}^{rs}&\equiv \left\{ \frac{\hbar}{4}  \int_p  \left[\frac{m^2-4E_p(E_p+m)-2E_p^2}{2m^2(E_p+m)}- 2 \frac{m^2-2E_p(E_p+m)-E_p^2}{m(E_p+m)} \frac{p_z^2}{E_p^2} \bxi \right]  f^{(0)}_\text{LE} \right\}^{-1}\n\\
    &\times 2 \int_p \left(\frac{p}{E_p}\right)^r \left(\frac{m}{E_p}\right)^s Y_2^1(\theta,\phi) \bxi p^z p^y f^{(0)}_\text{LE}\; ,\n\\
 \bar{\kappa}_{21}^{rs}&\equiv  -2\hbar \int_p \left(\frac{p}{E_p}\right)^r \left(\frac{m}{E_p}\right)^s Y_2^1(\theta,\phi) \bxi p^z p^y f^{(0)}_\text{LE} -\lambda_{21} \frac{\hbar^2}{2}  \int_{p} \left(\frac{p}{E_p}\right)^r \left(\frac{m}{E_p}\right)^s \frac{m^2-2E_p(E_p+m)-E_p^2}{m(E_p+m)} \bxi  \frac{p_z^2}{E_p^2} f^{(0)}_\text{LE}\; . 
 \label{lambkapodd}
\end{align}

\section{Asymptotic integrals}
\label{asympapp}

In this appendix, we provide auxiliary identities which follow from the properties of the associated Legendre polynomials and are useful to derive \eqs\eqref{kdd} and \eqref{kxy}.
For \eqs\eqref{asint1} and \eqref{asint2} we find
  \begin{align}
  \mathcal{K}^i_{n\ell,\infty}&\equiv 2\int_{p\ms} \ms^i \left(\frac{p}{E_p}\right)^2 \Pc_n^\ell(\cos\theta)\cos^2\theta\, e^{i\ell\phi} f_\infty\n\\
  &= 2\int_{p\ms} \ms^z \left(\frac{p}{E_p}\right)^2 \bigg[ \frac{(n+1-\ell)(n+2-\ell)}{(2n+1)(2n+3)} \mathcal{P}_{n+2}^\ell(\cos\theta)+\left(\frac{(n+1-\ell)(n+1+\ell)}{(2n+1)(2n+3)}+\frac{(n+\ell)(n-\ell)}{(2n+1)(2n-1)}\right)\n\\
    &\times \mathcal{P}^\ell_n(\cos\theta)+\frac{(n+\ell)(n-1+\ell)}{(2n+1)(2n-1)}\mathcal{P}^\ell_{n-2}(\cos\theta)\bigg] e^{i\ell\phi} f_\infty\n\\
    \mathcal{D}^x_{n\ell,\infty}&\equiv \int_{p\ms} \ms^x \Pc_n^\ell(\cos\theta)\left(\frac{p}{E_p}\right)^2 \cos\theta\, \sin\theta\, \cos\phi\, e^{i\ell\phi} f_\infty\; ,\n\\
    &= -\frac12\int_{p\ms} \ms^x \left(\frac{p}{E_p}\right)^2 \frac{1}{2n+1}\left\{\frac{1}{2n+3}[(n-\ell+1)\Pc^{\ell+1}_{n+2}(\cos\theta)+(n+\ell+2)\Pc^{\ell+1}_n(\cos\theta)]\right.\n\\
    &\left.-\frac{1}{2n-1}[(n-\ell-1)\Pc^{\ell+1}_{n}(\cos\theta)+(n+\ell)\Pc^{\ell+1}_{n-2}(\cos\theta)]\right\} e^{i(\ell+1)\phi} f_\infty\n\\
    &+\frac12\int_{p\ms} \ms^x \left(\frac{p}{E_p}\right)^2 \frac{1}{2n+1}\left\{ (n-\ell+1)(n-\ell+2)\frac{1}{2n+3}[(n-\ell+3)\Pc^{\ell-1}_{n+2}(\cos\theta)+(n+\ell)\Pc^{\ell-1}_n(\cos\theta)]\right.\n\\
     &\left.-(n+\ell-1)(n+\ell)\frac{1}{2n-1}[(n-\ell+1)\Pc^{\ell-1}_{n}(\cos\theta)+(n+\ell-2)\Pc^{\ell-1}_{n-2}]\right\}e^{i(\ell-1)\phi} f_\infty \; ,\n\\
    \bar{\mathcal{D}}^y_{n\ell,\infty}&\equiv \int_{p\ms} \ms^y \Pc_n^\ell(\cos\theta)\left(\frac{p}{E_p}\right)^2 \cos\theta\, \sin\theta\, \sin\phi\, e^{i\ell\phi} f_\infty\n\\
    &= \frac i2\int_{p\ms} \ms^y \left(\frac{p}{E_p}\right)^2 \frac{1}{2n+1}\left\{\frac{1}{2n+3}[(n-\ell+1)\Pc^{\ell+1}_{n+2}(\cos\theta)+(n+\ell+2)\Pc^{\ell+1}_n(\cos\theta)]\right.\n\\
    &\left.-\frac{1}{2n-1}[(n-\ell-1)\Pc^{\ell+1}_{n}(\cos\theta)+(n+\ell)\Pc^{\ell+1}_{n-2}(\cos\theta)]\right\} e^{i(\ell+1)\phi} f_\infty\n\\
    &+\frac i2\int_{p\ms} \ms^y \left(\frac{p}{E_p}\right)^2 \frac{1}{2n+1}\left\{ (n-\ell+1)(n-\ell+2)\frac{1}{2n+3}[(n-\ell+3)\Pc^{\ell-1}_{n+2}(\cos\theta)+(n+\ell)\Pc^{\ell-1}_n(\cos\theta)]\right.\n\\
     &\left.-(n+\ell-1)(n+\ell)\frac{1}{2n-1}[(n-\ell+1)\Pc^{\ell-1}_{n}(\cos\theta)+(n+\ell-2)\Pc^{\ell-1}_{n-2}(\cos\theta)]\right\}e^{i(\ell-1)\phi} f_\infty \; .
\end{align}

\section{Analysis of the equations of motion for the total angular momentum}
\label{littlejapp}

In this appendix, we demonstrate how the closed equations of motion for the total angular momentum are obtained. First, we derive the free equations of motion for $\tilde{j}^{zx}$, $\tilde{j}^{xz}$, $\tilde{j}^{z0}$, and $\tilde{j}_+^{0z}$ and discuss their behavior in the limit $\cos\theta\rightarrow0$. Then, we outline the derivation of the equations of motion for $\tilde{\mJ}^{zx}$, $\tilde{\mJ}^{zx}_+$, $\tilde{\mJ}^{z0}$, and $\tilde{\mJ}^{z0}_+$ including both the free-streaming and the late-time dynamics.
We obtain the equation of motion for $\tilde{j}^{zx}$ defined in \eq\eqref{littlej} from \eq\eqref{boltzbjorkcomp} in the free-streaming limit
\begin{align}
\partial_\tau \tilde{j}^{zx} =  &-\frac{1}{\tau}\frac{\hbar}{4m} \int_{p\ms} \frac{1}{E_p}\left\{\left(\frac{1}{E_p}-\frac{p_z^2}{E_p^3}\right) \left(1+\frac{E_p}{2(E_p+m)}\right)p^{z}\right.\n\\
&\left.+\frac{p_z}{E_p}\left[ \left(1+\frac{E_p}{2(E_p+m)}\right)+\frac{p^2_z}{2(E_p+m)}\left(\frac{1}{E_p}-\frac{1}{E_p+m}\right)\right]\right\} \ms^{x} f\n\\ 
    &\simeq -\frac{1}{\tau}\frac{\hbar}{4m} \int_{p\ms} \frac{1}{E_p}\left\{\frac{1}{E_p} \left(1+\frac{E_p}{2(E_p+m)}\right)p^{z} +\frac{p_z}{E_p} \left(1+\frac{E_p}{2(E_p+m)}\right)\right\} \ms^{x} f\n\\
    &=-2\frac{1}{\tau}\frac{\hbar}{4m} \int_{p\ms} \frac{p_z}{E_p^2} \left(1+\frac{E_p}{2(E_p+m)}\right) \ms^{x} f\n\\
    &=-2 \frac{1}{\tau} \tilde{j}^{zx}\; ,
    \label{dtaujzxfree}
\end{align}
where the symbol $\simeq$ means equality for $\cos\theta=0$. On the other hand, for  $\tilde{j}^{xz}$ we find
\begin{align}
\partial_\tau  \tilde{j}^{xz} &= -\frac{1}{\tau}\frac{\hbar}{4} \int_{p\ms} \frac{1}{E_p}\left\{\left(\frac{1}{E_p}-\frac{p_z^2}{E_p^3}\right) \frac{1}{m} \left(1+\frac{E_p}{2(E_p+m)}\right)p^{x} +\frac{p_z}{E_p}\frac{p_z}{2m(E_p+m)}\left(\frac{1}{E_p}-\frac{1}{E_p+m}\right)p^{x}\right\} \ms^{z} f\n\\ 
    & -\frac{\hbar}{4} \frac{1}{\tau}\int_{p\ms} \frac{p_z}{E_p^2} \frac{1}{m} \left(1+\frac{E_p}{2(E_p+m)}\right)p^{x} \frac{1}{m^2}\left( p^x\ms^x+p^y \ms^y+p^z \ms^z\right)
     f\n\\
     &\simeq-\frac{1}{\tau}\frac{\hbar}{4} \int_{p\ms} \frac{1}{E_p^2} \frac{1}{m} \left(1+\frac{E_p}{2(E_p+m)}\right)p^{x}  \ms^{z} f\n\\
     &=-\frac{1}{\tau} \tilde{j}^{xz}\; .
     \label{dtaujxzfree}
\end{align}
The $z$-$y$-components are obtained analogously. The equations of motion for $ \tilde{j}^{z0}$ are found to be
\begin{align}
\partial_\tau \tilde{j}^{z0} &= -\frac{1}{\tau}\frac{\hbar}{4} \int_{p\ms} \frac{1}{E_p}\left\{\left(\frac{1}{E_p}-\frac{p_z^2}{E_p^3}\right) \frac{1}{m} \left(1+\frac{E_p}{2(E_p+m)}\right)p^{z} +\frac{p_z}{E_p}\right.  \n\\
  &\left. \times\left[\frac{1}{m} \left(1+\frac{E_p}{2(E_p+m)}\right)+\frac{p_z}{2m(E_p+m)}\left(\frac{1}{E_p}-\frac{1}{E_p+m}\right)p^{z}\right]\right\} \ms^{0} f\n\\ 
    & -\frac{\hbar}{4} \frac{1}{\tau}\int_{p\ms} \frac{p_z}{E_p^2} \frac{1}{m} \left(1+\frac{E_p}{2(E_p+m)}\right)p^{z} \bigg\{\left(\frac{1}{m^2}-\frac{1}{E_p^2}\right)\frac{p_z}{E_p}\left( p^x\ms^x+p^y \ms^y+p^z \ms^z\right)
    +\frac{1}{E_p}\ms^z\bigg\} f\n\\
    &\simeq  -\frac{1}{\tau}\frac{\hbar}{4} \int_{p\ms} \frac{1}{E_p}\left\{\frac{1}{E_p} \frac{1}{m} \left(1+\frac{E_p}{2(E_p+m)}\right)p^{z} +\frac{p_z}{E_p} \frac{1}{m} \left(1+\frac{E_p}{2(E_p+m)}\right)\right\} \ms^{0} f\n\\
    &= -\frac{1}{\tau}2\frac{\hbar}{4m} \int_{p\ms} \frac{p^z}{E_p^2}\left(1+\frac{E_p}{2(E_p+m)}\right) \ms^{0} f\n\\
    &=-\frac{1}{\tau} 2 \tilde{j}^{z0}
    \label{dtaujz0free}
\end{align}
and for $ \tilde{j}^{0z}_+$ defined in \eq\eqref{littlej+} one obtains
\begin{align}
\partial_\tau \tilde{j}^{0z}_+    &=-\frac{1}{\tau}\frac{\hbar}{4} \int_{p\ms} \frac{1}{E_p}\left\{\left(\frac{1}{E_p}-\frac{p_z^2}{E_p^3}\right) \left[  -\frac{m}{2(E_p+m)} +\frac{1}{m} \left(1+\frac{E_p}{2(E_p+m)}\right)p^{0} \right]+\frac{p_z}{E_p}\right.  \n\\
  &\left. \times\left[\frac{m}{2(E_p+m)^2}  \frac{p_z}{E_p}+\frac{1}{m} \left(1+\frac{E_p}{2(E_p+m)}\right)\frac{p_z}{E_p}+\frac{p_z}{2m(E_p+m)}\left(\frac{1}{E_p}-\frac{1}{E_p+m}\right)p^{0}\right]\right\} \ms^{z} f\n\\ 
   & +\frac{\hbar}{4} \frac{1}{\tau}  \int_{p\ms} \frac{p_z}{E_p^4}  \frac{m}{2(E_p+m)} \left( p^x\ms^x+p^y \ms^y+p^z \ms^z\right)f\n\\ 
    & -\frac{\hbar}{4} \frac{1}{\tau}\int_{p\ms} \frac{p_z}{E_p^2} \frac{1}{m} \left(1+\frac{E_p}{2(E_p+m)}\right)p^{0} \frac{1}{m^2}\left( p^x\ms^x+p^y \ms^y+p^z \ms^z\right)
     f \n\\
     &\simeq-\frac{1}{\tau} \frac{\hbar}{4} \int_{p\ms} \frac{1}{E_p^2} \left[  -\frac{m}{2(E_p+m)} +\frac{1}{m} \left(1+\frac{E_p}{2(E_p+m)}\right)p^{0} \right] \ms^{z} f\n\\
     &=-\frac{1}{\tau} \tilde{j}^{0z}_+\; .
     \label{dtauj0zfree}
\end{align}
To obtain closed equations of motion which are valid at any time, we follow the same strategy as outlined for the spin moments in the main part of this paper and use an interpolation between free streaming and the hydrodynamic regime. Using \eq\eqref{dtaujzxfree} we find
\begin{align}
    \partial_w \tilde{j}^{zx}&= -\frac{2}{w} \tilde{j}^{zx}-\frac{1}{w}\left(1-e^{-w/2}\right)\frac{\hbar}{4m} \int_{p\ms} \frac{p_z^3}{E_p^2}\left\{-\frac{1}{E_p^2} \left(1+\frac{E_p}{2(E_p+m)}\right)+ \frac{1}{2(E_p+m)}\left(\frac{1}{E_p}-\frac{1}{E_p+m}\right)\right\} \ms^{x} f_\infty \n\\ 
    &-(\tilde{j}^{zx}-\tilde{j}^{zx}_\infty)\n\\
&= -\frac{2}{w} \tilde{j}^{zx}+\frac{1}{w}\left(1-e^{-w/2}\right)\frac{\hbar}{4m} \int_{p\ms} \frac{p_z^3}{E_p^2}\left[\frac{1}{E_p^2} + \frac{1}{2(E_p+m)^2}\right] \ms^{x} f_\infty -(\tilde{j}^{zx}-\tilde{j}^{zx}_\infty) 
\label{eomsmjxzapp}
\end{align}
and with \eq\eqref{dtaujxzfree} we obtain
\begin{align}
    \partial_w \tilde{j}^{xz}&=-\frac{1}{w}\tilde{j}^{xz}-\frac{1}{w}\left(1-e^{-w/2}\right)\frac{\hbar}{4} \int_{p\ms} \frac{p_z^2}{E_p^2}\left\{-\frac{1}{E_p^2} \frac{1}{m} \left(1+\frac{E_p}{2(E_p+m)}\right) +\frac{1}{2m(E_p+m)}\left(\frac{1}{E_p}-\frac{1}{E_p+m}\right)\right\} p^{x}\ms^{z} f_\infty\n\\ 
    & -\frac{\hbar}{4} \frac{1}{w}\left(1-e^{-w/2}\right)\int_{p\ms} \frac{p_z}{E_p} \frac{1}{m} \left(1+\frac{E_p}{2(E_p+m)}\right)p^{x} \frac{1}{m^2}\ms^0
     f_\infty-(\tilde{j}^{xz}-\tilde{j}^{xz}_\infty)\n\\
     &= -\frac{1}{w}\tilde{j}^{xz}+\frac{1}{w}\left(1-e^{-w/2}\right)\frac{\hbar}{4} \int_{p\ms} \frac{p_z^2}{E_p^2}\left[\frac{1}{E_p^2} \frac{1}{m}  +\frac{1}{2m(E_p+m)^2}\right] p^{x}\ms^{z} f_\infty-(\tilde{j}^{xz}-\tilde{j}^{xz}_\infty)\; ,
     \label{eomsmjzxapp}
\end{align}
where we used that the first term in the second line vanishes. With these results, we can derive \eqs\eqref{eomjplus} of the main part. Combining \eqs\eqref{eomsmjxzapp} and \eqref{eomsmjzxapp} we obtain for $\tilde{\mathcal{J}}^{zx}$ with \eq\eqref{splitjzx}
\begin{align}
    \partial_w \tilde{\mathcal{J}}^{zx}&=-\frac{1}{w} \frac12\left(3 \tilde{\mathcal{J}}^{zx}+ \tilde{\mathcal{J}}^{zx}_+ \right)+\frac{1}{w}\left(1-e^{-w/2}\right)\frac{\hbar}{4} \int_{p\ms} \frac{p_z^2}{E_p^2}\left[\frac{1}{E_p^2} \frac{1}{m}  +\frac{1}{2m(E_p+m)^2}\right] p^{[z}\ms^{x]} f_\infty\n\\
    &=-\frac{1}{w} \frac12\left(3 \tilde{\mathcal{J}}^{zx}+ \tilde{\mathcal{J}}^{zx}_+ \right)+\frac{1}{w}\left(1-e^{-w/2}\right) \Lambda_{x} \tilde{\mathcal{J}}^{zx}
\end{align}
and for $\tilde{\mathcal{J}}^{zx}_+$ defined in \eq\eqref{j+}
\begin{align}
    \partial_w \tilde{\mathcal{J}}^{zx}_+&=-\frac{1}{w} \frac12\left( \tilde{\mathcal{J}}^{zx}+ 3\tilde{\mathcal{J}}^{zx}_+ \right)+\frac{1}{w}\left(1-e^{-w/2}\right)\frac{\hbar}{4} \int_{p\ms} \frac{p_z^2}{E_p^2}\left[\frac{1}{E_p^2} \frac{1}{m}  +\frac{1}{2m(E_p+m)^2}\right] p^{(z}\ms^{x)} f_\infty- \tilde{\mathcal{J}}^{zx}_+\n\\
&=-\frac{1}{w} \frac12\left( \tilde{\mathcal{J}}^{zx}+ 3\tilde{\mathcal{J}}^{zx}_+ \right)- \tilde{\mathcal{J}}^{zx}_+ \; .
\label{eomjplusapp}
\end{align}
Here we used that
\begin{align}
    & \frac{\hbar}{4} \int_{p\ms} \frac{p_z^2}{E_p^2}\left[\frac{1}{E_p^2} \frac{1}{m}  +\frac{1}{2m(E_p+m)^2}\right] p^{[z}\ms^{x]} f_\infty\n\\
    &= -\hbar^2 \kappa_0^{y} \int_{p} \frac{1}{E_p} \frac{1}{m^2} \left[\frac{1}{E_p^2} +\frac{1}{2(E_p+m)^2}\right]  (1+m\bxi)   p_z^4 f^{(0)}_\text{LE}-\hbar^2 \int_{p} \frac{1}{E_p} \frac{1}{m} \left[\frac{1}{E_p^2}   +\frac{1}{2(E_p+m)^2}\right]  \bxi  p_z^4 \partial^{y} \beta u^0 f^{(0)}_\text{LE}\n\\
    &=\Lambda_x\tilde{\mathcal{J}}^{zx}\; ,
\end{align}
and defined
\begin{align}
    \Lambda_x&\equiv \left[ \int_{p} \frac{1}{E_p}  p_z^2\left(1+\frac{E_p}{2(E_p+m)}\right)    f^{(0)}_\text{LE}\right]^{-1} \int_{p} \frac{1}{E_p} p_z^4 \left[\frac{1}{E_p^2}  +\frac{1}{2(E_p+m)^2}\right]  f^{(0)}_\text{LE}\; .
    \label{lambdax}
\end{align}

Using \eq\eqref{dtaujz0free}, the equation of motion for $\tilde{j}^{z0}$ is obtained as
\begin{align}
    \partial_w \tilde{j}^{z0}&= -\frac2w \tilde{j}^{z0}-\frac{1}{w}\left(1-e^{-w/2}\right)\frac{\hbar}{4} \int_{p\ms} \frac{1}{E_p}\left\{-\frac{p_z^3}{E_p^3}\frac{1}{m} \left(1+\frac{E_p}{2(E_p+m)}\right) +\frac{p_z}{E_p} \frac{p_z^2}{2m(E_p+m)}\left(\frac{1}{E_p}-\frac{1}{E_p+m}\right)\right\} \ms^{0} f_\infty\n\\ 
    & -\frac{\hbar}{4} \frac{1}{w}\left(1-e^{-w/2}\right)\int_{p\ms} \frac{p_z^2}{E_p^2} \frac{1}{m} \left(1+\frac{E_p}{2(E_p+m)}\right) \bigg\{\left(\frac{1}{m^2}-\frac{1}{E_p^2}\right)p_z\ms^0
    +\frac{1}{E_p}\ms^z\bigg\} f_\infty-(\tilde{j}^{z0}-\tilde{j}^{z0}_\infty)\n\\
    &= -\frac2w \tilde{j}^{z0}-\frac{1}{w}\left(1-e^{-w/2}\right)\frac{\hbar}{4} \int_{p\ms} \frac{1}{E_p}\left\{-\frac{p_z^3}{E_p^3}\frac{1}{m} -\frac{1}{E_p}\frac{p_z^3}{2m(E_p+m)^2}\right\} \ms^{0} f_\infty\n\\ 
    & -\frac{\hbar}{4} \frac{1}{w}\left(1-e^{-w/2}\right)\int_{p\ms} \frac{1}{E_p}  \bigg\{ \frac{p_z^3}{E_p} \frac{1}{m} \left(1+\frac{E_p}{2(E_p+m)}\right)\left(\frac{1}{m^2}-\frac{1}{E_p^2}\right)\ms^0
    +\frac{p_z^2}{E_p^2} \frac{1}{m} \left(1+\frac{E_p}{2(E_p+m)}\right)\ms^z\bigg\} f_\infty\n\\
    &-(\tilde{j}^{zx}-\tilde{j}^{zx}_\infty)\n\\
  &= -\frac2w \tilde{j}^{z0}-\frac{1}{w}\left(1-e^{-w/2}\right)\frac{\hbar}{4} \int_{p\ms} \frac{p_z^3}{E_p^2}\frac1m\left[-\frac{2}{E_p^2}+\frac{1}{m^2} -\frac{1}{2(E_p+m)^2}+\frac{E_p}{2(E_p+m)}\left(\frac{1}{m^2}-\frac{1}{E_p^2}\right)\right] \ms^{0} f_\infty\n\\ 
    & -\frac{\hbar}{4} \frac{1}{w}\left(1-e^{-w/2}\right)\int_{p\ms} \frac{p_z^2}{E_p^3} \frac{1}{m} \left(1+\frac{E_p}{2(E_p+m)}\right)\ms^z f_\infty-(\tilde{j}^{z0}-\tilde{j}^{z0}_\infty)  
\end{align}
and for $\tilde{j}^{0z}_+$
\begin{align}
    \partial_w \tilde{j}^{0z}_+&=-\frac1w \tilde{j}_+^{0z}-\frac{1}{w}\left(1-e^{-w/2}\right)\frac{\hbar}{4} \int_{p\ms} \frac{1}{E_p}\left\{-\frac{p_z^2}{E_p^3} \left[  -\frac{m}{2(E_p+m)} +\frac{1}{m} \left(1+\frac{E_p}{2(E_p+m)}\right)p^{0} \right]+\frac{p_z}{E_p}\right.  \n\\
  &\left. \times\left[\frac{m}{2(E_p+m)^2}  \frac{p_z}{E_p}+\frac{1}{m} \left(1+\frac{E_p}{2(E_p+m)}\right)\frac{p_z}{E_p}+\frac{p_z}{2m(E_p+m)}\left(\frac{1}{E_p}-\frac{1}{E_p+m}\right)p^{0}\right]\right\} \ms^{z} f_\infty\n\\ 
   & +\frac{\hbar}{4} \frac{1}{w}\left(1-e^{-w/2}\right)  \int_{p\ms} \left[\frac{p_z}{E_p^3}  \frac{m}{2(E_p+m)}-\frac{p_z}{m^3} \left(1+\frac{E_p}{2(E_p+m)}\right) \right]\ms^0 f_\infty -(\tilde{j}^{0z}_+-\tilde{j}^{0z}_{+,\infty})\n\\
 &=-\frac1w \tilde{j}_+^{0z}-\frac{1}{w}\left(1-e^{-w/2}\right)\frac{\hbar}{4} \int_{p\ms} \frac{p_z^2}{E_p^2}\left\{   \frac{1}{E_p^2}\frac{m}{2(E_p+m)} -\frac{1}{m} \frac{1}{E_p}\left(1+\frac{E_p}{2(E_p+m)}\right) \right.  \n\\
  &\left. +\frac{m}{2(E_p+m)^2}  \frac{1}{E_p}+\frac{1}{m} \left(1+\frac{E_p}{2(E_p+m)}\right)\frac{1}{E_p}+\frac{1}{2m(E_p+m)}\left(1-\frac{E_p}{E_p+m}\right)\right\} \ms^{z} f_\infty\n\\ 
   & +\frac{\hbar}{4} \frac{1}{w}\left(1-e^{-w/2}\right)  \int_{p\ms} \left[\frac{p_z}{E_p^3}  \frac{m}{2(E_p+m)}-\frac{p_z}{m^3} \left(1+\frac{E_p}{2(E_p+m)}\right) \right]\ms^0 f_\infty -(\tilde{j}^{0z}_+-\tilde{j}^{0z}_{+,\infty})\n\\
 &=-\frac1w \tilde{j}_+^{0z}-\frac{1}{w}\left(1-e^{-w/2}\right)\frac{\hbar}{4} \int_{p\ms} \frac{p_z^2}{E_p^2}\left\{   \frac{1}{E_p^2}\frac{m}{2(E_p+m)}  +\frac{m}{2(E_p+m)^2}  \frac{1}{E_p}+\frac{1}{2m(E_p+m)}\left(1-\frac{E_p}{E_p+m}\right)\right\} \ms^{z} f_\infty\n\\ 
   & +\frac{\hbar}{4} \frac{1}{w}\left(1-e^{-w/2}\right)  \int_{p\ms} \left[\frac{p_z}{E_p^3}  \frac{m}{2(E_p+m)}-\frac{p_z}{m^3} \left(1+\frac{E_p}{2(E_p+m)}\right) \right]\ms^0 f_\infty -(\tilde{j}^{0z}_+-\tilde{j}^{0z}_{+,\infty}) \; . 
\end{align}
Using definition \eqref{jz0+} and noting that with \eq\eqref{finfty1}
\begin{align}
    \tilde{\mathcal{J}}^{z0}_{+,\infty}
    &= \frac\hbar4 \int_{p\ms} \frac{1}{E_p^2} \left[  -\frac{m^2-2E_p(E_p+m)-E_p^2}{2m(E_p+m)} \ms^z +\frac{1}{m} \left(1+\frac{E_p}{2(E_p+m)}\right)p^{z}\ms^0 \right] f_\infty\n\\
    &=   \frac\hbar4 \int_{p} \frac{1}{E_p}   \frac{m^2-2E_p(E_p+m)-E_p^2}{2m(E_p+m)} \left[ \frac{\hbar}{m}\tilde{\Omega}^{z0} E_p  -\xi\frac{2\hbar}{m(E_p+m)}\epsilon^{ij0 z}p_j  p^\nu (\Omega_{i\nu}+\partial_i \beta_\nu)\right] f_\text{LE}^{(0)}  \n\\
     &- \frac{\hbar^2}{4} \int_p \frac{1}{m^2} \frac{1}{E_p} \left(1+\frac{E_p}{2(E_p+m)}\right)p_z^2 \tilde{\Omega}^{0z}  f^{(0)}_\text{LE}\n\\
         &=   -\frac{\hbar^2}{2} \int_{p} \frac{1}{E_p}   \left[\frac{m^2-2E_p(E_p+m)-E_p^2}{2m^2(E_p+m)}E_p+\frac{p_z^2}{m^2}\left(1+\frac{E_p}{2(E_p+m)}\right)\right] \Omega^{xy}   f_\text{LE}^{(0)}  \n\\
     &-\frac{\hbar^2}{2} \int_{p} \frac{1}{E_p^2}   \frac{m^2-2E_p(E_p+m)-E_p^2}{2m(E_p+m)} \bxi \left[p_y  p^y (\Omega_{xy}+\partial_x \beta_y)-p_x  p^x (\Omega_{yx}+\partial_y \beta_x)\right] f_\text{LE}^{(0)} \n\\
    &= -\frac{\hbar^2}{2}  \Omega^{xy} \int_p  \left[\frac{m^2-2E_p(E_p+m)-E_p^2}{2m^2(E_p+m)}\left(1-\frac{p_z^2}{E_p^2}m\bxi\right)+\frac{p_z^2}{E_pm^2}\left(1+\frac{E_p}{2(E_p+m)}\right)\right]  f^{(0)}_\text{LE}\n\\
    &- \frac{\hbar^2}{2} \varpi^{xy} \int_{p}  \frac{m^2-2E_p(E_p+m)-E_p^2}{2m(E_p+m)} \bxi  \frac{p_z^2}{E_p^2} f^{(0)}_\text{LE}\n\\
     &=\Gamma_\Omega \tilde{\mathcal{J}}^{z0}+\Gamma_\varpi \varpi^{xy}
\end{align}
with
\begin{align}
    \Gamma_\Omega&\equiv - \left\{ \int_p  \left[\frac{m^2-2E_p(E_p+m)-E_p^2}{2m^2(E_p+m)}\left(1-\frac{p_z^2}{E_p^2}m\bxi\right)-\frac{p_z^2}{E_pm^2}\left(1+\frac{E_p}{2(E_p+m)}\right)\right]  f^{(0)}_\text{LE}\right\}^{-1} \n\\
    &\times \int_p  \left[\frac{m^2-2E_p(E_p+m)-E_p^2}{2m^2(E_p+m)}\left(1-\frac{p_z^2}{E_p^2}m\bxi\right)+\frac{p_z^2}{E_pm^2}\left(1+\frac{E_p}{2(E_p+m)}\right)\right]  f^{(0)}_\text{LE}\; ,\n\\
    \Gamma_\varpi&\equiv -(1+ \Gamma_\Omega) \frac{\hbar^2}{2} \int_{p}  \frac{m^2-2E_p(E_p+m)-E_p^2}{2m(E_p+m)} \bxi  \frac{p_z^2}{E_p^2} f^{(0)}_\text{LE} 
    \label{gamma}
\end{align}
we obtain \eqs\eqref{eomtjz0} of the main part as follows,
\begin{align}
  \partial_w  \tilde{\mathcal{J}}^{z0} &= -\frac{1}{w} \frac12\left(3 \tilde{\mathcal{J}}^{z0}+ \tilde{\mathcal{J}}^{z0}_+ \right)-\frac{1}{w}\left(1-e^{-w/2}\right)\frac{\hbar}{4m} \int_{p\ms} \frac{p_z^3}{E_p^2}\left[-\frac{2}{E_p^2}+\frac{1}{m^2} -\frac{1}{2(E_p+m)^2}+\frac{E_p}{2(E_p+m)}\left(\frac{1}{m^2}-\frac{1}{E_p^2}\right)\right] \n\\ 
    &\times \ms^{0} f_\infty -\frac{\hbar}{4m} \frac{1}{w}\left(1-e^{-w/2}\right)\int_{p\ms} \frac{p_z^2}{E_p^3} \left(1+\frac{E_p}{2(E_p+m)}\right)\ms^z f_\infty\n\\
    &+\frac{1}{w}\left(1-e^{-w/2}\right)\frac{\hbar}{4} \int_{p\ms} \frac{p_z^2}{E_p^2}\left\{   \frac{1}{E_p^2}\frac{m}{2(E_p+m)}  +\frac{m}{2(E_p+m)^2}  \frac{1}{E_p}+\frac{1}{2m(E_p+m)}\left(1-\frac{E_p}{E_p+m}\right)\right\} \ms^{z} f_\infty\n\\ 
   & -\frac{\hbar}{4} \frac{1}{w}\left(1-e^{-w/2}\right)  \int_{p\ms} \left[\frac{p_z}{E_p^3}  \frac{m}{2(E_p+m)}-\frac{p_z}{m^3} \left(1+\frac{E_p}{2(E_p+m)}\right) \right]\ms^0 f_\infty \n\\
&= -\frac{1}{w} \frac12\left(3 \tilde{\mathcal{J}}^{z0}+ \tilde{\mathcal{J}}^{z0}_+ \right)+\frac{1}{w}\left(1-e^{-w/2}\right) \left(\Lambda_0  \tilde{\mathcal{J}}^{z0}+\Lambda_\varpi \varpi^{xy} \right)
\end{align}
and
\begin{align}
  \partial_w  \tilde{\mathcal{J}}_+^{z0} &= -\frac{1}{w} \frac12\left( \tilde{\mathcal{J}}^{z0}+ 3\tilde{\mathcal{J}}^{z0}_+ \right)-\frac{1}{w}\left(1-e^{-w/2}\right)\frac{\hbar}{4m} \int_{p\ms} \frac{p_z^3}{E_p^2}\left[-\frac{2}{E_p^2}+\frac{1}{m^2} -\frac{1}{2(E_p+m)^2}+\frac{E_p}{2(E_p+m)}\left(\frac{1}{m^2}-\frac{1}{E_p^2}\right)\right] \n\\ 
    &\times \ms^{0} f_\infty -\frac{\hbar}{4m} \frac{1}{w}\left(1-e^{-w/2}\right)\int_{p\ms} \frac{p_z^2}{E_p^3} \left(1+\frac{E_p}{2(E_p+m)}\right)\ms^z f_\infty\n\\
    &-\frac{1}{w}\left(1-e^{-w/2}\right)\frac{\hbar}{4} \int_{p\ms} \frac{p_z^2}{E_p^2}\left\{   \frac{1}{E_p^2}\frac{m}{2(E_p+m)}  +\frac{m}{2(E_p+m)^2}  \frac{1}{E_p}+\frac{1}{2m(E_p+m)}\left(1-\frac{E_p}{E_p+m}\right)\right\} \ms^{z} f_\infty\n\\ 
   & +\frac{\hbar}{4} \frac{1}{w}\left(1-e^{-w/2}\right)  \int_{p\ms} \left[\frac{p_z}{E_p^3}  \frac{m}{2(E_p+m)}-\frac{p_z}{m^3} \left(1+\frac{E_p}{2(E_p+m)}\right) \right]\ms^0 f_\infty-(\tilde{\mathcal{J}}_+^{z0}-\tilde{\mathcal{J}}_{+,\infty}^{z0}) \n\\
&= -\frac{1}{w} \frac12\left( \tilde{\mathcal{J}}^{z0}+3 \tilde{\mathcal{J}}^{z0}_+ \right)+\frac{1}{w}\left(1-e^{-w/2}\right) \left(\bar{\Lambda}_0  \tilde{\mathcal{J}}^{z0}+\bar{\Lambda}_\varpi \varpi^{xy} \right)-\left(\tilde{\mathcal{J}}^{z0}_+- \Gamma_\Omega \tilde{\mathcal{J}}^{z0}-\Gamma_\varpi \varpi^{xy}\right)\; .
\end{align}
Here we used that
\begin{align}
&-\frac{\hbar}{4m} \int_{p\ms} \bigg\{\frac{p_z^2}{E_p^2}\left[-\frac{2}{E_p^2}+\frac{1}{m^2} -\frac{1}{2(E_p+m)^2}+\frac{E_p}{2(E_p+m)}\left(\frac{1}{m^2}-\frac{1}{E_p^2}\right)\right]+\frac{1}{E_p^3}  \frac{m^2}{2(E_p+m)}\n\\
&-\frac{1}{m^2} \left(1+\frac{E_p}{2(E_p+m)}\right)\bigg\} p^z \ms^{0} f_\infty  \n\\
&-\frac{\hbar}{4m}\int_{p\ms} \frac{p_z^2}{E_p^3} \bigg[1+\frac{E_p}{2(E_p+m)}- \frac{1}{E_p}\frac{m^2}{2(E_p+m)}  -\frac{m^2}{2(E_p+m)^2} -\frac{E_p}{2(E_p+m)}\left(1-\frac{E_p}{E_p+m}\right)\bigg]\ms^z f_\infty\n\\
&= \frac{\hbar^2}{2m}  \Omega^{xy} \int_p  \bigg\{\frac{p_z^2}{mE_p} \bigg(1- \frac{1}{E_p}\frac{m^2}{2(E_p+m)}  +\frac{E_p^2-m^2}{2(E_p+m)^2}\bigg)\left(1-\frac{p_z^2}{E_p^2}m\bxi\right)\n\\
&+ \frac{p_z^4}{mE_p}\left[-\frac{2}{E_p^2}+\frac{1}{m^2} -\frac{1}{2(E_p+m)^2}+\frac{E_p}{2(E_p+m)}\left(\frac{1}{m^2}-\frac{1}{E_p^2}\right)\right]+\frac{p_z^2}{E_p^2}  \frac{m}{2(E_p+m)}-\frac{E_pp_z^2}{m^3} \left(1+\frac{E_p}{2(E_p+m)}\right)  \bigg\}  f^{(0)}_\text{LE}\n\\
    &+ \frac{\hbar^2}{2m} \varpi^{xy} \int_{p}  \frac{p_z^4}{mE_p^3} \bigg(1- \frac{1}{E_p}\frac{m^2}{2(E_p+m)}  +\frac{E_p^2-m^2}{2(E_p+m)^2}\bigg) \bxi   f^{(0)}_\text{LE}
\end{align}
and 
\begin{align}
    &-\frac{\hbar}{4m} \int_{p\ms} \bigg\{\frac{p_z^2}{E_p^2}\left[-\frac{2}{E_p^2}+\frac{1}{m^2} -\frac{1}{2(E_p+m)^2}+\frac{E_p}{2(E_p+m)}\left(\frac{1}{m^2}-\frac{1}{E_p^2}\right)\right]-\frac{1}{E_p^3}  \frac{m^2}{2(E_p+m)}\n\\
&+\frac{1}{m^2} \left(1+\frac{E_p}{2(E_p+m)}\right)\bigg\} p^z \ms^{0} f_\infty  \n\\
&-\frac{\hbar}{4m}\int_{p\ms} \frac{p_z^2}{E_p^3} \bigg[1+\frac{E_p}{2(E_p+m)}+ \frac{1}{E_p}\frac{m^2}{2(E_p+m)}  +\frac{m^2}{2(E_p+m)^2} +\frac{E_p}{2(E_p+m)}\left(1-\frac{E_p}{E_p+m}\right)\bigg]\ms^z f_\infty\n\\
&= \frac{\hbar^2}{2m}  \Omega^{xy} \int_p  \bigg\{\frac{p_z^2}{mE_p} \bigg(1+\frac{E_p}{E_p+m}+ \frac{1}{E_p}\frac{m^2}{2(E_p+m)}  -\frac{E_p^2-m^2}{2(E_p+m)^2}\bigg)\left(1-\frac{p_z^2}{E_p^2}m\bxi\right)\n\\
&+ \frac{p_z^4}{mE_p}\left[-\frac{2}{E_p^2}+\frac{1}{m^2} -\frac{1}{2(E_p+m)^2}+\frac{E_p}{2(E_p+m)}\left(\frac{1}{m^2}-\frac{1}{E_p^2}\right)\right]-\frac{p_z^2}{E_p^2}  \frac{m}{2(E_p+m)}+\frac{E_pp_z^2}{m^3} \left(1+\frac{E_p}{2(E_p+m)}\right)  \bigg\}  f^{(0)}_\text{LE}\n\\
    &+ \frac{\hbar^2}{2m} \varpi^{xy} \int_{p}  \frac{p_z^4}{mE_p^3} \bigg(1+\frac{E_p}{E_p+m}+ \frac{1}{E_p}\frac{m^2}{2(E_p+m)}  -\frac{E_p^2-m^2}{2(E_p+m)^2}\bigg) \bxi   f^{(0)}_\text{LE}
\end{align}
We also defined
\begin{align}
    \Lambda_0&\equiv \left\{ \int_p  \left[\frac{m^2-2E_p(E_p+m)-E_p^2}{2m^2(E_p+m)}\left(1-\frac{p_z^2}{E_p^2}m\bxi\right)-\frac{p_z^2}{E_pm^2}\left(1+\frac{E_p}{2(E_p+m)}\right)\right]  f^{(0)}_\text{LE}\right\}^{-1}\n\\
    &\times \frac1m\int_p  \bigg\{\frac{p_z^2}{mE_p} \bigg(1- \frac{1}{E_p}\frac{m^2}{2(E_p+m)}  +\frac{E_p^2-m^2}{2(E_p+m)^2}\bigg)\left(1-\frac{p_z^2}{E_p^2}m\bxi\right)\n\\
&+ \frac{p_z^4}{mE_p}\left[-\frac{2}{E_p^2}+\frac{1}{m^2} -\frac{1}{2(E_p+m)^2}+\frac{E_p}{2(E_p+m)}\left(\frac{1}{m^2}-\frac{1}{E_p^2}\right)\right]+\frac{p_z^2}{E_p^2}  \frac{m}{2(E_p+m)}-\frac{E_pp_z^2}{m^3} \left(1+\frac{E_p}{2(E_p+m)}\right)  \bigg\}  f^{(0)}_\text{LE} \; ,\n\\
    \Lambda_\varpi& \equiv  \frac{\hbar^2}{2m}  \int_{p}  \frac{p_z^4}{mE_p^3} \bigg(1- \frac{1}{E_p}\frac{m^2}{2(E_p+m)}  +\frac{E_p^2-m^2}{2(E_p+m)^2}\bigg) \bxi   f^{(0)}_\text{LE} - \Lambda_0 \frac{\hbar^2}{2} \int_{p}  \frac{m^2-2E_p(E_p+m)-E_p^2}{2m(E_p+m)} \bxi  \frac{p_z^2}{E_p^2} f^{(0)}_\text{LE} \; ,\n\\
 \bar{\Lambda}_0&\equiv \left\{ \int_p  \left[\frac{m^2-2E_p(E_p+m)-E_p^2}{2m^2(E_p+m)}\left(1-\frac{p_z^2}{E_p^2}m\bxi\right)-\frac{p_z^2}{E_pm^2}\left(1+\frac{E_p}{2(E_p+m)}\right)\right]  f^{(0)}_\text{LE}\right\}^{-1} \n\\
 &\times \frac1m \int_p  \bigg\{\frac{p_z^2}{mE_p} \bigg(1+\frac{E_p}{E_p+m}+ \frac{1}{E_p}\frac{m^2}{2(E_p+m)}  -\frac{E_p^2-m^2}{2(E_p+m)^2}\bigg)\left(1-\frac{p_z^2}{E_p^2}m\bxi\right)\n\\
&+ \frac{p_z^4}{mE_p}\left[-\frac{2}{E_p^2}+\frac{1}{m^2} -\frac{1}{2(E_p+m)^2}+\frac{E_p}{2(E_p+m)}\left(\frac{1}{m^2}-\frac{1}{E_p^2}\right)\right]-\frac{p_z^2}{E_p^2}  \frac{m}{2(E_p+m)}+\frac{E_pp_z^2}{m^3} \left(1+\frac{E_p}{2(E_p+m)}\right)  \bigg\}  f^{(0)}_\text{LE}\; ,\n\\
    \bar{\Lambda}_\varpi& \equiv -\frac{\hbar^2}{2m} \int_{p}  \frac{p_z^4}{mE_p^3} \bigg(1+\frac{E_p}{E_p+m}+ \frac{1}{E_p}\frac{m^2}{2(E_p+m)}  -\frac{E_p^2-m^2}{2(E_p+m)^2}\bigg) \bxi   f^{(0)}_\text{LE}- \bar{\Lambda}_0 \frac{\hbar^2}{2} \int_{p}  \frac{m^2-2E_p(E_p+m)-E_p^2}{2m(E_p+m)} \bxi  \frac{p_z^2}{E_p^2} f^{(0)}_\text{LE}  \; .
    \label{biglambda}
\end{align}

\section{Final equations of motion}
\label{finapp}

In this appendix, we collect the complete set of closed equations of motion for all dynamical quantities.
Inserting \eqs\eqref{kdd} into \eqs\eqref{eomgzfin} we obtain for the longitudinal spin moments
\begin{align}
   \partial_w  \mG^z_{000}&= -\frac1w\left( \zeta_{00} \mG^z_{00}+\xi_{00} \mG^z_{200} \right)-\frac1w\left(1-e^{-w/2}\right)\Big( a_{00} \tilde{\mathcal{J}}^{z0}+b_{00}\varpi^{xy}+c_{00}\sigma\Big)-\left(\mG^z_{000}-\lambda_{00}^{00} \tilde{\mathcal{J}}^{z0}-\kappa_{00}^{00} \varpi^{xy}\right)\; ,\n\\
   \partial_w  \mG^z_{200}&= -\frac1w\left( \zeta_{20} \mG^z_{200}+\chi_{20} \mG^z_{000}+\xi_{20} \mG^z_{400} \right)-\frac1w\left(1-e^{-w/2}\right)\Big( a_{20} \tilde{\mathcal{J}}^{z0}+b_{20}\varpi^{xy}+c_{20}\sigma \Big)\n\\
   &-\left(\mG^z_{200}-\lambda_{20}^{00} \tilde{\mathcal{J}}^{z0}-\kappa_{20}^{00} \varpi^{xy}\right)\; ,\n\\
   \partial_w  \mG^z_{220}&= -\frac1w\left( \zeta_{22} \mG^z_{220}+\xi_{22} \mG^z_{420} \right)-\frac1w\left(1-e^{-w/2}\right)\Big(a_{22} \tilde{\mathcal{J}}^{z0}+b_{22}\varpi^{xy}+c_{22}\sigma\Big)-\mG^z_{220}\; ,\n\\
  \partial_w  \mG^z_{400}&= -\frac1w\left( \hat{\zeta}_{40} \mG^z_{400}+\chi_{40} \mG^z_{200} \right)
   -\frac1w\left(1-e^{-w/2}\right)\Big( a_{40} \tilde{\mathcal{J}}^{z0}+b_{40}\varpi^{xy}+c_{40}\sigma \Big)-\mG^z_{400}\; ,\n\\
   \partial_w  \mG^z_{420}&= -\frac1w\left( \hat{\zeta}_{42} \mG^z_{420}+\chi_{42} \mG^z_{220} \right)
   -\frac1w\left(1-e^{-w/2}\right)\Big( a_{42} \tilde{\mathcal{J}}^{z0}+b_{42}\varpi^{xy}+c_{42}\sigma \Big)-\mG^z_{420}\; ,\n\\
   \partial_w  \mG^z_{110}&= -\frac1w\left( \zeta_{11} \mG^z_{110}+\xi_{11} \mG^z_{310} \right)-\frac1w\left(1-e^{-w/2}\right)\left( \mathfrak{a}_{11} \tilde{\mathcal{J}}^{xz}+\bar{\mathfrak{a}}_{11}\tilde{\mathcal{J}}^{yz} \right)-\left(\mG^z_{110}-\lambda_{11}^{00} \tilde{\mathcal{J}}^{xz}-\bar{\lambda}_{11}^{00} \tilde{\mathcal{J}}^{yz}\right)\; ,\n\\
  \partial_w  \mG^z_{310}&= -\frac1w\left( \hat{\zeta}_{31} \mG^z_{310}+\chi_{31} \mG^z_{110} \right)
   -\frac1w\left(1-e^{-w/2}\right)\left(  \mathfrak{a}_{n1} \tilde{\mathcal{J}}^{xz}+\bar{\mathfrak{a}}_{31}\tilde{\mathcal{J}}^{yz}\right)-\mG^z_{310}\; .
   \label{eomgzfinfin}
\end{align}
Analogously one finds
\begin{align}
   \partial_w  \mI^z_{000}&= -\frac1w\left( \zeta_{00} \mI^z_{00}+\xi_{00} \mI^z_{200} \right)-\frac1w\left(1-e^{-w/2}\right)\Big( a_{00}^\prime \tilde{\mathcal{J}}^{z0}+b_{00}^\prime\varpi^{xy}+c_{00}^\prime\sigma \Big)-\left(\mI^z_{000}-\lambda_{00}^{01} \tilde{\mathcal{J}}^{z0}-\kappa_{00}^{01} \varpi^{xy}\right)\; ,\n\\
   \partial_w  \mI^z_{200}&= -\frac1w\left( \zeta_{20} \mI^z_{200}+\chi_{20} \mI^z_{000}+\xi_{20} \mI^z_{400} \right)-\frac1w\left(1-e^{-w/2}\right)\Big( a^\prime_{20} \tilde{\mathcal{J}}^{z0}+b^\prime_{20}\varpi^{xy}+c^\prime_{20}\sigma\Big)\n\\
   &-\left(\mI^z_{200}-\lambda_{20}^{01} \tilde{\mathcal{J}}^{z0}-\kappa_{20}^{01} \varpi^{xy}\right)\; ,\n\\
   \partial_w  \mI^z_{220}&= -\frac1w\left( \zeta_{22} \mI^z_{220}+\xi_{22} \mI^z_{420} \right)-\frac1w\left(1-e^{-w/2}\right)\Big( a_{22}^\prime \tilde{\mathcal{J}}^{z0}+b^\prime_{22}\varpi^{xy}+c_{22}^\prime\sigma\Big)-\mI^z_{220}\; ,\n\\
  \partial_w  \mI^z_{400}&= -\frac1w\left( \hat{\zeta}_{40} \mI^z_{400}+\chi_{40} \mI^z_{200} \right)
   -\frac1w\left(1-e^{-w/2}\right)\Big( a^\prime_{40} \tilde{\mathcal{J}}^{z0}+b^\prime_{40}\varpi^{xy}+c^\prime_{40}\sigma \Big)-\mI^z_{400}\; ,\n\\
   \partial_w  \mI^z_{420}&= -\frac1w\left( \hat{\zeta}_{42} \mI^z_{420}+\chi_{42} \mI^z_{220} \right)
   -\frac1w\left(1-e^{-w/2}\right)\Big( a^\prime_{42} \tilde{\mathcal{J}}^{z0}+b^\prime_{42}\varpi^{xy}+c^\prime_{42}\sigma\Big)-\mI^z_{420}\; ,\n\\
   \partial_w  \mI^z_{110}&= -\frac1w\left( \zeta_{11} \mI^z_{110}+\xi_{11} \mI^z_{310} \right)-\frac1w\left(1-e^{-w/2}\right)\left( \mathfrak{a}^\prime_{n1} \tilde{\mathcal{J}}^{xz}+\bar{\mathfrak{a}}^\prime_{11}\tilde{\mathcal{J}}^{yz}\right)-\left(\mI^z_{110}-\lambda_{11}^{01} \tilde{\mathcal{J}}^{xz}-\bar{\lambda}_{11}^{01} \tilde{\mathcal{J}}^{yz}\right)\; ,\n\\
  \partial_w  \mI^z_{310}&= -\frac1w\left( \hat{\zeta}_{31} \mI^z_{310}+\chi_{31} \mI^z_{110} \right)
   -\frac1w\left(1-e^{-w/2}\right)\left(  \mathfrak{a}_{n1}^\prime \tilde{\mathcal{J}}^{xz}+\bar{\mathfrak{a}}_{31}^\prime\tilde{\mathcal{J}}^{yz} \right)-\mI^z_{310}\; .
   \label{eomizfinfin}
\end{align}
The coefficients in \eqs\eqref{eomgzfinfin} and \eq\eqref{eomizfinfin} are defined in \eqs\eqref{zetachixi}, \eqref{abc}, \eqref{abcprime}, and \eqref{lambkapevapp}.

Furthermore, inserting \eqs\eqref{kxy} into \eqs\eqref{eomgxfin}, we obtain for the transverse spin moments
\begin{align}
 \partial_w \mG_{100}^x&=-\frac{1}{w} \left( \bar{a}_{100} \mG^x_{100}+\bar{c}_{100} \mG^x_{300}\right)-\frac1w\left(1-e^{-w/2}\right)\frac12\alpha_{10} \tilde{\mathcal{J}}^{xz}-\left(\mG_{100}^x-\lambda_{10}^{00} \tilde{\mathcal{J}}^{xz}  \right)\; ,\n\\
 \partial_w \mG_{300}^x&=-\frac{1}{w} \left( \hat{a}_{300} \mG^x_{300}+\bar{b}_{300}\mG^x_{100}\right)-\frac1w\left(1-e^{-w/2}\right) \frac12\alpha_{30} \tilde{\mathcal{J}}^{xz}-\mG_{300}^x \; ,\n\\
 \partial_w \mG_{210}^x&=-\frac{1}{w} \left( \bar{a}_{210} \mG^x_{210}+\bar{c}_{210} \mG^x_{410}\right)-\frac1w\left(1-e^{-w/2}\right) \frac12\left(\alpha_{21} \tilde{\mathcal{J}}^{z0}+\eta_{21}\varpi^{xy}-\tilde{\alpha}_{21}\sigma\right)\n\\
 &-\left(\mG_{210}^x- \lambda_{21}^{00} \tilde{\mathcal{J}}^{z0}-\kappa_{21}^{00}\varpi^{xy}+\tilde{\kappa}_{21}^{00}\sigma\right) \; ,\n\\
 \partial_w \mG_{410}^x&=-\frac{1}{w} \left( \hat{a}_{410} \mG^x_{410}+\bar{b}_{410}\mG^x_{210}\right)-\frac1w\left(1-e^{-w/2}\right) \frac12\left(\alpha_{41} \tilde{\mathcal{J}}^{z0}+\eta_{41}\varpi^{xy}-\tilde{\alpha}_{41}\sigma \right)-\mG_{410}^x \; , \n\\
  \partial_w \mG_{100}^y&=-\frac{1}{w} \left( \bar{a}_{100} \mG^y_{100}+\bar{c}_{100} \mG^y_{300}\right)-\frac1w\left(1-e^{-w/2}\right)\frac12\alpha_{10} \tilde{\mathcal{J}}^{yz}-\left(\mG_{100}^y-\lambda_{10}^{00} \tilde{\mathcal{J}}^{yz} \right)\; ,\n\\
 \partial_w \mG_{300}^y&=-\frac{1}{w} \left( \hat{a}_{300} \mG^y_{300}+\bar{b}_{300}\mG^y_{100}\right)-\frac1w\left(1-e^{-w/2}\right) \frac12\alpha_{30} \tilde{\mathcal{J}}^{yz}-\mG_{300}^y \; ,\n\\
 \partial_w \mG_{210}^y&=-\frac{1}{w} \left( \bar{a}_{210} \mG^y_{210}+\bar{c}_{210} \mG^y_{410}\right)-\frac1w\left(1-e^{-w/2}\right) \frac12\left(\alpha_{21} \tilde{\mathcal{J}}^{z0}+\eta_{21}\varpi^{xy}+\hat{\alpha}_{21}\sigma\right)\n\\
 &-\left(\mG_{210}^y- \bar{\lambda}_{21}^{00} \tilde{\mathcal{J}}^{z0}-\bar{\kappa}_{21}^{00}\varpi^{xy}-\hat{\kappa}_{21}^{00}\sigma\right) \; ,\n\\
 \partial_w \mG_{410}^y&=-\frac{1}{w} \left( \hat{a}_{410} \mG^y_{410}+\bar{b}_{410}\mG^y_{210}\right)-\frac1w\left(1-e^{-w/2}\right) \frac12\left(\alpha_{41} \tilde{\mathcal{J}}^{z0}+\eta_{41}\varpi^{xy}+\hat{\alpha}_{41}\sigma \right)-\mG_{410}^y 
 \label{eomgxyfinfin}
\end{align}
and, analogously,
\begin{align}
 \partial_w \mI_{100}^x&=-\frac{1}{w} \left( \bar{a}_{100} \mI^x_{100}+\bar{c}_{100} \mI^x_{300}\right)-\left(\mI_{100}^x-\lambda_{10}^{01} \tilde{\mathcal{J}}^{xz}  \right)\; ,\n\\
 \partial_w \mI_{300}^x&=-\frac{1}{w} \left( \hat{a}_{300} \mI^x_{300}+\bar{b}_{300}\mI^x_{100}\right)-\mI_{300}^x \; ,\n\\
 \partial_w \mI_{210}^x&=-\frac{1}{w} \left( \bar{a}_{210} \mI^x_{210}+\bar{c}_{210} \mI^x_{410}\right)-\left(\mI_{210}^x- \lambda_{21}^{01} \tilde{\mathcal{J}}^{z0}-\kappa_{21}^{01}\varpi^{xy}+\tilde{\kappa}_{21}^{01}\sigma\right) \; ,\n\\
 \partial_w \mI_{410}^x&=-\frac{1}{w} \left( \hat{a}_{410} \mI^x_{410}+\bar{b}_{410}\mI^x_{210}\right)-\mI_{410}^x \; ,\n\\
  \partial_w \mI_{100}^y&=-\frac{1}{w} \left( \bar{a}_{100} \mI^y_{100}+\bar{c}_{100} \mI^y_{300}\right)-\left(\mI_{100}^y-\lambda_{10}^{01} \tilde{\mathcal{J}}^{yz} \right)\; ,\n\\
 \partial_w \mI_{300}^y&=-\frac{1}{w} \left( \hat{a}_{300} \mI^y_{300}+\bar{b}_{300}\mI^y_{100}\right)-\mI_{300}^y \; ,\n\\
 \partial_w \mI_{210}^y&=-\frac{1}{w} \left( \bar{a}_{210} \mI^y_{210}+\bar{c}_{210} \mI^y_{410}\right)-\left(\mI_{210}^y- \bar{\lambda}_{21}^{01} \tilde{\mathcal{J}}^{z0}-\bar{\kappa}_{21}^{01}\varpi^{xy}-\hat{\kappa}_{21}^{01}\sigma\right) \; ,\n\\
 \partial_w \mI_{410}^y&=-\frac{1}{w} \left( \hat{a}_{410} \mI^y_{410}+\bar{b}_{410}\mI^y_{210}\right)-\mI_{410}^y \; .
 \label{eomixyfinfin}
\end{align}
The coefficients in \eqs\eqref{eomgxyfinfin} and \eqref{eomixyfinfin} are given in \eqs\eqref{abcdefbar}, \eqref{alphaeta}, and \eqref{lambkapodd}.

Finally, the equations of motion for the total angular momentum are given by \eqs\eqref{eomjplus} and \eqref{eomtjz0},
\begin{align}
    \partial_w \tilde{\mathcal{J}}^{zx}
    &=-\frac{1}{w} \frac12\left(3 \tilde{\mathcal{J}}^{zx}+ \tilde{\mathcal{J}}^{zx}_+ \right)+\frac{1}{w}\left(1-e^{-w/2}\right) \Lambda_{x} \tilde{\mathcal{J}}^{zx}\; ,\n\\
    \partial_w \tilde{\mathcal{J}}^{zx}_+
&=-\frac{1}{w} \frac12\left( \tilde{\mathcal{J}}^{zx}+ 3\tilde{\mathcal{J}}^{zx}_+ \right)- \tilde{\mathcal{J}}^{zx}_+ \; , \n\\
\partial_w \tilde{\mathcal{J}}^{zy}
    &=-\frac{1}{w} \frac12\left(3 \tilde{\mathcal{J}}^{zy}+ \tilde{\mathcal{J}}^{zy}_+ \right)+\frac{1}{w}\left(1-e^{-w/2}\right) \Lambda_{x} \tilde{\mathcal{J}}^{zy}\; ,\n\\
    \partial_w \tilde{\mathcal{J}}^{zy}_+
&=-\frac{1}{w} \frac12\left( \tilde{\mathcal{J}}^{zy}+ 3\tilde{\mathcal{J}}^{zy}_+ \right)- \tilde{\mathcal{J}}^{zy}_+ \; , \n\\
\partial_w  \tilde{\mathcal{J}}^{z0} 
&= -\frac{1}{w} \frac12\left(3 \tilde{\mathcal{J}}^{z0}+ \tilde{\mathcal{J}}^{z0}_+ \right)+\frac{1}{w}\left(1-e^{-w/2}\right) \left(\Lambda_0  \tilde{\mathcal{J}}^{z0}+\Lambda_\varpi \varpi^{xy} \right)\; , \n\\
  \partial_w  \tilde{\mathcal{J}}_+^{z0} 
&= -\frac{1}{w} \frac12\left( \tilde{\mathcal{J}}^{z0}+ 3\tilde{\mathcal{J}}^{z0}_+ \right)+\frac{1}{w}\left(1-e^{-w/2}\right) \left(\bar{\Lambda}_0  \tilde{\mathcal{J}}^{z0}+\bar{\Lambda}_\varpi \varpi^{xy} \right)-\left(\tilde{\mathcal{J}}^{z0}_+- \Gamma_\Omega \tilde{\mathcal{J}}^{z0}-\Gamma_\varpi \varpi^{xy}\right)\; 
\end{align}
with the coefficients defined in \eqs\eqref{lambdax}, \eqref{gamma} and \eqref{biglambda}.

\end{appendix}

\bibliography{biblio_paper_long}{}

\begin{thebibliography}{62}
\expandafter\ifx\csname natexlab\endcsname\relax\def\natexlab#1{#1}\fi
\expandafter\ifx\csname bibnamefont\endcsname\relax
  \def\bibnamefont#1{#1}\fi
\expandafter\ifx\csname bibfnamefont\endcsname\relax
  \def\bibfnamefont#1{#1}\fi
\expandafter\ifx\csname citenamefont\endcsname\relax
  \def\citenamefont#1{#1}\fi
\expandafter\ifx\csname url\endcsname\relax
  \def\url#1{\texttt{#1}}\fi
\expandafter\ifx\csname urlprefix\endcsname\relax\def\urlprefix{URL }\fi
\providecommand{\bibinfo}[2]{#2}
\providecommand{\eprint}[2][]{\url{#2}}

\bibitem[{\citenamefont{Becattini
  et~al.}(2013{\natexlab{a}})\citenamefont{Becattini, Csernai, and
  Wang}}]{Becattini:2013vja}
\bibinfo{author}{\bibfnamefont{F.}~\bibnamefont{Becattini}},
  \bibinfo{author}{\bibfnamefont{L.}~\bibnamefont{Csernai}}, \bibnamefont{and}
  \bibinfo{author}{\bibfnamefont{D.~J.} \bibnamefont{Wang}},
  \bibinfo{journal}{Phys. Rev.} \textbf{\bibinfo{volume}{C88}},
  \bibinfo{pages}{034905} (\bibinfo{year}{2013}{\natexlab{a}}),
  \bibinfo{note}{[Erratum: Phys. Rev.C93,no.6,069901(2016)]},
  \eprint{1304.4427}.

\bibitem[{\citenamefont{Becattini
  et~al.}(2013{\natexlab{b}})\citenamefont{Becattini, Chandra, Del~Zanna, and
  Grossi}}]{Becattini:2013fla}
\bibinfo{author}{\bibfnamefont{F.}~\bibnamefont{Becattini}},
  \bibinfo{author}{\bibfnamefont{V.}~\bibnamefont{Chandra}},
  \bibinfo{author}{\bibfnamefont{L.}~\bibnamefont{Del~Zanna}},
  \bibnamefont{and} \bibinfo{author}{\bibfnamefont{E.}~\bibnamefont{Grossi}},
  \bibinfo{journal}{Annals Phys.} \textbf{\bibinfo{volume}{338}},
  \bibinfo{pages}{32} (\bibinfo{year}{2013}{\natexlab{b}}), \eprint{1303.3431}.

\bibitem[{\citenamefont{Becattini et~al.}(2015)\citenamefont{Becattini,
  Inghirami, Rolando, Beraudo, Del~Zanna, De~Pace, Nardi, Pagliara, and
  Chandra}}]{Becattini:2015ska}
\bibinfo{author}{\bibfnamefont{F.}~\bibnamefont{Becattini}},
  \bibinfo{author}{\bibfnamefont{G.}~\bibnamefont{Inghirami}},
  \bibinfo{author}{\bibfnamefont{V.}~\bibnamefont{Rolando}},
  \bibinfo{author}{\bibfnamefont{A.}~\bibnamefont{Beraudo}},
  \bibinfo{author}{\bibfnamefont{L.}~\bibnamefont{Del~Zanna}},
  \bibinfo{author}{\bibfnamefont{A.}~\bibnamefont{De~Pace}},
  \bibinfo{author}{\bibfnamefont{M.}~\bibnamefont{Nardi}},
  \bibinfo{author}{\bibfnamefont{G.}~\bibnamefont{Pagliara}}, \bibnamefont{and}
  \bibinfo{author}{\bibfnamefont{V.}~\bibnamefont{Chandra}},
  \bibinfo{journal}{Eur. Phys. J.} \textbf{\bibinfo{volume}{C75}},
  \bibinfo{pages}{406} (\bibinfo{year}{2015}), \bibinfo{note}{[Erratum: Eur.
  Phys. J.C78,no.5,354(2018)]}, \eprint{1501.04468}.

\bibitem[{\citenamefont{Becattini et~al.}(2017)\citenamefont{Becattini,
  Karpenko, Lisa, Upsal, and Voloshin}}]{Becattini:2016gvu}
\bibinfo{author}{\bibfnamefont{F.}~\bibnamefont{Becattini}},
  \bibinfo{author}{\bibfnamefont{I.}~\bibnamefont{Karpenko}},
  \bibinfo{author}{\bibfnamefont{M.}~\bibnamefont{Lisa}},
  \bibinfo{author}{\bibfnamefont{I.}~\bibnamefont{Upsal}}, \bibnamefont{and}
  \bibinfo{author}{\bibfnamefont{S.}~\bibnamefont{Voloshin}},
  \bibinfo{journal}{Phys. Rev.} \textbf{\bibinfo{volume}{C95}},
  \bibinfo{pages}{054902} (\bibinfo{year}{2017}), \eprint{1610.02506}.

\bibitem[{\citenamefont{Karpenko and Becattini}(2017)}]{Karpenko:2016jyx}
\bibinfo{author}{\bibfnamefont{I.}~\bibnamefont{Karpenko}} \bibnamefont{and}
  \bibinfo{author}{\bibfnamefont{F.}~\bibnamefont{Becattini}},
  \bibinfo{journal}{Eur. Phys. J.} \textbf{\bibinfo{volume}{C77}},
  \bibinfo{pages}{213} (\bibinfo{year}{2017}), \eprint{1610.04717}.

\bibitem[{\citenamefont{Pang et~al.}(2016)\citenamefont{Pang, Petersen, Wang,
  and Wang}}]{Pang:2016igs}
\bibinfo{author}{\bibfnamefont{L.-G.} \bibnamefont{Pang}},
  \bibinfo{author}{\bibfnamefont{H.}~\bibnamefont{Petersen}},
  \bibinfo{author}{\bibfnamefont{Q.}~\bibnamefont{Wang}}, \bibnamefont{and}
  \bibinfo{author}{\bibfnamefont{X.-N.} \bibnamefont{Wang}},
  \bibinfo{journal}{Phys. Rev. Lett.} \textbf{\bibinfo{volume}{117}},
  \bibinfo{pages}{192301} (\bibinfo{year}{2016}), \eprint{1605.04024}.

\bibitem[{\citenamefont{Xie et~al.}(2017)\citenamefont{Xie, Wang, and
  Csernai}}]{Xie:2017upb}
\bibinfo{author}{\bibfnamefont{Y.}~\bibnamefont{Xie}},
  \bibinfo{author}{\bibfnamefont{D.}~\bibnamefont{Wang}}, \bibnamefont{and}
  \bibinfo{author}{\bibfnamefont{L.~P.} \bibnamefont{Csernai}},
  \bibinfo{journal}{Phys. Rev.} \textbf{\bibinfo{volume}{C95}},
  \bibinfo{pages}{031901} (\bibinfo{year}{2017}), \eprint{1703.03770}.

\bibitem[{\citenamefont{Adamczyk et~al.}(2017)}]{STAR:2017ckg}
\bibinfo{author}{\bibfnamefont{L.}~\bibnamefont{Adamczyk}} \bibnamefont{et~al.}
  (\bibinfo{collaboration}{STAR}), \bibinfo{journal}{Nature}
  \textbf{\bibinfo{volume}{548}}, \bibinfo{pages}{62} (\bibinfo{year}{2017}),
  \eprint{1701.06657}.

\bibitem[{\citenamefont{Adam et~al.}(2018)}]{Adam:2018ivw}
\bibinfo{author}{\bibfnamefont{J.}~\bibnamefont{Adam}} \bibnamefont{et~al.}
  (\bibinfo{collaboration}{STAR}), \bibinfo{journal}{Phys. Rev.}
  \textbf{\bibinfo{volume}{C98}}, \bibinfo{pages}{014910}
  (\bibinfo{year}{2018}), \eprint{1805.04400}.

\bibitem[{\citenamefont{Acharya et~al.}(2020)}]{ALICE:2019aid}
\bibinfo{author}{\bibfnamefont{S.}~\bibnamefont{Acharya}} \bibnamefont{et~al.}
  (\bibinfo{collaboration}{ALICE}), \bibinfo{journal}{Phys. Rev. Lett.}
  \textbf{\bibinfo{volume}{125}}, \bibinfo{pages}{012301}
  (\bibinfo{year}{2020}), \eprint{1910.14408}.

\bibitem[{\citenamefont{Adam et~al.}(2019)}]{STAR:2019erd}
\bibinfo{author}{\bibfnamefont{J.}~\bibnamefont{Adam}} \bibnamefont{et~al.}
  (\bibinfo{collaboration}{STAR}), \bibinfo{journal}{Phys. Rev. Lett.}
  \textbf{\bibinfo{volume}{123}}, \bibinfo{pages}{132301}
  (\bibinfo{year}{2019}), \eprint{1905.11917}.

\bibitem[{\citenamefont{Liu and Yin}(2021)}]{Liu:2021uhn}
\bibinfo{author}{\bibfnamefont{S.~Y.~F.} \bibnamefont{Liu}} \bibnamefont{and}
  \bibinfo{author}{\bibfnamefont{Y.}~\bibnamefont{Yin}},
  \bibinfo{journal}{JHEP} \textbf{\bibinfo{volume}{07}}, \bibinfo{pages}{188}
  (\bibinfo{year}{2021}), \eprint{2103.09200}.

\bibitem[{\citenamefont{Fu et~al.}(2021)\citenamefont{Fu, Liu, Pang, Song, and
  Yin}}]{Fu:2021pok}
\bibinfo{author}{\bibfnamefont{B.}~\bibnamefont{Fu}},
  \bibinfo{author}{\bibfnamefont{S.~Y.~F.} \bibnamefont{Liu}},
  \bibinfo{author}{\bibfnamefont{L.}~\bibnamefont{Pang}},
  \bibinfo{author}{\bibfnamefont{H.}~\bibnamefont{Song}}, \bibnamefont{and}
  \bibinfo{author}{\bibfnamefont{Y.}~\bibnamefont{Yin}},
  \bibinfo{journal}{Phys. Rev. Lett.} \textbf{\bibinfo{volume}{127}},
  \bibinfo{pages}{142301} (\bibinfo{year}{2021}), \eprint{2103.10403}.

\bibitem[{\citenamefont{Becattini
  et~al.}(2021{\natexlab{a}})\citenamefont{Becattini, Buzzegoli, and
  Palermo}}]{Becattini:2021suc}
\bibinfo{author}{\bibfnamefont{F.}~\bibnamefont{Becattini}},
  \bibinfo{author}{\bibfnamefont{M.}~\bibnamefont{Buzzegoli}},
  \bibnamefont{and} \bibinfo{author}{\bibfnamefont{A.}~\bibnamefont{Palermo}},
  \bibinfo{journal}{Phys. Lett. B} \textbf{\bibinfo{volume}{820}},
  \bibinfo{pages}{136519} (\bibinfo{year}{2021}{\natexlab{a}}),
  \eprint{2103.10917}.

\bibitem[{\citenamefont{Becattini
  et~al.}(2021{\natexlab{b}})\citenamefont{Becattini, Buzzegoli, Inghirami,
  Karpenko, and Palermo}}]{Becattini:2021iol}
\bibinfo{author}{\bibfnamefont{F.}~\bibnamefont{Becattini}},
  \bibinfo{author}{\bibfnamefont{M.}~\bibnamefont{Buzzegoli}},
  \bibinfo{author}{\bibfnamefont{G.}~\bibnamefont{Inghirami}},
  \bibinfo{author}{\bibfnamefont{I.}~\bibnamefont{Karpenko}}, \bibnamefont{and}
  \bibinfo{author}{\bibfnamefont{A.}~\bibnamefont{Palermo}},
  \bibinfo{journal}{Phys. Rev. Lett.} \textbf{\bibinfo{volume}{127}},
  \bibinfo{pages}{272302} (\bibinfo{year}{2021}{\natexlab{b}}),
  \eprint{2103.14621}.

\bibitem[{\citenamefont{Weickgenannt et~al.}(2019)\citenamefont{Weickgenannt,
  Sheng, Speranza, Wang, and Rischke}}]{Weickgenannt:2019dks}
\bibinfo{author}{\bibfnamefont{N.}~\bibnamefont{Weickgenannt}},
  \bibinfo{author}{\bibfnamefont{X.-L.} \bibnamefont{Sheng}},
  \bibinfo{author}{\bibfnamefont{E.}~\bibnamefont{Speranza}},
  \bibinfo{author}{\bibfnamefont{Q.}~\bibnamefont{Wang}}, \bibnamefont{and}
  \bibinfo{author}{\bibfnamefont{D.~H.} \bibnamefont{Rischke}},
  \bibinfo{journal}{Phys. Rev.} \textbf{\bibinfo{volume}{D100}},
  \bibinfo{pages}{056018} (\bibinfo{year}{2019}), \eprint{1902.06513}.

\bibitem[{\citenamefont{Gao and Liang}(2019)}]{Gao:2019znl}
\bibinfo{author}{\bibfnamefont{J.-H.} \bibnamefont{Gao}} \bibnamefont{and}
  \bibinfo{author}{\bibfnamefont{Z.-T.} \bibnamefont{Liang}},
  \bibinfo{journal}{Phys. Rev.} \textbf{\bibinfo{volume}{D100}},
  \bibinfo{pages}{056021} (\bibinfo{year}{2019}), \eprint{1902.06510}.

\bibitem[{\citenamefont{Hattori
  et~al.}(2019{\natexlab{a}})\citenamefont{Hattori, Hidaka, and
  Yang}}]{Hattori:2019ahi}
\bibinfo{author}{\bibfnamefont{K.}~\bibnamefont{Hattori}},
  \bibinfo{author}{\bibfnamefont{Y.}~\bibnamefont{Hidaka}}, \bibnamefont{and}
  \bibinfo{author}{\bibfnamefont{D.-L.} \bibnamefont{Yang}},
  \bibinfo{journal}{Phys. Rev.} \textbf{\bibinfo{volume}{D100}},
  \bibinfo{pages}{096011} (\bibinfo{year}{2019}{\natexlab{a}}),
  \eprint{1903.01653}.

\bibitem[{\citenamefont{Wang et~al.}(2019)\citenamefont{Wang, Guo, Shi, and
  Zhuang}}]{Wang:2019moi}
\bibinfo{author}{\bibfnamefont{Z.}~\bibnamefont{Wang}},
  \bibinfo{author}{\bibfnamefont{X.}~\bibnamefont{Guo}},
  \bibinfo{author}{\bibfnamefont{S.}~\bibnamefont{Shi}}, \bibnamefont{and}
  \bibinfo{author}{\bibfnamefont{P.}~\bibnamefont{Zhuang}},
  \bibinfo{journal}{Phys. Rev. D} \textbf{\bibinfo{volume}{100}},
  \bibinfo{pages}{014015} (\bibinfo{year}{2019}), \eprint{1903.03461}.

\bibitem[{\citenamefont{Weickgenannt
  et~al.}(2021{\natexlab{a}})\citenamefont{Weickgenannt, Speranza, Sheng, Wang,
  and Rischke}}]{Weickgenannt:2020aaf}
\bibinfo{author}{\bibfnamefont{N.}~\bibnamefont{Weickgenannt}},
  \bibinfo{author}{\bibfnamefont{E.}~\bibnamefont{Speranza}},
  \bibinfo{author}{\bibfnamefont{X.-l.} \bibnamefont{Sheng}},
  \bibinfo{author}{\bibfnamefont{Q.}~\bibnamefont{Wang}}, \bibnamefont{and}
  \bibinfo{author}{\bibfnamefont{D.~H.} \bibnamefont{Rischke}},
  \bibinfo{journal}{Phys. Rev. Lett.} \textbf{\bibinfo{volume}{127}},
  \bibinfo{pages}{052301} (\bibinfo{year}{2021}{\natexlab{a}}),
  \eprint{2005.01506}.

\bibitem[{\citenamefont{Liu et~al.}(2020)\citenamefont{Liu, Mameda, and
  Huang}}]{Liu:2020flb}
\bibinfo{author}{\bibfnamefont{Y.-C.} \bibnamefont{Liu}},
  \bibinfo{author}{\bibfnamefont{K.}~\bibnamefont{Mameda}}, \bibnamefont{and}
  \bibinfo{author}{\bibfnamefont{X.-G.} \bibnamefont{Huang}},
  \bibinfo{journal}{Chin. Phys. C} \textbf{\bibinfo{volume}{44}},
  \bibinfo{pages}{094101} (\bibinfo{year}{2020}), \bibinfo{note}{[Erratum:
  Chin.Phys.C 45, 089001 (2021)]}, \eprint{2002.03753}.

\bibitem[{\citenamefont{Weickgenannt
  et~al.}(2021{\natexlab{b}})\citenamefont{Weickgenannt, Speranza, Sheng, Wang,
  and Rischke}}]{Weickgenannt:2021cuo}
\bibinfo{author}{\bibfnamefont{N.}~\bibnamefont{Weickgenannt}},
  \bibinfo{author}{\bibfnamefont{E.}~\bibnamefont{Speranza}},
  \bibinfo{author}{\bibfnamefont{X.-l.} \bibnamefont{Sheng}},
  \bibinfo{author}{\bibfnamefont{Q.}~\bibnamefont{Wang}}, \bibnamefont{and}
  \bibinfo{author}{\bibfnamefont{D.~H.} \bibnamefont{Rischke}},
  \bibinfo{journal}{Phys. Rev. D} \textbf{\bibinfo{volume}{104}},
  \bibinfo{pages}{016022} (\bibinfo{year}{2021}{\natexlab{b}}),
  \eprint{2103.04896}.

\bibitem[{\citenamefont{Sheng et~al.}(2021)\citenamefont{Sheng, Weickgenannt,
  Speranza, Rischke, and Wang}}]{Sheng:2021kfc}
\bibinfo{author}{\bibfnamefont{X.-L.} \bibnamefont{Sheng}},
  \bibinfo{author}{\bibfnamefont{N.}~\bibnamefont{Weickgenannt}},
  \bibinfo{author}{\bibfnamefont{E.}~\bibnamefont{Speranza}},
  \bibinfo{author}{\bibfnamefont{D.~H.} \bibnamefont{Rischke}},
  \bibnamefont{and} \bibinfo{author}{\bibfnamefont{Q.}~\bibnamefont{Wang}},
  \bibinfo{journal}{Phys. Rev. D} \textbf{\bibinfo{volume}{104}},
  \bibinfo{pages}{016029} (\bibinfo{year}{2021}), \eprint{2103.10636}.

\bibitem[{\citenamefont{Wagner et~al.}(2022)\citenamefont{Wagner, Weickgenannt,
  and Rischke}}]{Wagner:2022amr}
\bibinfo{author}{\bibfnamefont{D.}~\bibnamefont{Wagner}},
  \bibinfo{author}{\bibfnamefont{N.}~\bibnamefont{Weickgenannt}},
  \bibnamefont{and} \bibinfo{author}{\bibfnamefont{D.~H.}
  \bibnamefont{Rischke}}, \bibinfo{journal}{Phys. Rev. D}
  \textbf{\bibinfo{volume}{106}}, \bibinfo{pages}{116021}
  (\bibinfo{year}{2022}), \eprint{2210.06187}.

\bibitem[{\citenamefont{Wagner et~al.}(2023)\citenamefont{Wagner, Weickgenannt,
  and Speranza}}]{Wagner:2023cct}
\bibinfo{author}{\bibfnamefont{D.}~\bibnamefont{Wagner}},
  \bibinfo{author}{\bibfnamefont{N.}~\bibnamefont{Weickgenannt}},
  \bibnamefont{and} \bibinfo{author}{\bibfnamefont{E.}~\bibnamefont{Speranza}}
  (\bibinfo{year}{2023}), \eprint{2306.05936}.

\bibitem[{\citenamefont{Weickgenannt and
  Blaizot}(2024{\natexlab{a}})}]{Weickgenannt:2024ibf}
\bibinfo{author}{\bibfnamefont{N.}~\bibnamefont{Weickgenannt}}
  \bibnamefont{and} \bibinfo{author}{\bibfnamefont{J.-P.}
  \bibnamefont{Blaizot}} (\bibinfo{year}{2024}{\natexlab{a}}),
  \eprint{2409.11045}.

\bibitem[{\citenamefont{Wang and Lin}(2024)}]{Wang:2024lis}
\bibinfo{author}{\bibfnamefont{Z.}~\bibnamefont{Wang}} \bibnamefont{and}
  \bibinfo{author}{\bibfnamefont{S.}~\bibnamefont{Lin}} (\bibinfo{year}{2024}),
  \eprint{2411.19550}.

\bibitem[{\citenamefont{Florkowski
  et~al.}(2018{\natexlab{a}})\citenamefont{Florkowski, Friman, Jaiswal, and
  Speranza}}]{Florkowski:2017ruc}
\bibinfo{author}{\bibfnamefont{W.}~\bibnamefont{Florkowski}},
  \bibinfo{author}{\bibfnamefont{B.}~\bibnamefont{Friman}},
  \bibinfo{author}{\bibfnamefont{A.}~\bibnamefont{Jaiswal}}, \bibnamefont{and}
  \bibinfo{author}{\bibfnamefont{E.}~\bibnamefont{Speranza}},
  \bibinfo{journal}{Phys. Rev.} \textbf{\bibinfo{volume}{C97}},
  \bibinfo{pages}{041901} (\bibinfo{year}{2018}{\natexlab{a}}),
  \eprint{1705.00587}.

\bibitem[{\citenamefont{Florkowski
  et~al.}(2018{\natexlab{b}})\citenamefont{Florkowski, Friman, Jaiswal,
  Ryblewski, and Speranza}}]{Florkowski:2017dyn}
\bibinfo{author}{\bibfnamefont{W.}~\bibnamefont{Florkowski}},
  \bibinfo{author}{\bibfnamefont{B.}~\bibnamefont{Friman}},
  \bibinfo{author}{\bibfnamefont{A.}~\bibnamefont{Jaiswal}},
  \bibinfo{author}{\bibfnamefont{R.}~\bibnamefont{Ryblewski}},
  \bibnamefont{and} \bibinfo{author}{\bibfnamefont{E.}~\bibnamefont{Speranza}},
  \bibinfo{journal}{Phys. Rev.} \textbf{\bibinfo{volume}{D97}},
  \bibinfo{pages}{116017} (\bibinfo{year}{2018}{\natexlab{b}}),
  \eprint{1712.07676}.

\bibitem[{\citenamefont{Florkowski et~al.}(2019)\citenamefont{Florkowski,
  Ryblewski, and Kumar}}]{Florkowski:2018fap}
\bibinfo{author}{\bibfnamefont{W.}~\bibnamefont{Florkowski}},
  \bibinfo{author}{\bibfnamefont{R.}~\bibnamefont{Ryblewski}},
  \bibnamefont{and} \bibinfo{author}{\bibfnamefont{A.}~\bibnamefont{Kumar}},
  \bibinfo{journal}{Prog. Part. Nucl. Phys.} \textbf{\bibinfo{volume}{108}},
  \bibinfo{pages}{103709} (\bibinfo{year}{2019}), \eprint{1811.04409}.

\bibitem[{\citenamefont{Montenegro and Torrieri}(2019)}]{Montenegro:2018bcf}
\bibinfo{author}{\bibfnamefont{D.}~\bibnamefont{Montenegro}} \bibnamefont{and}
  \bibinfo{author}{\bibfnamefont{G.}~\bibnamefont{Torrieri}},
  \bibinfo{journal}{Phys. Rev. D} \textbf{\bibinfo{volume}{100}},
  \bibinfo{pages}{056011} (\bibinfo{year}{2019}), \eprint{1807.02796}.

\bibitem[{\citenamefont{Hattori
  et~al.}(2019{\natexlab{b}})\citenamefont{Hattori, Hongo, Huang, Matsuo, and
  Taya}}]{Hattori:2019lfp}
\bibinfo{author}{\bibfnamefont{K.}~\bibnamefont{Hattori}},
  \bibinfo{author}{\bibfnamefont{M.}~\bibnamefont{Hongo}},
  \bibinfo{author}{\bibfnamefont{X.-G.} \bibnamefont{Huang}},
  \bibinfo{author}{\bibfnamefont{M.}~\bibnamefont{Matsuo}}, \bibnamefont{and}
  \bibinfo{author}{\bibfnamefont{H.}~\bibnamefont{Taya}},
  \bibinfo{journal}{Phys. Lett.} \textbf{\bibinfo{volume}{B795}},
  \bibinfo{pages}{100} (\bibinfo{year}{2019}{\natexlab{b}}),
  \eprint{1901.06615}.

\bibitem[{\citenamefont{Bhadury et~al.}(2021)\citenamefont{Bhadury, Florkowski,
  Jaiswal, Kumar, and Ryblewski}}]{Bhadury:2020puc}
\bibinfo{author}{\bibfnamefont{S.}~\bibnamefont{Bhadury}},
  \bibinfo{author}{\bibfnamefont{W.}~\bibnamefont{Florkowski}},
  \bibinfo{author}{\bibfnamefont{A.}~\bibnamefont{Jaiswal}},
  \bibinfo{author}{\bibfnamefont{A.}~\bibnamefont{Kumar}}, \bibnamefont{and}
  \bibinfo{author}{\bibfnamefont{R.}~\bibnamefont{Ryblewski}},
  \bibinfo{journal}{Phys. Lett. B} \textbf{\bibinfo{volume}{814}},
  \bibinfo{pages}{136096} (\bibinfo{year}{2021}), \eprint{2002.03937}.

\bibitem[{\citenamefont{Singh et~al.}(2021)\citenamefont{Singh, Sophys, and
  Ryblewski}}]{Singh:2020rht}
\bibinfo{author}{\bibfnamefont{R.}~\bibnamefont{Singh}},
  \bibinfo{author}{\bibfnamefont{G.}~\bibnamefont{Sophys}}, \bibnamefont{and}
  \bibinfo{author}{\bibfnamefont{R.}~\bibnamefont{Ryblewski}},
  \bibinfo{journal}{Phys. Rev. D} \textbf{\bibinfo{volume}{103}},
  \bibinfo{pages}{074024} (\bibinfo{year}{2021}), \eprint{2011.14907}.

\bibitem[{\citenamefont{Montenegro and Torrieri}(2020)}]{Montenegro:2020paq}
\bibinfo{author}{\bibfnamefont{D.}~\bibnamefont{Montenegro}} \bibnamefont{and}
  \bibinfo{author}{\bibfnamefont{G.}~\bibnamefont{Torrieri}},
  \bibinfo{journal}{Phys. Rev. D} \textbf{\bibinfo{volume}{102}},
  \bibinfo{pages}{036007} (\bibinfo{year}{2020}), \eprint{2004.10195}.

\bibitem[{\citenamefont{Gallegos et~al.}(2021)\citenamefont{Gallegos, G\"ursoy,
  and Yarom}}]{Gallegos:2021bzp}
\bibinfo{author}{\bibfnamefont{A.~D.} \bibnamefont{Gallegos}},
  \bibinfo{author}{\bibfnamefont{U.}~\bibnamefont{G\"ursoy}}, \bibnamefont{and}
  \bibinfo{author}{\bibfnamefont{A.}~\bibnamefont{Yarom}},
  \bibinfo{journal}{SciPost Phys.} \textbf{\bibinfo{volume}{11}},
  \bibinfo{pages}{041} (\bibinfo{year}{2021}), \eprint{2101.04759}.

\bibitem[{\citenamefont{Fukushima and Pu}(2021)}]{Fukushima:2020ucl}
\bibinfo{author}{\bibfnamefont{K.}~\bibnamefont{Fukushima}} \bibnamefont{and}
  \bibinfo{author}{\bibfnamefont{S.}~\bibnamefont{Pu}}, \bibinfo{journal}{Phys.
  Lett. B} \textbf{\bibinfo{volume}{817}}, \bibinfo{pages}{136346}
  (\bibinfo{year}{2021}), \eprint{2010.01608}.

\bibitem[{\citenamefont{Li et~al.}(2021)\citenamefont{Li, Stephanov, and
  Yee}}]{Li:2020eon}
\bibinfo{author}{\bibfnamefont{S.}~\bibnamefont{Li}},
  \bibinfo{author}{\bibfnamefont{M.~A.} \bibnamefont{Stephanov}},
  \bibnamefont{and} \bibinfo{author}{\bibfnamefont{H.-U.} \bibnamefont{Yee}},
  \bibinfo{journal}{Phys. Rev. Lett.} \textbf{\bibinfo{volume}{127}},
  \bibinfo{pages}{082302} (\bibinfo{year}{2021}), \eprint{2011.12318}.

\bibitem[{\citenamefont{Wang et~al.}(2021)\citenamefont{Wang, Fang, and
  Pu}}]{Wang:2021ngp}
\bibinfo{author}{\bibfnamefont{D.-L.} \bibnamefont{Wang}},
  \bibinfo{author}{\bibfnamefont{S.}~\bibnamefont{Fang}}, \bibnamefont{and}
  \bibinfo{author}{\bibfnamefont{S.}~\bibnamefont{Pu}}, \bibinfo{journal}{Phys.
  Rev. D} \textbf{\bibinfo{volume}{104}}, \bibinfo{pages}{114043}
  (\bibinfo{year}{2021}), \eprint{2107.11726}.

\bibitem[{\citenamefont{Hu}(2022)}]{Hu:2021pwh}
\bibinfo{author}{\bibfnamefont{J.}~\bibnamefont{Hu}}, \bibinfo{journal}{Phys.
  Rev. D} \textbf{\bibinfo{volume}{105}}, \bibinfo{pages}{076009}
  (\bibinfo{year}{2022}), \eprint{2111.03571}.

\bibitem[{\citenamefont{Hongo et~al.}(2021)\citenamefont{Hongo, Huang,
  Kaminski, Stephanov, and Yee}}]{Hongo:2021ona}
\bibinfo{author}{\bibfnamefont{M.}~\bibnamefont{Hongo}},
  \bibinfo{author}{\bibfnamefont{X.-G.} \bibnamefont{Huang}},
  \bibinfo{author}{\bibfnamefont{M.}~\bibnamefont{Kaminski}},
  \bibinfo{author}{\bibfnamefont{M.}~\bibnamefont{Stephanov}},
  \bibnamefont{and} \bibinfo{author}{\bibfnamefont{H.-U.} \bibnamefont{Yee}},
  \bibinfo{journal}{JHEP} \textbf{\bibinfo{volume}{11}}, \bibinfo{pages}{150}
  (\bibinfo{year}{2021}), \eprint{2107.14231}.

\bibitem[{\citenamefont{Daher et~al.}(2022)\citenamefont{Daher, Das,
  Florkowski, and Ryblewski}}]{Daher:2022xon}
\bibinfo{author}{\bibfnamefont{A.}~\bibnamefont{Daher}},
  \bibinfo{author}{\bibfnamefont{A.}~\bibnamefont{Das}},
  \bibinfo{author}{\bibfnamefont{W.}~\bibnamefont{Florkowski}},
  \bibnamefont{and} \bibinfo{author}{\bibfnamefont{R.}~\bibnamefont{Ryblewski}}
  (\bibinfo{year}{2022}), \eprint{2202.12609}.

\bibitem[{\citenamefont{Weickgenannt
  et~al.}(2022{\natexlab{a}})\citenamefont{Weickgenannt, Wagner, Speranza, and
  Rischke}}]{Weickgenannt:2022zxs}
\bibinfo{author}{\bibfnamefont{N.}~\bibnamefont{Weickgenannt}},
  \bibinfo{author}{\bibfnamefont{D.}~\bibnamefont{Wagner}},
  \bibinfo{author}{\bibfnamefont{E.}~\bibnamefont{Speranza}}, \bibnamefont{and}
  \bibinfo{author}{\bibfnamefont{D.~H.} \bibnamefont{Rischke}},
  \bibinfo{journal}{Phys. Rev. D} \textbf{\bibinfo{volume}{106}},
  \bibinfo{pages}{096014} (\bibinfo{year}{2022}{\natexlab{a}}),
  \eprint{2203.04766}.

\bibitem[{\citenamefont{Weickgenannt
  et~al.}(2022{\natexlab{b}})\citenamefont{Weickgenannt, Wagner, and
  Speranza}}]{Weickgenannt:2022jes}
\bibinfo{author}{\bibfnamefont{N.}~\bibnamefont{Weickgenannt}},
  \bibinfo{author}{\bibfnamefont{D.}~\bibnamefont{Wagner}}, \bibnamefont{and}
  \bibinfo{author}{\bibfnamefont{E.}~\bibnamefont{Speranza}},
  \bibinfo{journal}{Phys. Rev. D} \textbf{\bibinfo{volume}{105}},
  \bibinfo{pages}{116026} (\bibinfo{year}{2022}{\natexlab{b}}),
  \eprint{2204.01797}.

\bibitem[{\citenamefont{Gallegos et~al.}(2022)\citenamefont{Gallegos, Gursoy,
  and Yarom}}]{Gallegos:2022jow}
\bibinfo{author}{\bibfnamefont{A.~D.} \bibnamefont{Gallegos}},
  \bibinfo{author}{\bibfnamefont{U.}~\bibnamefont{Gursoy}}, \bibnamefont{and}
  \bibinfo{author}{\bibfnamefont{A.}~\bibnamefont{Yarom}}
  (\bibinfo{year}{2022}), \eprint{2203.05044}.

\bibitem[{\citenamefont{Cao et~al.}(2022)\citenamefont{Cao, Hattori, Hongo,
  Huang, and Taya}}]{Cao:2022aku}
\bibinfo{author}{\bibfnamefont{Z.}~\bibnamefont{Cao}},
  \bibinfo{author}{\bibfnamefont{K.}~\bibnamefont{Hattori}},
  \bibinfo{author}{\bibfnamefont{M.}~\bibnamefont{Hongo}},
  \bibinfo{author}{\bibfnamefont{X.-G.} \bibnamefont{Huang}}, \bibnamefont{and}
  \bibinfo{author}{\bibfnamefont{H.}~\bibnamefont{Taya}}
  (\bibinfo{year}{2022}), \eprint{2205.08051}.

\bibitem[{\citenamefont{Weickgenannt
  et~al.}(2022{\natexlab{c}})\citenamefont{Weickgenannt, Wagner, Speranza, and
  Rischke}}]{Weickgenannt:2022qvh}
\bibinfo{author}{\bibfnamefont{N.}~\bibnamefont{Weickgenannt}},
  \bibinfo{author}{\bibfnamefont{D.}~\bibnamefont{Wagner}},
  \bibinfo{author}{\bibfnamefont{E.}~\bibnamefont{Speranza}}, \bibnamefont{and}
  \bibinfo{author}{\bibfnamefont{D.~H.} \bibnamefont{Rischke}},
  \bibinfo{journal}{Phys. Rev. D} \textbf{\bibinfo{volume}{106}},
  \bibinfo{pages}{L091901} (\bibinfo{year}{2022}{\natexlab{c}}),
  \eprint{2208.01955}.

\bibitem[{\citenamefont{Biswas et~al.}(2023)\citenamefont{Biswas, Daher, Das,
  Florkowski, and Ryblewski}}]{Biswas:2023qsw}
\bibinfo{author}{\bibfnamefont{R.}~\bibnamefont{Biswas}},
  \bibinfo{author}{\bibfnamefont{A.}~\bibnamefont{Daher}},
  \bibinfo{author}{\bibfnamefont{A.}~\bibnamefont{Das}},
  \bibinfo{author}{\bibfnamefont{W.}~\bibnamefont{Florkowski}},
  \bibnamefont{and} \bibinfo{author}{\bibfnamefont{R.}~\bibnamefont{Ryblewski}}
  (\bibinfo{year}{2023}), \eprint{2304.01009}.

\bibitem[{\citenamefont{Weickgenannt}(2023)}]{Weickgenannt:2023btk}
\bibinfo{author}{\bibfnamefont{N.}~\bibnamefont{Weickgenannt}},
  \bibinfo{journal}{Phys. Rev. D} \textbf{\bibinfo{volume}{108}},
  \bibinfo{pages}{076011} (\bibinfo{year}{2023}), \eprint{2307.13561}.

\bibitem[{\citenamefont{Weickgenannt and
  Blaizot}(2024{\natexlab{b}})}]{Weickgenannt:2023bss}
\bibinfo{author}{\bibfnamefont{N.}~\bibnamefont{Weickgenannt}}
  \bibnamefont{and} \bibinfo{author}{\bibfnamefont{J.-P.}
  \bibnamefont{Blaizot}}, \bibinfo{journal}{Phys. Rev. D}
  \textbf{\bibinfo{volume}{109}}, \bibinfo{pages}{056019}
  (\bibinfo{year}{2024}{\natexlab{b}}), \eprint{2312.05917}.

\bibitem[{\citenamefont{Daher et~al.}(2024)\citenamefont{Daher, Florkowski,
  Ryblewski, and Taghinavaz}}]{Daher:2024bah}
\bibinfo{author}{\bibfnamefont{A.}~\bibnamefont{Daher}},
  \bibinfo{author}{\bibfnamefont{W.}~\bibnamefont{Florkowski}},
  \bibinfo{author}{\bibfnamefont{R.}~\bibnamefont{Ryblewski}},
  \bibnamefont{and}
  \bibinfo{author}{\bibfnamefont{F.}~\bibnamefont{Taghinavaz}},
  \bibinfo{journal}{Phys. Rev. D} \textbf{\bibinfo{volume}{109}},
  \bibinfo{pages}{114001} (\bibinfo{year}{2024}), \eprint{2403.04711}.

\bibitem[{\citenamefont{Wagner et~al.}(2024)\citenamefont{Wagner, Shokri, and
  Rischke}}]{Wagner:2024fhf}
\bibinfo{author}{\bibfnamefont{D.}~\bibnamefont{Wagner}},
  \bibinfo{author}{\bibfnamefont{M.}~\bibnamefont{Shokri}}, \bibnamefont{and}
  \bibinfo{author}{\bibfnamefont{D.~H.} \bibnamefont{Rischke}}
  (\bibinfo{year}{2024}), \eprint{2405.00533}.

\bibitem[{\citenamefont{Drogosz
  et~al.}(2024{\natexlab{a}})\citenamefont{Drogosz, Florkowski, and
  Hontarenko}}]{Drogosz:2024gzv}
\bibinfo{author}{\bibfnamefont{Z.}~\bibnamefont{Drogosz}},
  \bibinfo{author}{\bibfnamefont{W.}~\bibnamefont{Florkowski}},
  \bibnamefont{and}
  \bibinfo{author}{\bibfnamefont{M.}~\bibnamefont{Hontarenko}}
  (\bibinfo{year}{2024}{\natexlab{a}}), \eprint{2408.03106}.

\bibitem[{\citenamefont{Wagner}(2024)}]{Wagner:2024fry}
\bibinfo{author}{\bibfnamefont{D.}~\bibnamefont{Wagner}}
  (\bibinfo{year}{2024}), \eprint{2409.07143}.

\bibitem[{\citenamefont{Drogosz
  et~al.}(2024{\natexlab{b}})\citenamefont{Drogosz, Florkowski, \L{}ygan, and
  Ryblewski}}]{Drogosz:2024lkx}
\bibinfo{author}{\bibfnamefont{Z.}~\bibnamefont{Drogosz}},
  \bibinfo{author}{\bibfnamefont{W.}~\bibnamefont{Florkowski}},
  \bibinfo{author}{\bibfnamefont{N.}~\bibnamefont{\L{}ygan}}, \bibnamefont{and}
  \bibinfo{author}{\bibfnamefont{R.}~\bibnamefont{Ryblewski}}
  (\bibinfo{year}{2024}{\natexlab{b}}), \eprint{2411.06154}.

\bibitem[{\citenamefont{Liang and Wang}(2005)}]{Liang:2004ph}
\bibinfo{author}{\bibfnamefont{Z.-T.} \bibnamefont{Liang}} \bibnamefont{and}
  \bibinfo{author}{\bibfnamefont{X.-N.} \bibnamefont{Wang}},
  \bibinfo{journal}{Phys. Rev. Lett.} \textbf{\bibinfo{volume}{94}},
  \bibinfo{pages}{102301} (\bibinfo{year}{2005}), \bibinfo{note}{[Erratum:
  Phys. Rev. Lett.96,039901(2006)]}, \eprint{nucl-th/0410079}.

\bibitem[{\citenamefont{Voloshin}(2004)}]{Voloshin:2004ha}
\bibinfo{author}{\bibfnamefont{S.~A.} \bibnamefont{Voloshin}}
  (\bibinfo{year}{2004}), \eprint{nucl-th/0410089}.

\bibitem[{\citenamefont{Betz et~al.}(2007)\citenamefont{Betz, Gyulassy, and
  Torrieri}}]{Betz:2007kg}
\bibinfo{author}{\bibfnamefont{B.}~\bibnamefont{Betz}},
  \bibinfo{author}{\bibfnamefont{M.}~\bibnamefont{Gyulassy}}, \bibnamefont{and}
  \bibinfo{author}{\bibfnamefont{G.}~\bibnamefont{Torrieri}},
  \bibinfo{journal}{Phys. Rev. C} \textbf{\bibinfo{volume}{76}},
  \bibinfo{pages}{044901} (\bibinfo{year}{2007}), \eprint{0708.0035}.

\bibitem[{\citenamefont{Becattini et~al.}(2008)\citenamefont{Becattini,
  Piccinini, and Rizzo}}]{Becattini:2007sr}
\bibinfo{author}{\bibfnamefont{F.}~\bibnamefont{Becattini}},
  \bibinfo{author}{\bibfnamefont{F.}~\bibnamefont{Piccinini}},
  \bibnamefont{and} \bibinfo{author}{\bibfnamefont{J.}~\bibnamefont{Rizzo}},
  \bibinfo{journal}{Phys. Rev.} \textbf{\bibinfo{volume}{C77}},
  \bibinfo{pages}{024906} (\bibinfo{year}{2008}), \eprint{0711.1253}.

\bibitem[{\citenamefont{Jaiswal et~al.}(2022)\citenamefont{Jaiswal, Blaizot,
  Bhalerao, Chen, Jaiswal, and Yan}}]{Jaiswal:2022udf}
\bibinfo{author}{\bibfnamefont{S.}~\bibnamefont{Jaiswal}},
  \bibinfo{author}{\bibfnamefont{J.-P.} \bibnamefont{Blaizot}},
  \bibinfo{author}{\bibfnamefont{R.~S.} \bibnamefont{Bhalerao}},
  \bibinfo{author}{\bibfnamefont{Z.}~\bibnamefont{Chen}},
  \bibinfo{author}{\bibfnamefont{A.}~\bibnamefont{Jaiswal}}, \bibnamefont{and}
  \bibinfo{author}{\bibfnamefont{L.}~\bibnamefont{Yan}},
  \bibinfo{journal}{Phys. Rev. C} \textbf{\bibinfo{volume}{106}},
  \bibinfo{pages}{044912} (\bibinfo{year}{2022}), \eprint{2208.02750}.

\bibitem[{\citenamefont{Weickgenannt and
  Blaizot}(2024{\natexlab{c}})}]{Weickgenannt:2023nge}
\bibinfo{author}{\bibfnamefont{N.}~\bibnamefont{Weickgenannt}}
  \bibnamefont{and} \bibinfo{author}{\bibfnamefont{J.-P.}
  \bibnamefont{Blaizot}}, \bibinfo{journal}{Phys. Rev. D}
  \textbf{\bibinfo{volume}{109}}, \bibinfo{pages}{056012}
  (\bibinfo{year}{2024}{\natexlab{c}}), \eprint{2311.15817}.

\bibitem[{\citenamefont{Buzzegoli}(2022)}]{Buzzegoli:2021wlg}
\bibinfo{author}{\bibfnamefont{M.}~\bibnamefont{Buzzegoli}},
  \bibinfo{journal}{Phys. Rev. C} \textbf{\bibinfo{volume}{105}},
  \bibinfo{pages}{044907} (\bibinfo{year}{2022}), \eprint{2109.12084}.

\end{thebibliography}

\end{document}